\documentclass{emulateapj}
\pdfoutput=1
\usepackage{graphicx}
\usepackage{natbib}

\slugcomment{Accepted for publication in The Astronomical Journal}

\shorttitle{PCA analysis of simulated jets}
\shortauthors{Cerqueira et al.}

\begin{document}


\title{Principal Component Analysis of computed emission lines from proto-stellar jets\altaffilmark{1}}


\author{A.H. Cerqueira\altaffilmark{2,3};
        J. Reyes-Iturbide\altaffilmark{2,4,5}; 
        F. De Colle\altaffilmark{5} and M.J.
        Vasconcelos\altaffilmark{2,3}}

\altaffiltext{1}{
{\it This is an author-created, un-copyedited version of an article
accepted for publication in The Astronomical Journal. IOP
Publishing Ltd is not responsible for any errors or omissions in
this version of the manuscript or any version derived from it.}
}

\altaffiltext{2}{Laborat\'orio de Astrof\'{i}sica Te\'orica e
Observacional, Universidade Estadual de Santa Cruz, Rod. Jorge
Amado, km 16, Ilh\'eus, BA, Brazil - CEP 45662-900}

\altaffiltext{3}{Institut de Plan\'etologie et d'Astrophysique de Grenoble,
              Universit\'e Grenoble Alpes, BP, 53, 38041 Grenoble, France}

\altaffiltext{4}{Tecnol\'ogico de Estudios Superiores de Tianguistenco,
Carretera Tenango - La Marquesa Km 22,
Santiago Tianguistenco, Estado de Mexico, Mexico}

\altaffiltext{5}{Instituto de Ciencias Nucleares, Universidad
Nacional Aut\'onoma de M\'exico, Apdo. Postal 70-543, D.F.  Mexico}

\begin{abstract}

A very important issue concerning protostellar jets is the mechanism
behind their formation. Obtaining information on the region
at the base of a jet can shed light into the subject and some years
ago this has been done through a search for a rotational
signature at the jet line spectrum. The existence of such
signatures, however, remains controversial. In order to contribute
to the clarification of this issue, in this paper we show that the
Principal Component Analysis (PCA) can potentially help to distinguish
between rotation and precession effects in protostellar jet
images.  This method reduces the dimensions of the data, facilitating
the efficient extraction of information from large datasets
as those arising from Integral Field Spectroscopy. The PCA transforms
the system of correlated coordinates into a system of uncorrelated
coordinates, the eigenvectors, ordered by principal components
of decreasing variance. The projection of the data on these
coordinates produces images called tomograms, while
eigenvectors can be displayed as eigenspectra. The combined
analysis of both can allow the identification of patterns correlated
to a particular physical property that would otherwise remain hidden,
and can help separating in the data the effect of physically
uncorrelated phenomena. These are for example, rotation and
precession in the kinematics of a stellar jet. In order to show the
potential of the PCA analysis, we apply it to synthetic spectro-imaging
datacubes generated as an output of numerical simulations of
protostellar jets. In this way we generate a benchmark to which a
PCA diagnostics of real observations can be confronted. Using the
computed emission line profiles for [\ion{O}{1}]$\lambda$6300 and
[\ion{S}{2}]$\lambda$6716, we recover and analyze the effects of
rotation and precession in tomograms generated by PCA.  We show
that different combinations of the eigenvectors can be used to
enhance and to identify the rotation features present in the
data.  Our results indicate that the PCA can be useful for
disentangling rotation from precession in jets with an
inclination of the jet with respect to the plane of the sky
as high as $45^{\circ}$. We have been able to recover the
initially imposed rotation jet profile for models at moderate
inclination angle ($\phi \le 15^{\circ}$) and without precession.

\end{abstract}


\section{Introduction}

Recent observations of protostellar jets show, in some cases, a
systematic side-to-side (with respect to the jet axis) shift in
radial velocities in several emission lines in the optical and UV.
\citet{davi00} and \citet{bacc02} reported the first
results of this kind, in a molecular and atomic jet, respectively.
In particular, \citet{bacc02} presented an analysis performed on
high resolution Hubble Space Telescope (HST) long-slit spectra
taken at different positions, both along and across the DG Tau
jet, in the region close to its driving source.  They found a
systematic side-to-side difference in the radial velocity of the
jet with respect to its axis, which they interpreted as evidence
of jet rotation, what was later found to be in agreement with the
sense of rotation of the circumstellar disk of DG Tau \citep{testi02}.

This observational finding has been followed by other studies which
tried to look for the same kind of pattern in other jet sources.
\citet{coff04}, using the Space Telescope Imaging Spectrograph
(STIS) of the HST found similar results for the jets in the T Tauri
stars TH 28, RW Aur and LkH$\alpha$ 321.  Consistently with the
predictions of the magnetocentrifugally driven mechanism for jet
launching \cite[e.g.,][]{blan82, ferr97}, the bipolar collimated
outflows in TH 28 and RW Aur showed the same signs of velocity
asymmetry in both red- and blue-shift lobes, implying the same sense
of rotation in both. However, the sense of the jet rotation in RW
Aur was later inferred to be the opposite of the disk rotation
\citep{cabr06}. Furthermore, recent observations by \citet{coff12}
showed a radial velocity shift of order of $\sim 10$ km/s in the
near-UV, corresponding to a rotation in the opposite sense of that
determined by the optical observation.  In addition, no shift in
velocity was detected in the observation of RW Aur with the same
instrument six months later \citep{coff12}.

The CW Tau and HH 30 jets were observed in the near ultraviolet
(NUV) and optical wavelengths, using STIS by \citet{coff07}, who
also presented new observations of the DG Tau and TH 28 jets. A 
systematic radial velocity shift pattern was observed in the CW Tau
jet spectra in the same range of
velocities, of $\sim 10-20$ km s$^{-1}$  for the
optical shifts, with slightly smaller shifts in the NUV ($\sim 5-10$
km s$^{-1}$), while no significant shift in radial velocity was
observed for the HH 30 jet. For this jet, \citet{pety06} did not
observe also any rotation signature at millimeter wavelengths.  As
noted by, e.g., \citet{cai08}, the lack of any rotational evidence
for HH 30 jet is unexpected since the rotation signature should be
maximum for jets moving in the plane of the sky as is the case of
HH 30 (inclination angle $\lesssim 1^\circ$, \citealt{mund90}).

These optical and NUV observations of a sample of jets have been
followed by near-infrared (NIR) long slit spectroscopy of HH 212
carried out by \citet{corr09}. HH 212 is one of the bona fide
examples of jet symmetry (between both lobes) and the first HH jet
for which evidence of rotation was found \citep{davi00}.  In
their work, \citet{corr09} argued that the combined effect of
rotation and precession of the jet axis would be responsible for
the observed pattern, and that rotation alone would not be able to
account for the data. The HH 212 jet has also been observed by
\citet{code07}, who did not find evidence of velocity gradients
compatible with rotation in SiO observations.

Along with these efforts, attempts have been made to observe rotation
of the jet axis through ground based, world class telescopes.
\citet{coff11} used the Gemini South Telescope to observe HH 34,
HH 111 and HH 212. They found evidences for rotation in some knots
of HH 111 and 212. For HH 111, they noted that the sense of rotation
obtained for the observed knot is in opposite sense to the disk
rotation, while for HH 212 they agree (in both receding and approaching
lobes, and the disk itself).  \cite{coff15} looked for
radial velocity shifts in the RY Tau system, by using
Gemini NIFIS+ALTAIR observations of the jet, combined with radio
observations of the disk using the Plateau de Bure.  Although a
Keplerian rotation pattern for the disk was clearly obtained, the
value of the radial velocity shifts remained below the 3$\sigma$
detection limit. However, the authors considered the obtained radial
velocity values as upper limits and assuming steady state, constrained
the launching region to be below 0.45 AU. In \citet{fend11} a
comprehensive list and a detailed discussion of HH objects for which
there is evidence for jet rotation are presented.

If the rotation interpretation is correct, these studies  open an
important new window to investigate the physics behind the jet
production and launching mechanisms.  The measurement of jet rotation
has implications in the estimate of the angular momentum flux
lost by the system, on the determination of the footpoints of the
jet launching region, and could help to discriminate among different
jet production models (see, for instance, \citealt{pudr04}).  The
comparison between the expected values for the radial velocity
asymmetry, from magnetohydrodynamics (MHD) disk wind models, and
the values obtained for DG Tau by \citet{bacc02} was presented by
\citet{pese04}. They found that both classes of MHD disk wind models,
the so-called cold and warm solutions \citep[see][]{cass00a, cass00b},
are able to adjust the observed trend for DG Tau transverse radial
velocity, but only the warm solution is able to reproduce the
velocity shifts.

The indication of rotation in DG Tau, TH 28 and LkH$\alpha$128 is
also consistent with a wind launched at the innermost part of the
accretion disk. It also reproduces the onion like structure (in
radial velocities) expected for a wind launched by an accretion
disk. \citet{ande03} found for the DG Tau jet, using
\citet{bacc02} data, a launching radius of $r_0 \sim 0.3-4$ AU from
the YSO, thus excluding the X-wind model \citep[e.g.,][]{cai08} as
a jet launching mechanism.  The observational determination of the
launching region is, unarguably, a key piece to advance further and
to gather all the information in a consistent and complete
model.

However, the interpretation of side-to-side velocity shifts
as evidence of rotation is still controversial.  Side to side
asymmetries in the velocity along the main jet axis of order
of 10\% of the observed jet speed would produce an effect on
the radial velocity similar to that observed in the works
cited above.  In this sense, the lack of velocity shifts in HH 30
and the presence of clear side-to-side asymmetries in the electron
and hydrogen densities of some of the jets where the radial velocity
shifts are measured (e.g., Th28, see \citealt{coff08}) are consistent
with the possibility that at least part of the observed radial
velocity shifts  originates from side-to-side asymmetries in the
velocity of the jet along  the direction of the jet propagation.

\citet{cerq06} discussed the possibility that the precession of the
jet axis could also give rise to the same radial velocity asymmetries
observed on jet spectra by, for example, \citet{bacc02}. They also
investigated the combined effect of precession and rotation, which
seem to operate together in the DG Tau jet.  While it is clear that
the models presented by \citet{cerq06} do not apply for jets with
small ($\lesssim 5^{\circ}$) precession angles, as mentioned by
\citet{coff07}, it is also evident that at least for those systems
that show evidence for precession (e.g., DG Tau), its presence
should be considered properly.

We propose here a method that could be used to distinguish between
rotation and precession and in general any asymmetry in the velocity
components of the jet when interpreting a given spectra from a
Fabry-Perot or an Integral Field Unit (IFU)-like data.  For this
purpose, we perform a Principal Component Analysis (PCA) of  synthetic
spectra generated from the numerical simulations performed in
\citet{cerq06}, intended here as idealized reproductions of the
observations. Following this approach, we construct 3D synthetic
datacubes in two relevant emission lines ([\ion{S}{2}]$\lambda$6716
$~$ and [\ion{O}{1}]$\lambda$6300\footnote{We have actually
applied the PCA technique also to datacubes from [\ion{O}{1}]$\lambda$6363
and [\ion{S}{2}]$\lambda$6731 emission lines.  We found, however,
that the results for the two oxygen lines are the same, and that
the results for the two sulfur lines are very similar. Therefore,
throughout the paper we will discuss the results for only one
emission line for each atom.}) and then apply the PCA technique
to split the spectra into its components ranked by the variance.
The main goal of this work is to show how the PCA can help disentangling
the different mechanisms that cause asymmetries at the jet lines
for rotating and precessing jets.

The paper is organized as follows. In Section \ref{sec:2} we discuss
the PCA technique as applied in our numerical simulations, which
are briefly reviewed in Section \ref{sec:num}.  In Section
\ref{apenda} we discuss tomograms and eigenspectra and their physical
interpretation. In Section \ref{sec:res} the results based
on the reconstructed process of the treated datacube are shown and
in Section \ref{sec:con} the main conclusions are presented.
In the Appendix \ref{APP2} we present Tables for eigenvalues and
in the Appendix \ref{apendb} we show how the analysis change when
we take into account noisy data and how we can extract the desired
information through the use of the enhancement factor in PCA.

\section{The PCA technique applied to a datacube}
\label{sec:2}

The Principal Component Analysis (PCA) is a method to extract
information from large, multidimensional datasets, to identify
peculiar patterns in the data that would otherwise remain hidden
or mutually combined. The isolated patterns can in many cases be
associated to physical properties that would remain undetected in
traditional spectro-imaging diagnostics. This is done through the
construction of a new system of ``natural" uncorrelated coordinates
ordered by decreasing variance with respect to the average
image/spectrum.  This new system describes the dataset in a more
efficient way, as the projection of the data on each coordinate
isolates one relevant information.

In the presentation of the PCA technique, we will use the same
formalism  of \cite{stei09}, \cite{ricc11} and \cite{men14}.  These
authors developed a PCA-based tool to investigate the properties
of datacubes obtained from observations made with the Gemini
Multi-Object Spectrograph in its Integral Field Unit mode
(GMOS-IFU) for active galaxies (NGC 4736, NGC 7097 and NGC 3115,
respectively).  First, they use PCA, together with other cleaning
techniques, to remove instrumental spatial fingerprints (see
also \citealt{cerq15}, who applied the same cleaning technique to
GMOS-IFU observations of the HH 111 jet).  As we shall see below,
in the PCA technique a given datacube is firstly reduced to a
particular bi-dimensional array from which one calculates the
covariance matrix. Eigenvectors and eigenvalues are then determined
for this matrix, and ordered by decreasing variance. The eigenvectors
constitute the new orthogonal basis, and the data can be projected
on this basis to form the tomograms.  Each tomogram will highlight
a particular pattern in the data. The {\em tomograms}, that are
bi-dimensional maps which describe the data in an orthogonal basis
of uncorrelated coordinates, can then be interpreted and can reveal
physical properties associated with the obtained pattern.  Tomograms
can be combined linearly to reconstruct cleaned 2D images in the
normal space in which that particular property is now evident.

\subsection{Preparing the data for the PCA analysis}
\label{sec:pca}

\begin{figure}
\centerline{
\includegraphics[scale=0.35]{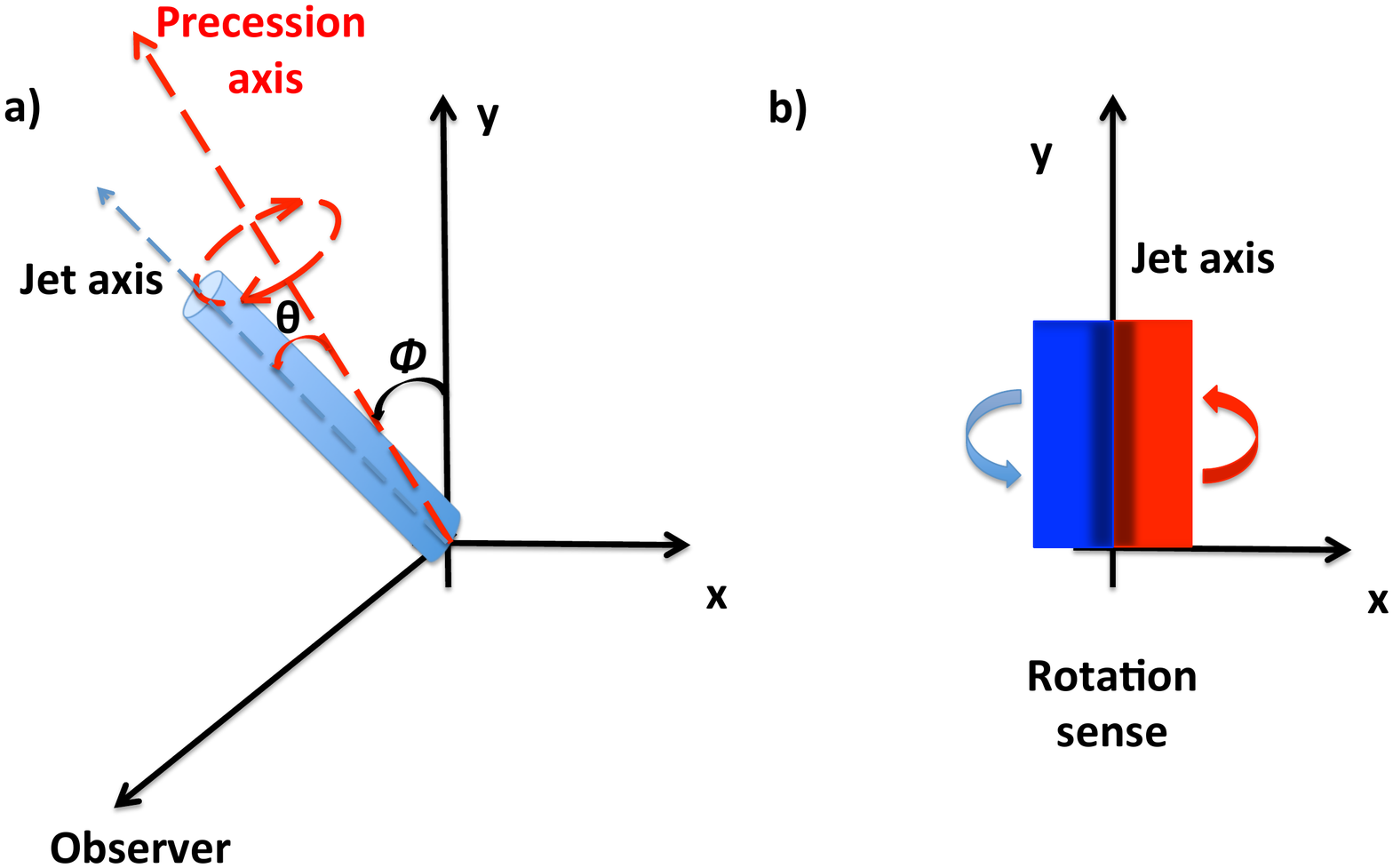}
}
\vskip -1.0truecm
\caption{Sketch of the simulation setup. In a) we show the general
case in which we have a precessing jet observed at a given inclination
angle $\phi$, which is the angle between the precessing axis with
respect to the ``plane of the sky" (the $x-y$ plane), whereas
$\theta$ is the precessing angle of the jet axis measured from the
precessing axis.  The sense of the precession is also shown.  In
b) we show the sense of the jet rotation as seen by the observer
for a rotating jet model in the plane of the sky.}
\label{fig01}
\end{figure}

In the more general case, we simulate a jet that can precess
and rotate (see Figure \ref{fig01}). The axis of precession makes
an angle $\phi$ with respect to the plane of the sky, assumed here
to be the (x,y) plane. The axis of the rotating/precessing jet lies
along a cone that makes an angle $\theta$ with respect to the axis
of precession. A non precessing, but rotating jet has $\theta =0$.
We compute as an output of the numerical simulation the flux of
each emission line arranged in a tridimensional datacube with two
spatial and one spectral dimensions. Each element of the datacube
collects the flux produced in a square region of 5.74 AU (the spatial
resolution of the simulation) in (x,y), in the radial velocity
interval from -400 km s$^{-1}$ to +100 km s$^{-1}$ (corresponding
to a spectral width of $\sim 10$ \AA $~$ for [\ion{O}{1}]$\lambda$6300
and $\sim 11$ \AA $~$ for [\ion{S}{2}]$\lambda$6716), with a spectral
sampling of 10 km s$^{-1}$.  The datacube will have $n = \mu \times
\nu$ = 128$\times$128 pixels in space and $m$ = 50 pixels in
wavelength.  Our data is equivalent to those generated by an IFU:
from the simulation, we can obtain velocity channel maps around a
given emission line for a fixed wavelength. By varying the wavelength,
we obtain a datacube \citep[see][]{cerq06}.

For each value of $m$\footnote {The wavelength is transformed
in radial velocity with respect to the rest frame of the star in
our velocity channel maps. In \cite{stei09}, $m$ refers to the
wavelength, which is equivalent.} we can define an intensity,
$(\mathrm{I}_{ij\lambda})_0$, with $i=1, \dots, \mu$; $j=1, \dots,
\nu$; $\lambda=1, \dots, m$.  The ``0'' index refers to the original
intensity extracted from the data (see \citealt{ricc11} and
\citealt{stei09}), i.e., the original intensity at each velocity
channel map.  Since ${\mu}$ and $\nu$ are the total number of pixels
in the $x$ and $y$ spatial directions we can define the average
spectrum of the datacube as the intensity { averaged} over the two
spatial dimensions:

\begin{equation}\label{ql}
Q_{\lambda}=\frac{1}{n}\sum_{i=1}^{\mu}\sum_{j=1}^{\nu}(I_{ij\lambda})_0 \;,
\end{equation}

\noindent where $n=\mu\times\nu$. This 
mean emission is used to redefine the datacube as:

\begin{equation}\label{intsub}
I_{ij\lambda}=(I_{ij\lambda})_0 - Q_{\lambda} \;,
\label{eq2}
\end{equation}

\noindent which is the excess (or deficit) of flux in comparison
to the average flux at each pixel at a given wavelength.

Since we are looking for patterns of variation in the data,
the subtraction of the average spectra is an important procedure.
As pointed out by \cite{stei09}, this also eliminates components
of emission in the spectra that have null spatial variance at a
given wavelength, as is the case of the sky emission that could be
eventually constant in the observed field in real (i.e., not
synthetic) datacubes.

We can now transform the datacube in an {\em intensity matrix},
${\bf I}_{\beta \lambda}$, of $n$ rows (the spatial pixels) and $m$
columns (the spectral pixels)\footnote{The spatial pixels represent
the \emph{objects} while the spectral pixels their \emph{properties}
\citep[see][]{stei09}.}. The spatial pixel in the intensity
matrix ($\beta$) is related to the datacube spatial pixels
($i$ and $j$) through the relation

\begin{equation}\label{betarel}
\beta = \mu(i-1)+j\,{\rm .}
\label{eq3}
\end{equation}

\noindent This univocal relation is used to i) build up the intensity
matrix (${\bf I}_{\beta \lambda}$) mentioned before and ii) to
recover the datacube intensity for a given projection (see Section
\S \ref{orth} below), which can contain the whole data or just a
fraction of it. In the simulations that we are going to present in
this paper, $\mu=\nu=128$ and $m = 50$ (and so $n = 16\,384$).

The matrix ${\bf I}_{\beta \lambda}$ is then ready to be processed
using the PCA technique. In the next section we discuss the PCA
pipeline as described in \cite{stei09}.  We present here the same
approach but for simulated datacubes. Our aim, as already emphasized,
is to use controlled inputs, including different physical mechanisms
like precession and rotation, to see their imprints in the tomograms.
Here, we want to test  the potential of PCA in disentangling
the effects of rotation and precession in real data of protostellar
jets.

\subsection{An orthogonal basis of uncorrelated coordinates}
\label{orth}

The main goal of the PCA technique is to represent the data in a
new set of mutually orthogonal basis formed finding the
eigenvectors and eigenvalues of the covariance matrix of the modified
dataset ${\bf I}_{\beta \lambda}$, defined as:

\begin{equation}\label{covar}
{\bf C}_{\rm {cov}} = \frac{[{\bf I}_{\beta \lambda}]^{\rm T}
\cdot {\bf I}_{\beta \lambda}}{n - 1} \,{\rm ,}
\end{equation}

\noindent \citep[see][]{stei09} where $[{\bf I}_{\beta
\lambda}]^{\rm T}$ is the transpose of the intensity matrix, ${\bf
I}_{\beta \lambda}$. The covariance matrix has $m$ eigenvalues,
representing the variance of the data for the associated eigenvector
$E_k$ (where $k$ is the order of the eigenvector), that can be
ranked in the new basis with decreasing value of the variance, i.e.
from the largest ($\Lambda_1$) to the smallest ($\Lambda_m$) variance,
as just the ones corresponding to the largest variances will contain
relevant information. The variances can be represented in
a normalized way in terms of percentages, and they are the diagonal
elements of the covariance matrix; the sum of all variances must
return $\sum_{k=1}^m \Lambda_k=1$ (or 100\%).

The eigenvectors represent a new orthogonal basis in which the data
can be described.  These eigenvectors are used to build {\it the
characteristic matrix}, ${\bf E}_{\lambda k}$, which has $m$
eigenvectors in its columns, sorted by decreasing variance. In this
new orthogonal basis, the data can be decomposed as:

\begin{equation}\label{tbk}
{\bf T}_{\beta k}={\bf I}_{\beta \lambda} \cdot {\bf E}_{\lambda k} \,,
\end{equation}

\noindent where ${\bf T}_{\beta k}$ is the intensity matrix expressed
in the new basis. In this basis, the covariance matrix ${\bf D}_{\rm
{cov}}$, defined as:

\begin{equation}
{\bf D}_{\rm {cov}} = \frac{[{\bf T}_{\beta k}]^{\rm T} 
\cdot {\bf T}_{\beta k}}{n - 1} \,{\rm ,}
\end{equation}

\noindent is diagonal by construction (orthogonality property),
which means that it has null covariance between different
coordinates and its diagonal elements are the eigenvalues.
Since the basis is orthogonal, the projection ${\bf T}_{\beta k}$
on the space of eigenvectors represents a pattern that is {\em
uncorrelated} for each $k$, and one can refer to uncorrelated
physical properties that can be separated and recognized.

A way to visualize the projection ${\bf T}_{\beta k}$ of the
data onto the new basis is to produce from it 2D images called {\em
tomograms}. The tomogram is the quantity ${\bf T}_{i,j,k}$ obtained
by ${\bf T}_{\beta k}$ by selecting a given $k = k^{\prime}$ and
by {\em unpacking} $\beta$ into the two indexes $i, j$ by inverting
equation (\ref{betarel}). It is important to emphasize that the
tomograms are images of the 3D data in the new system of coordinate.
As in \cite{stei09}, the tomogram for the $k-$eigenvector is an
image that arises when we use equations (\ref{betarel}) and (\ref{tbk})
for a given $k$, to assign a value for the intensity for each pixel.
The generated image will have the same spatial dimension of the
original one.  Also, it will be independent of $\lambda$ (or the
radial velocity), since this ``index" is summed up in equation
(\ref{tbk}). Each tomogram corresponds to an {\em eigenspectrum},
$E_k$, which are essentially the components of the eigenvector plotted
against the wavelength (or radial velocity). The tomogram should
be analyzed together with its associated eigenspectrum, which can
help to identify the physical process behind the observed patterns.
In the Section \ref{apenda} we present and discuss several tomograms
and eigenspectra.

The datacube in the real space $(i,j,\lambda)$ can be reconstructed
by inverting the equation (\ref{tbk}):

\begin{equation}\label{rec1}
{\bf I}_{i,j,\lambda} \equiv {\bf I}_{\beta \lambda}={\bf T}_{\beta k}
\cdot[{\bf E}_{\lambda k}]^T \,.
\end{equation}

\noindent where we have used also the equation (\ref{betarel}).

The characteristic matrix, ${\bf E}_{\lambda k}$, can be
manipulated to keep, for example, just the eigenvectors of higher
relevance, suppressing the remaining ones.  The usefulness of such
a reconstruction is the possibility to obtain a datacube in which
some physical properties can be emphasized, and some undesirable
features presented in the data, like instrumental fingerprints or
even noise can be removed \citep{stei09,men14,cerq15}.  This procedure
is well documented in \cite{stei09}, but we will rewrite their
equation (8) here for the sake of clarity:

\begin{equation}\label{steiner8}
{\bf I}^{\prime}_{\beta \lambda}(\le k_{\rm max})={\bf T}_{\beta k} (\le k_{\rm max})
\cdot[{\bf E}_{\lambda k}(\le k_{\rm max})]^T \,.
\end{equation}


\begin{table}
\begin{center}
\caption{Simulated models}\label{tab1}
\bigskip
\begin{tabular}{cccccc}
\tableline
\tableline
Model  &  $\tau_p$\tablenotemark{1} & $\tau_{\theta}$\tablenotemark{2}  & ${\theta}$\tablenotemark{3}   & Rotation? & Notes \\
       &      (yr) &             (yr) & ($^{\circ}$) &                   \\
\tableline

M1     & 8 & - & - & no & Reference model\\
M2     & 8 & 8 & 5 & no & Precessing \\
M3     & 8 & - & - & yes & Rotating  \\
M4     & 8 & 8 & 5 & yes & Precessing/rotating\\
\tableline
\end{tabular}
\tablenotetext{1}{$\tau_p$ is the pulsation period of the jet variability.}
\tablenotetext{2}{$\tau_{\theta}$ is the period of precession of the jet axis.}
\tablenotetext{3}{$\theta$ is the angle the jet makes with the precession axis (the precession angle).}
\end{center}
\end{table}

\noindent In this equation, $k_{\rm max}$ is the limit of the relevance
of the eigenvector that the user wants to keep in both, the
characteristic matrix, ${\bf E}_{\lambda k}(\le k_{\rm max})$, and in the
data matrix (in the new coordinate system), ${\bf T}_{\beta k} (\le
k_{\rm max})$, in order to construct the new datacube ${\bf I}^{\prime}_{\beta
\lambda}(\le k_{\rm max})$. In general, we can discard a number of high order
eigenvectors that, gathered together, contribute to a small fraction
of the variance in the dataset. The method that we will use in order
to find the number of eigenvectors we must consider in the analysis,
which means all of the first eigenvectors with $k \le k_{\rm max}$, is the
{\em scree test} \citep[see the Section \ref{apenda}; see
also][]{stei09}.  It is among those eigenvectors limited by the
scree test that we will search the presence of signatures of
precession and rotation.  Once found, these properties can be
emphasized in the datacube reconstruction process: instead of
limiting the relevance as suggested by equation (\ref{steiner8}),
that will only discard the less relevant eigenvectors in the
reconstruction procedure in order to produce a ``cleaned" datacube,
we can pick only those that contain a given physical property and
proceed with the reconstruction process (see the Appendix \ref{apendb},
Section \ref{enhan} for more details). We will explore this in the
Section \ref{sec:res}, where we will discuss reconstructed datacube
putting in evidence properties like the jet precession and/or
rotation.

As a final comment, the mean spectrum that has been subtracted
from the datacube ($Q_{\lambda}$) can be reincorporated in the
reconstructed datacube to recover the flux-calibrated data. The
final product is a sequence of images (as a function of wavelength)
in the real space:

\begin{equation}\label{steiner9}
(I^{\prime}_{ij \lambda}(\le k_{\rm max}))_0= I^{\prime}_{ij \lambda}(\le k_{\rm max})
+Q_{\lambda}
\end{equation} 

\noindent from which one can reconstruct 2D images 
integrated in wavelength or select spectra from a given 
position in the images.

We will apply this formalism to our numerical simulations in the
next section.  We use PCA to mine data from numerical simulations
as has been done for instance by \cite{heyer97}, \cite{brunt09}
and \cite{carr10}.  We will show that rotation and/or precession
can appear in independent eigenvectors, and the PCA technique can
therefore be used in the study of jet rotation to disentangle, in
real data, signatures of rotation from other physical mechanisms,
such as shocks and velocity asymmetries.

\section{The numerical simulations}
\label{sec:num}

We apply the PCA analysis to the study of rotation in stellar jets
by using synthetic emission maps generated by numerical simulations.
The simulations presented here are the same shown previously in
\citet{cerq06}. The simulations were performed with the 3D hydrodynamic
code YGUAZ\'U-a \citep[e.g.][]{raga00}.  Briefly, we run 4
models of jets with pulsation period of $\tau_p = 8$ years
(see Table~\ref{tab1} and Figure \ref{fig01}).  The M1 model,
our \emph{reference jet} model, is intermittent, with a jet
velocity profile given by:

\begin{equation}
\label{jetvel}
v_j=v_0\bigg[ 1+A{\rm sin} \bigg( \frac{2 \pi}{\tau_p}t \bigg)
\bigg] \,,
\end{equation}

\noindent where $v_0=300$ km s$^{-1}$ and $A=0.33$. As discussed
in \citet{cerq06}, these values were chosen in order to reproduce
the physical conditions of the DG Tau jet. Besides pulsation (
model M1), we include precession in M2 model, rotation
in M3 model and precession and rotation in
M4 model. For models with precession we adopt a period of
precession equal to $\tau_{\theta} = 8$ yr and an angle of 
precession $\theta=5^{\circ}$.  The models with rotation have a
rotational velocity profile given by:

\begin{equation} \label{vrot}
v_{\phi}=8\;{\rm km}\;{\rm s}^{-1} \frac{R_j}{R}\,,
\label{eq:vphi}
\end{equation}

\noindent for $R > 0.15 R_j$, where $R_j$ is the jet radius
(1$R_j = 37.4$ AU) and $R$ is the cylindrical
radius. In order to avoid a singularity at the jet axis, the
rotational velocity is constant for $R \le 0.15R_j$, and equal to
$v_{\phi} = 55$ km s$^{-1}$.

\begin{figure}
\centerline{
\includegraphics[scale=0.37]{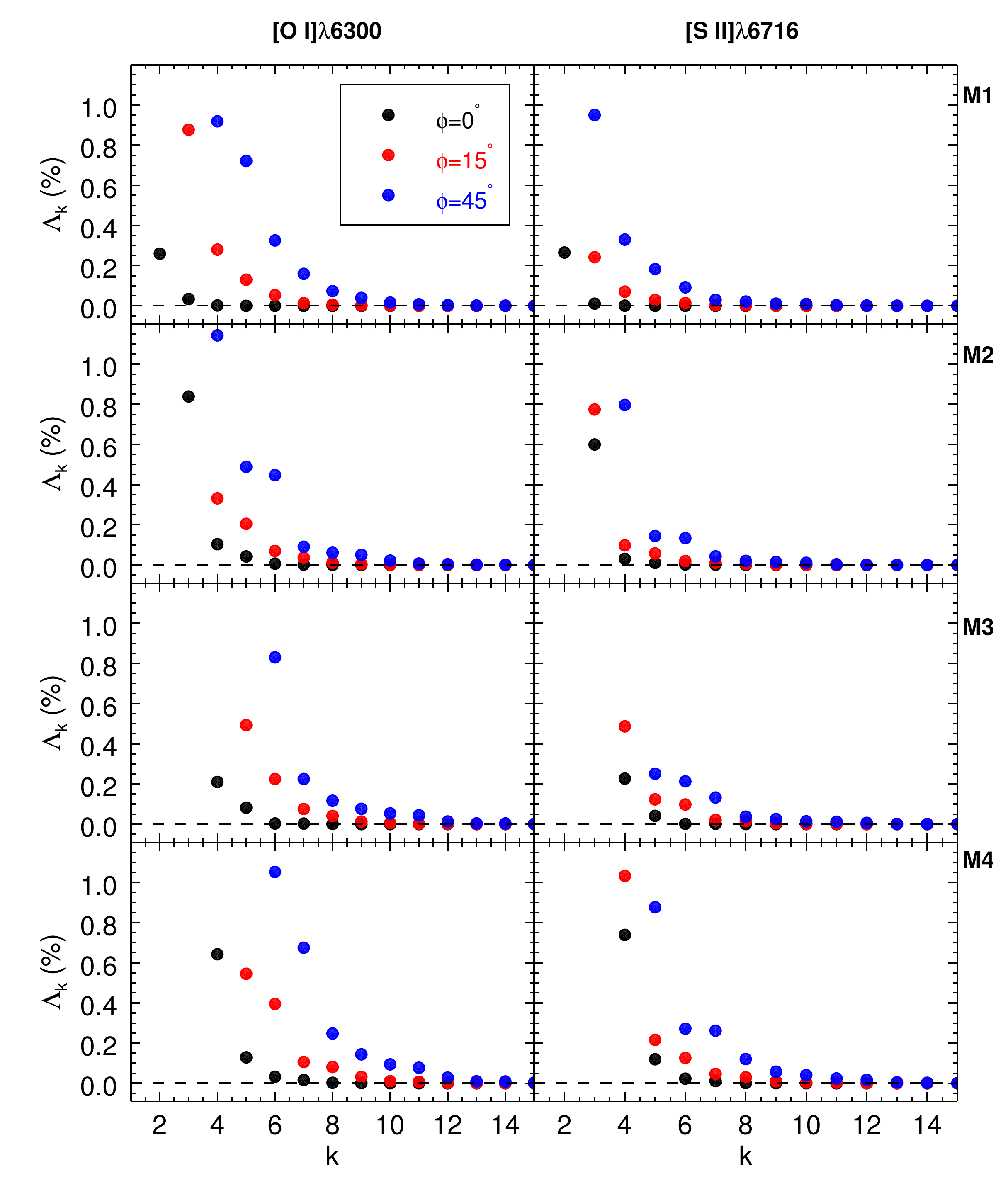}
}
\caption{Eigenvalues (in \% of the variance) as a function
of eigenvector's order for models M1 to M4 (from top to bottom,
respectively), for [\ion{O}{1}]$\lambda$6300 (left) and
[\ion{S}{2}]$\lambda$6716 (right) emission lines.  In each panel
we show also the different inclinations considered: $\phi = 0^{\circ}$
(black dots), $\phi = 15^{\circ}$ (red dots) and $\phi = 45^{\circ}$
(blue dots). The ordinate has been limited to 1.2\% and the first
eigenvalues are out of scale. However, they are given in Table \ref{taba1},
where we also provide all values until $k = 10$.
In each panel we have a horizontal, dashed black line, that was used to indicate
the $k_{\rm max}$ for each curve (see the text for discussion).}\label{fig02}
\end{figure}

The computational domain is a three-dimensional Cartesian box
with spatial dimensions of $x = z = 128$ and $y = 512$ pixels.  Each
pixel has a physical dimension of 5.74 AU.  We will present maps
only for $i \le 128$ and $j \le 128$\footnote{$i$ and $j$ are indexes
for the $x$ and $y$ computational cells, respectively.}, since we
are interested in the region near the jet inlet (the jet injection
point at the computational domain).

\begin{figure}
\centerline{
\includegraphics[scale=0.30]{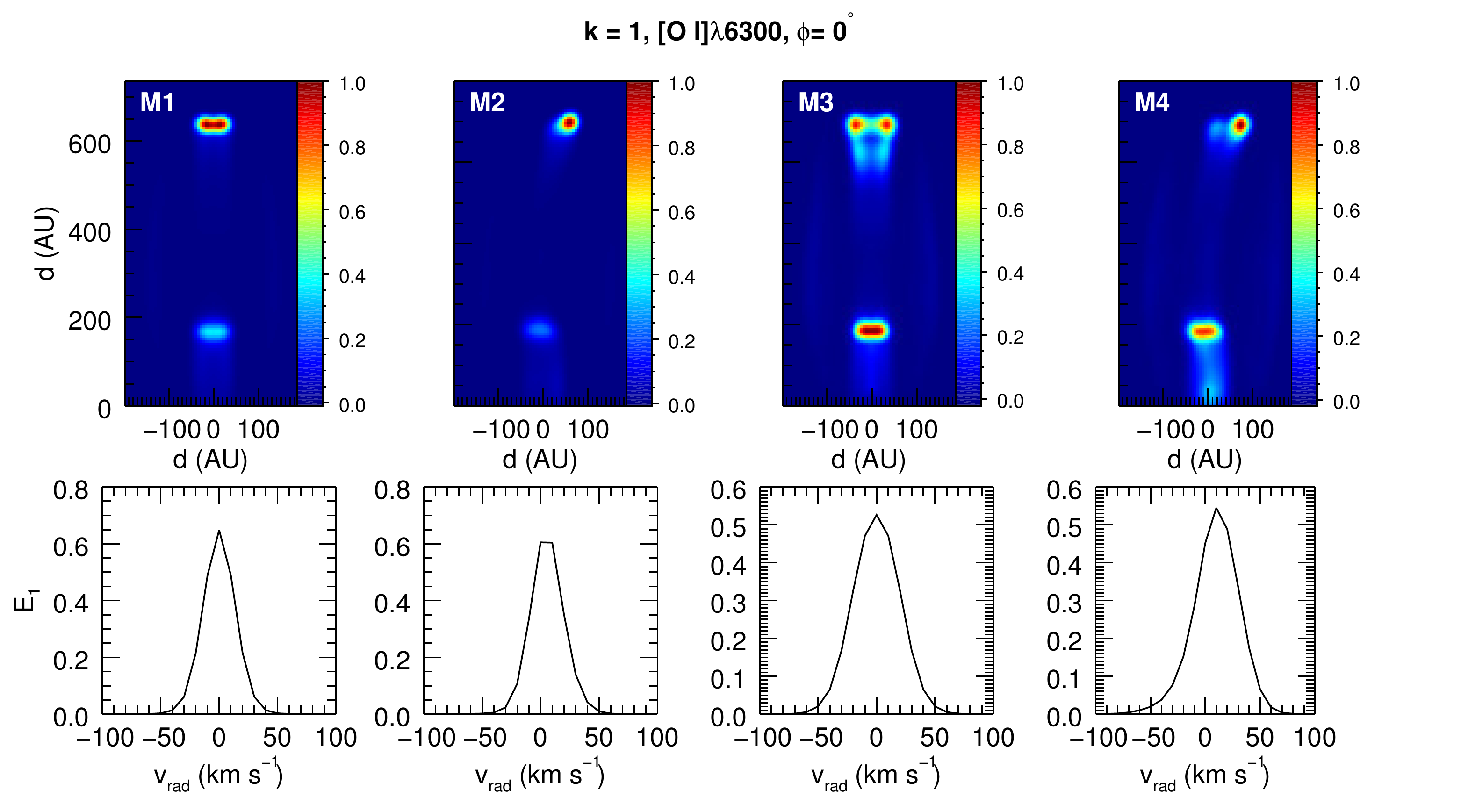}
}
\centerline{
\includegraphics[scale=0.30]{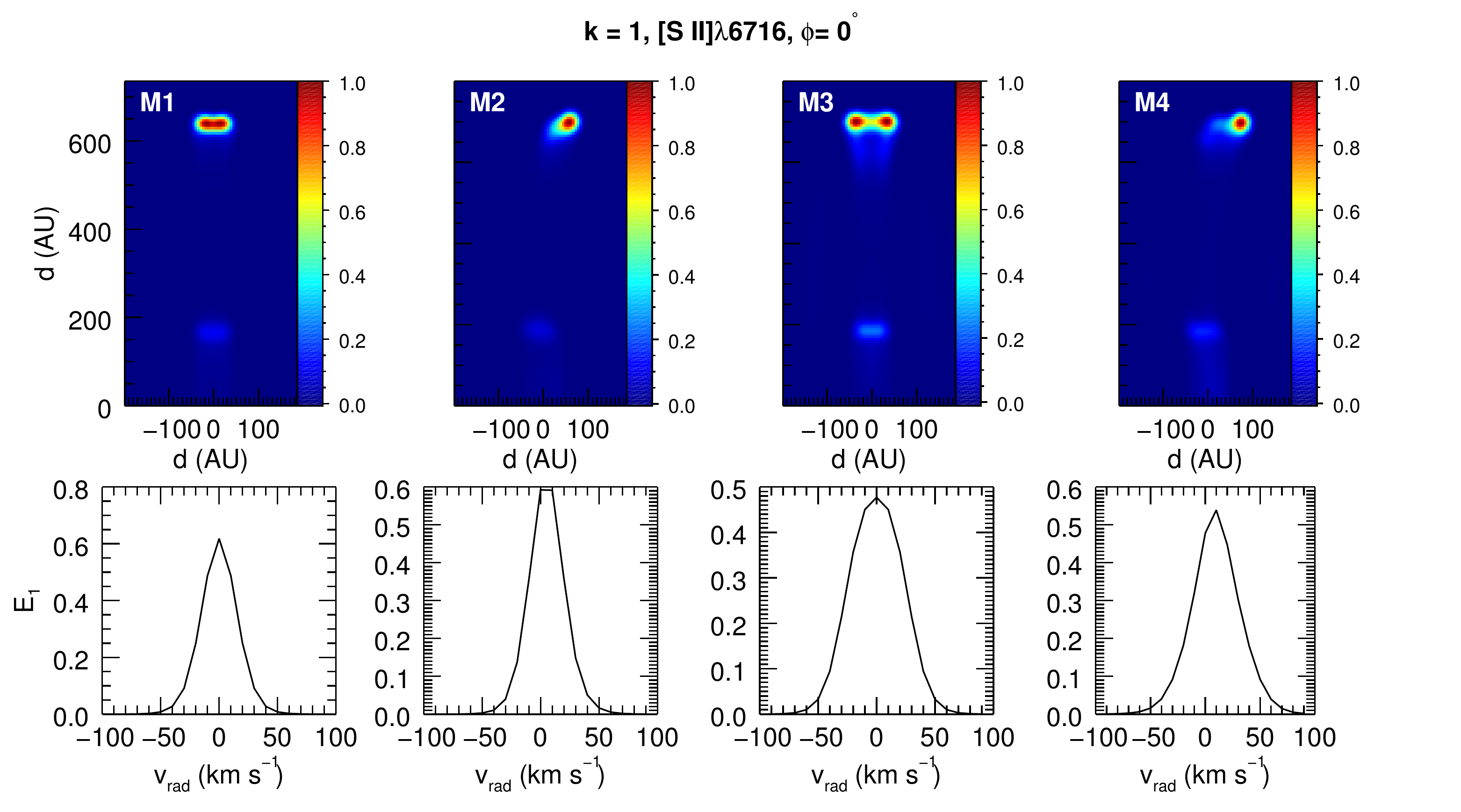}
}
\caption{
First tomogram (${\bf T}_{i,j,k}$, for $k = 1$)
and eigenspectrum ($E_k$, for $k=1$) for models M1 to M4 (from left
to right, respectively), for [\ion{O}{1}]$\lambda$6300 (two
topmost panel's sequence) and [\ion{S}{2}]$\lambda$6716 (two bottommost
panels) emission lines at $\phi = 0^{\circ}$. In each tomogram
the distance from the jet inlet, in AU, is indicated. The color bar
indicate the level of the (normalized) intensity. The
eigenspectrum of each model is plotted below  the respective
tomogram.}\label{fig03}
\end{figure}

Table \ref{tab1} summarizes the main parameters of the models.
To build the intensity matrix (equation \ref{intsub}) we construct
velocity channel maps (VCM) around a given wavelength (see
\citealt{cerq06} for the details).  For the forbidden emission lines
[\ion{O}{1}]$\lambda\lambda$6300,6363 and
[\ion{S}{2}]$\lambda\lambda$6716,6731\footnote{As we have already
mentioned and discussed, results will be presented only for the
[\ion{O}{1}]$\lambda$6300 and for the [\ion{S}{2}]$\lambda$6716
emission lines.} we calculate VCM from $v_{rad} = -400$ km s$^{-1}$
to +100 km s$^{-1}$, with a sampling of 10 km s$^{-1}$.  Each channel
map is one of the $m$ slices in $\lambda$ of the datacube, and we
can build the intensity matrix $(I_{ij\lambda})_0$ for each $\lambda$
(or, alternatively, $m$ or $v_{rad}$).

In order to consider the effect of finite angular resolution
we convolved each VCM with a Gaussian PSF with FWHM of $\sim 5$
pixels ($\sim$ 28.7 AU), or $0\farcs21$ for a jet at a distance
of Taurus \citep[140 pc; see][]{ken94}, for example.


\section{Eigenvalues, tomograms and eigenspectra}
\label{apenda}

\subsection{Eigenvalues and the scree test.}
\label{screet}

\begin{figure}
\centerline{
\includegraphics[scale=0.30]{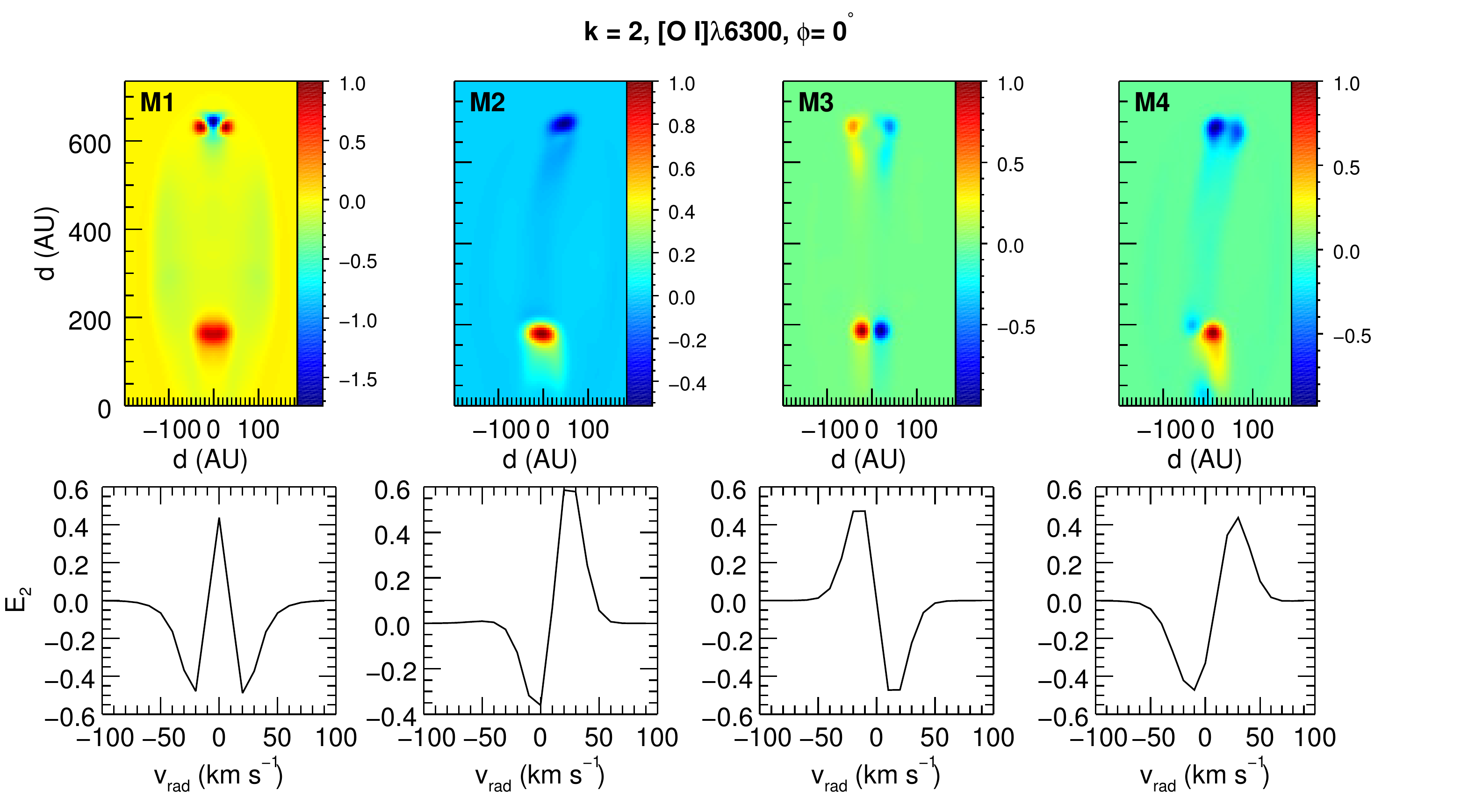}
}
\centerline{
\includegraphics[scale=0.30]{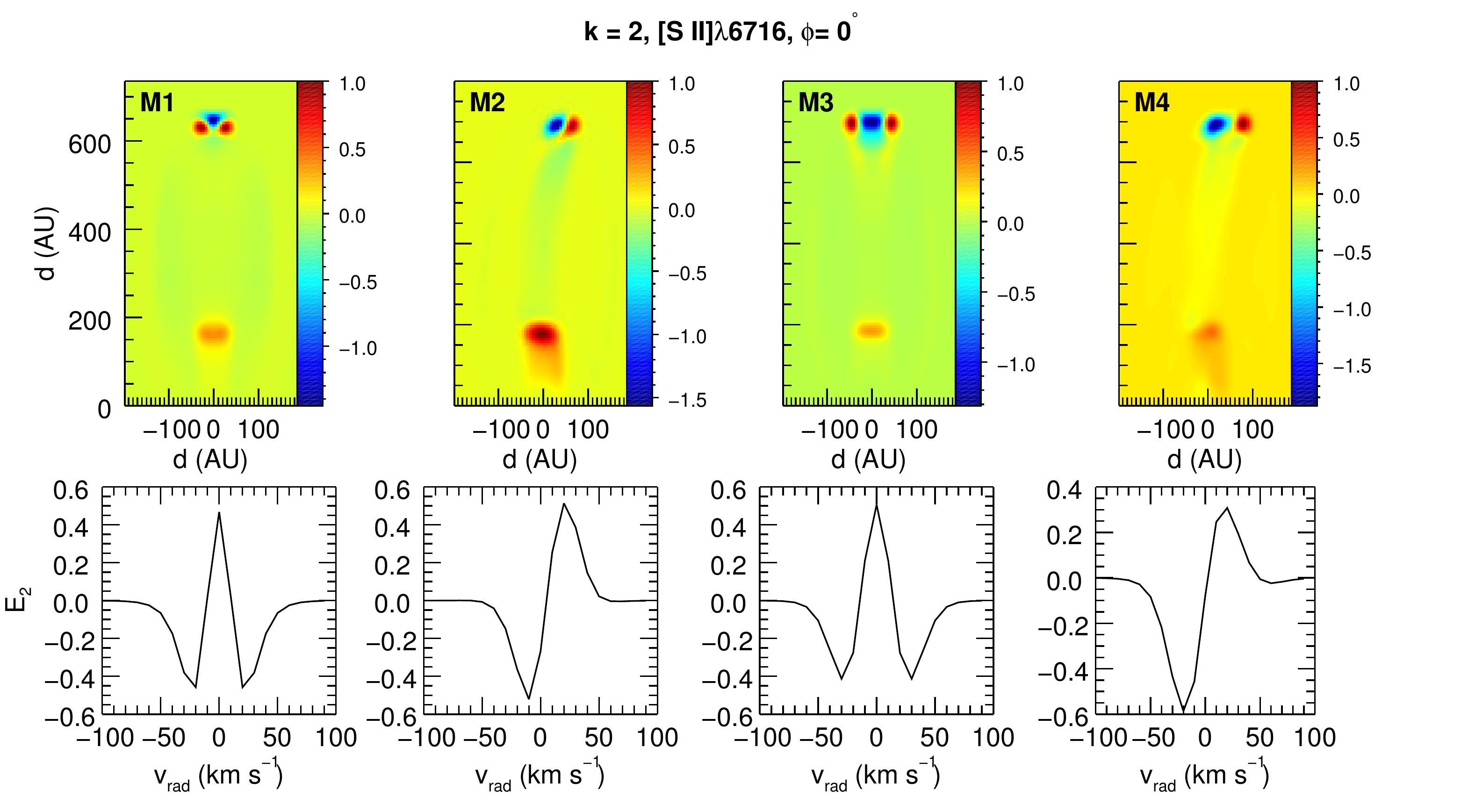}
}
\caption{The same as in Figure \ref{fig03}, but for
$k = 2$ (the second tomogram, ${\bf T}_{i,j,k}$, for $k = 2$,
and its respective eigenspectrum, $E_k$, for $k=2$).}\label{fig04}
\end{figure}

In the PCA, it is important to know how many eigenvectors one
must consider in the analysis. One way to determine this is to apply
the {\em scree test} \citep{stei09}, which is a {\em visual} test
that is used to find the order $k$ where the eigenvalue levels off.
To look at these $k$'s for our different models, emission lines and
inclination angles with respect to the plane of the sky, we built
diagrams of eigenvalues $\Lambda_k$ as a function of $k$.  In Figure
\ref{fig02} we show the eigenvalues for models M1, M2, M3 and M4
(from top to bottom, respectively), for [O I]$\lambda$6300 (left)
and [S II]$\lambda$6716 (right). The dots of different colors are
associated with a specific inclination angle: black for $\phi =
0^{\circ}$; red for $\phi = 15^{\circ}$; blue for $\phi = 45^{\circ}$.
We note from Figure \ref{fig02} that the eigenvalues go
to $\sim 0$ after a given $k = k_{\rm max}$. In order to help to find
these $k_{\rm max}$, we have drawn a horizontal line
in each plot at a constant $\Lambda_{{\rm threshold}} = 10^{-3}$ \%. The
first point (for each model, inclination and emission line) intercepted
by this horizontal line defines the $k_{\rm max}$.
In Appendix \ref{APP2} we present in Table \ref{taba1} the first
ten eigenvalues for each one of the models. We present also the
number of relevant eigenvectors as determined by the criteria of
the scree test (see Table \ref{taba5}). This means that for each
combination of parameters (model, emission line and inclination
angle) we have an already predefined number of eigenvectors that
we must look in searching for relevant uncorrelated phenomena.
However, we used higher order eigenvectors to perform a complete
analysis of the problem, although they will not be presented here
due to the lack of space.

\subsection{Tomograms and eigenspectra}
\label{tomsandspec}

\begin{figure}
\centerline{
\includegraphics[scale=0.30]{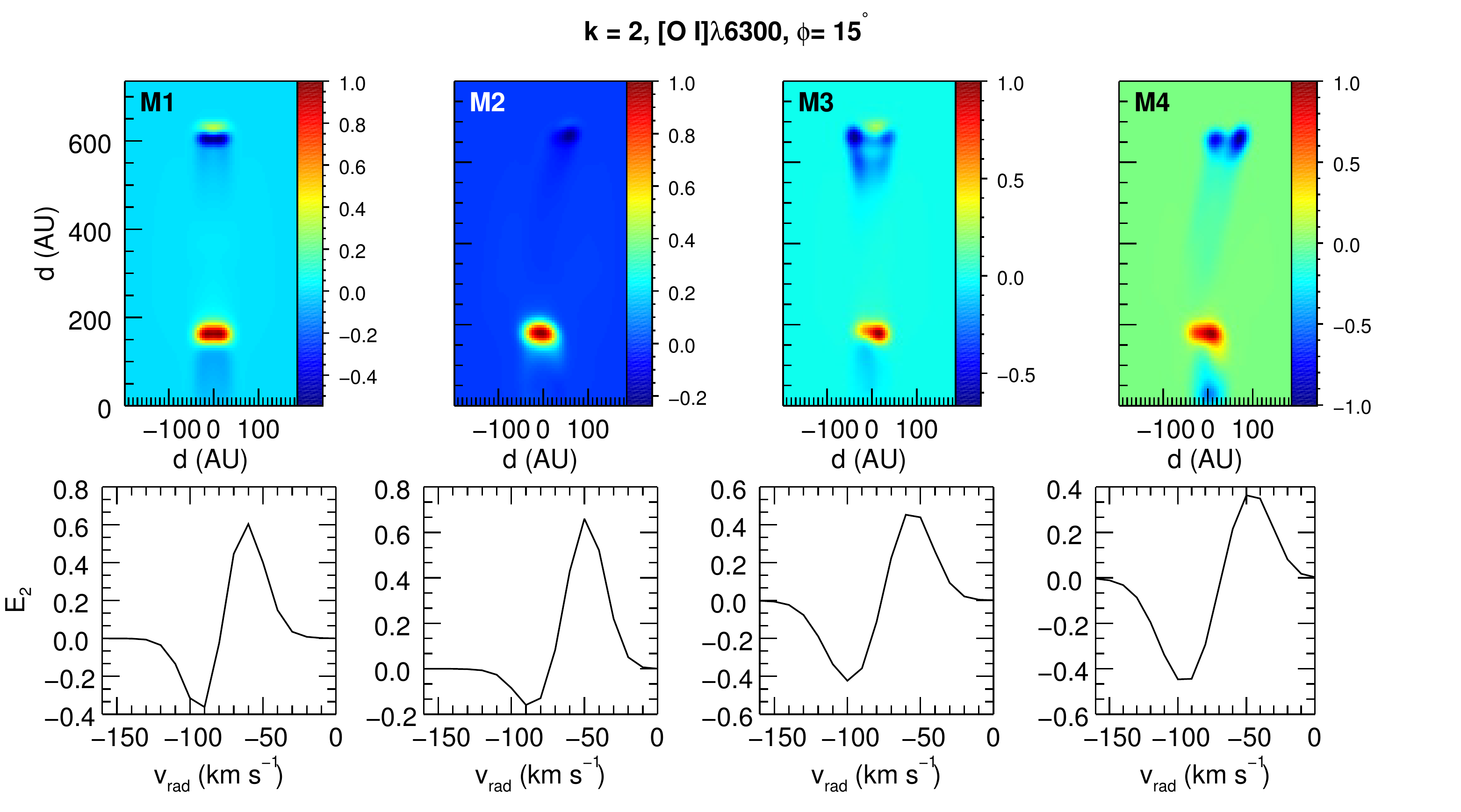}
}
\centerline{
\includegraphics[scale=0.30]{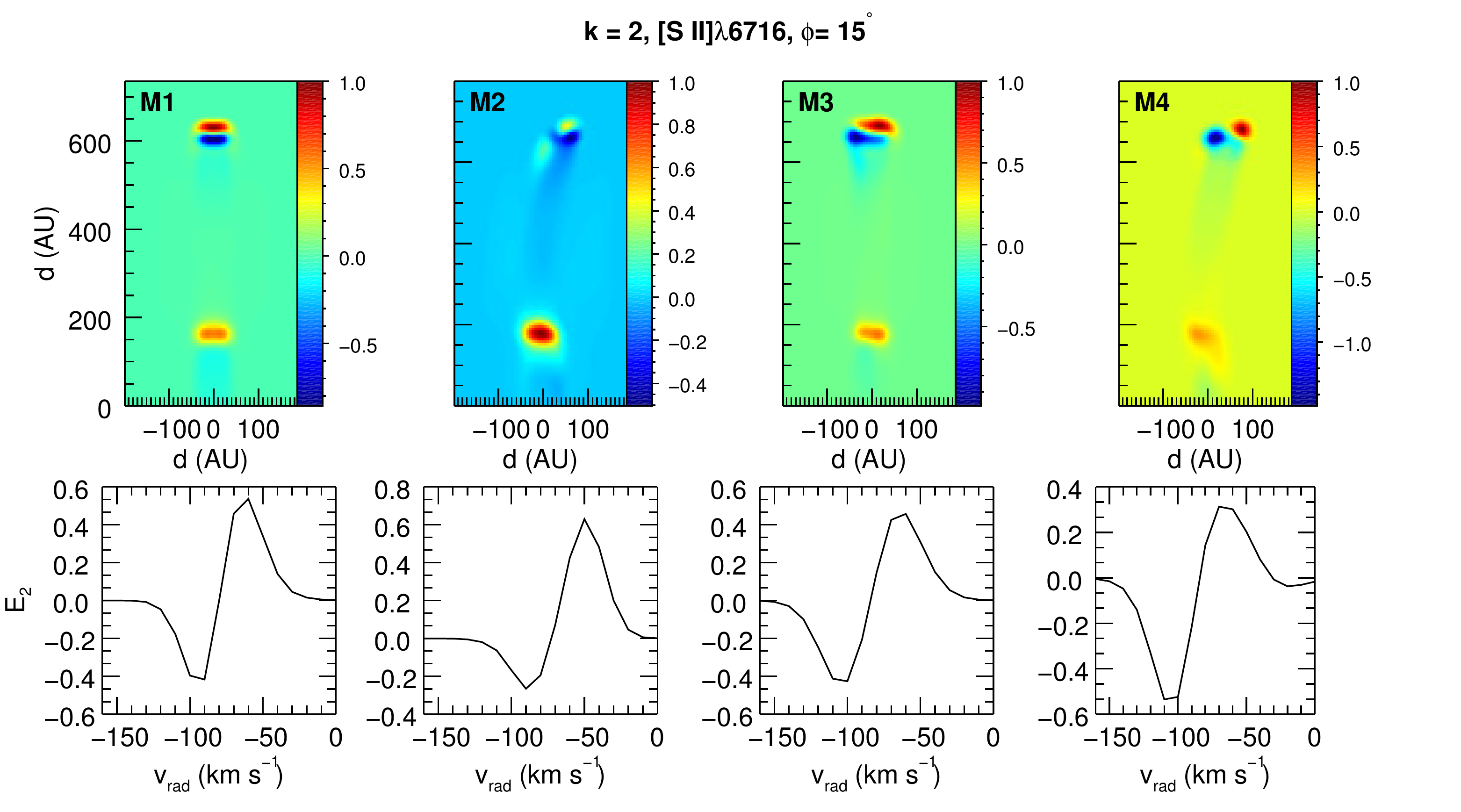}
}
\caption{
Second tomogram (${\bf T}_{i,j,k}$, for $k = 2$)
and eigenspectrum  ($E_k$, for $k=2$) for models M1 to M4 (from left
to right, respectively), for [\ion{O}{1}]$\lambda$6300 (two
topmost panel's sequence) and [\ion{S}{2}]$\lambda$6716 (two bottommost
panels) emission lines at $\phi = 15^{\circ}$. In each tomogram
the distance from the jet inlet, in AU, is indicated. The color bar
indicate the level of the (normalized) intensity.
The eigenspectrum of each model is plotted below the respective
tomogram.}\label{fig05}
\end{figure}

To obtain an image in the new coordinate system, or tomogram
(or also an {\em eigenimage}; see \citealt{heyer97}), we use equations
(\ref{betarel}) and (\ref{tbk}). A column in the data matrix (see
equation \ref{tbk}) can be then transformed in a image, that is the
projection of the datacube (in the new coordinate system) onto the
chosen eigenvector (of order $k$).  The eigenvector, on the other
hand, is a linear combination of the original (spectral) coordinates,
and its coefficients, or weights, can be both positive or
negative\footnote{The only constraint is that their sum in quadrature
must be equal to the unity, because the eigenvectors are normalized.}.
Viewed as a plot of weight versus radial velocity, the eigenvector
is called {\em eigenspectrum}.  Therefore, a tomogram can displays
either positive or negative {\em regions}, that are related with
the weights in the corresponding eigenspectrum.  This will allow
us to investigate both and identify some kinematical ``properties",
like jet precession or rotation.

\subsubsection{Jet or precessing axes in the plane of the sky ($\phi = 0^{\circ}$)}
\label{phi0}

\begin{figure}
\centerline{
\includegraphics[scale=0.35]{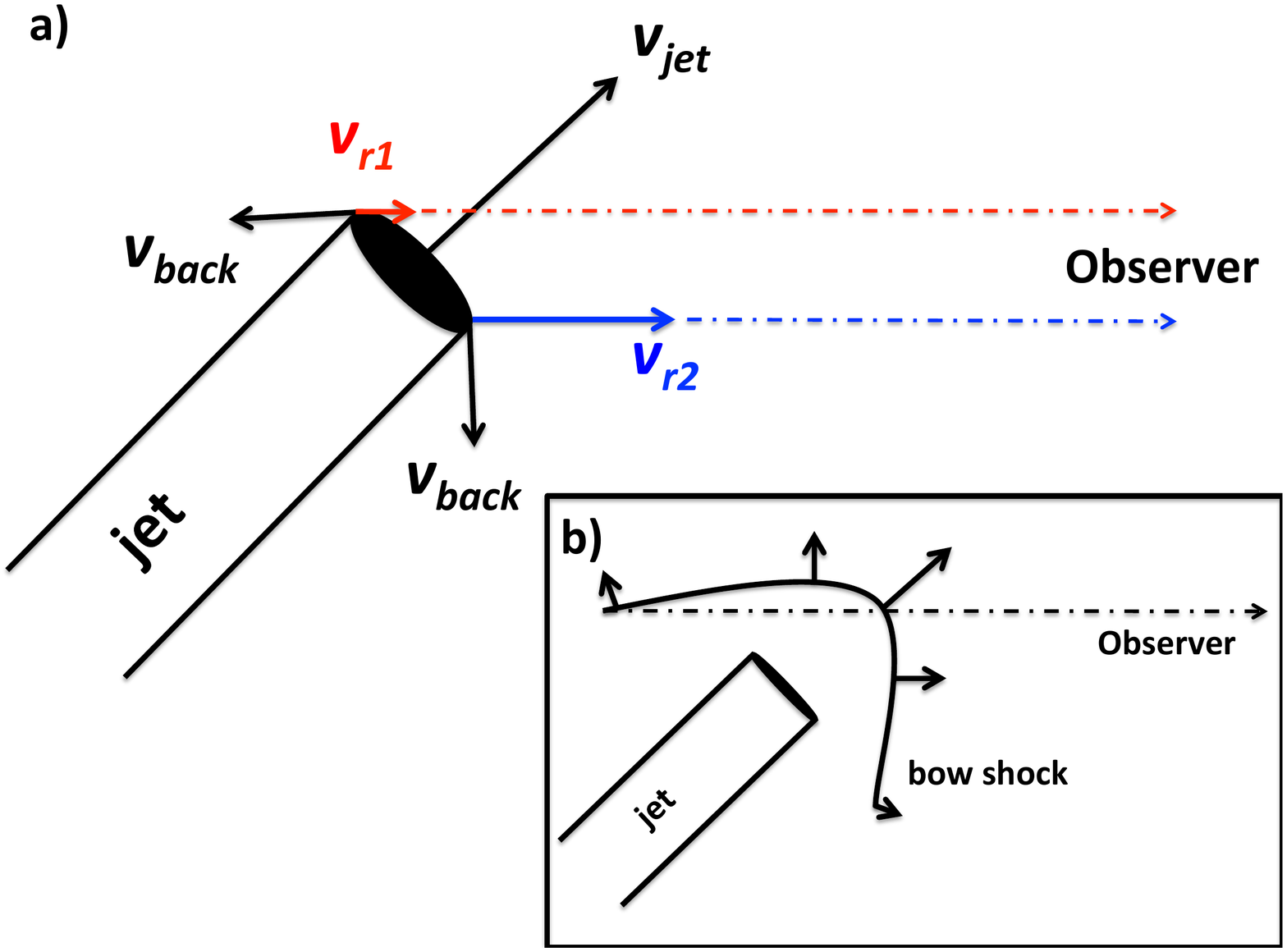}
}
\caption{a) Geometry for a jet inclined towards
the observer. Lateral backflowing material with velocities
$v_{back}$ compose with the jet velocity and favors
high radial velocities in the half bottom of the jet
cross section (as seen by the observer; indicated by the
blue arrow and with velocity $v_{r2}$) in comparison with
the half top part of the jet cross section (red arrow with
velocity $v_{r1}$). Both are blueshifted velocities but appear
in the tomogram of the model M1 in of Figure \ref{fig05}
as a red and blue regions. b) An inclined bow shock can
also explain the tomograms/eigenspectra of the model M1
and cannot be ruled out in the case of the internal
working surface, where the internal bow shock propagates
faster than the leading one due to the non-stationarity
of the ambient medium ahead of it.}\label{fig06}
\end{figure}

In Figure \ref{fig03} we show the first tomogram ($k = 1$) and
its associated eigenspectrum for models M1 to M4 (from left to
right), for [\ion{O}{1}]$\lambda$6300 (top) and [\ion{S}{2}]$\lambda$6716
(bottom) emission lines, at $\phi = 0^{\circ}$. The color bar in
the tomograms (from here and after) represent a non-dimensional
quantity, since pixel's intensities have been normalized to the
maximum value in each tomogram. Distances, in AU, are indicated in
the ordinate of the leftmost tomogram, as well as in each abscissa.
For the eigenspectra, the weights are plotted as a function of the
radial velocity in km s$^{-1}$. These $k = 1$ eigenvectors for each
model (see Table \ref{taba1}) present tomograms that are similar
to an integrated image in the respective emission line. They emphasize
the presence of the two working surfaces: one internal near to the
jet inlet, and the leading one at the tip of the jet, where most
of the emission comes from the radiative losses behind the shocks.
This eigenvector traces the most important shocks in the system.
The respective eigenspectra peak at $\sim$ 0 km s$^{-1}$ in the
case of models M1 and M3, and at $\sim$ 10 km s$^{-1}$ in the
precessing cases M2 and M4 (M4 also has rotation besides precession).

In Figure \ref{fig04} we show the second tomogram ($k = 2$)
and its associated eigenspectrum for models M1 to M4 (from left to
right), for [\ion{O}{1}]$\lambda$6300 (top) and [\ion{S}{2}]$\lambda$6716
(bottom) emission lines, at $\phi = 0^{\circ}$.

For M1 model (panels in the first column), this eigenvector
contributes to 0.2597\% and 0.2655\% of the variance (for [\ion{O}{1}]
and [\ion{S}{2}], respectively; see Table \ref{taba1}) and traces
velocity gradients at the jet head (for both emission lines; their
tomograms and eigenspectra are equivalent) that are perpendicular
to the jet axis, but symmetric with respect to it.

\begin{figure}
\centerline{
\includegraphics[scale=0.30]{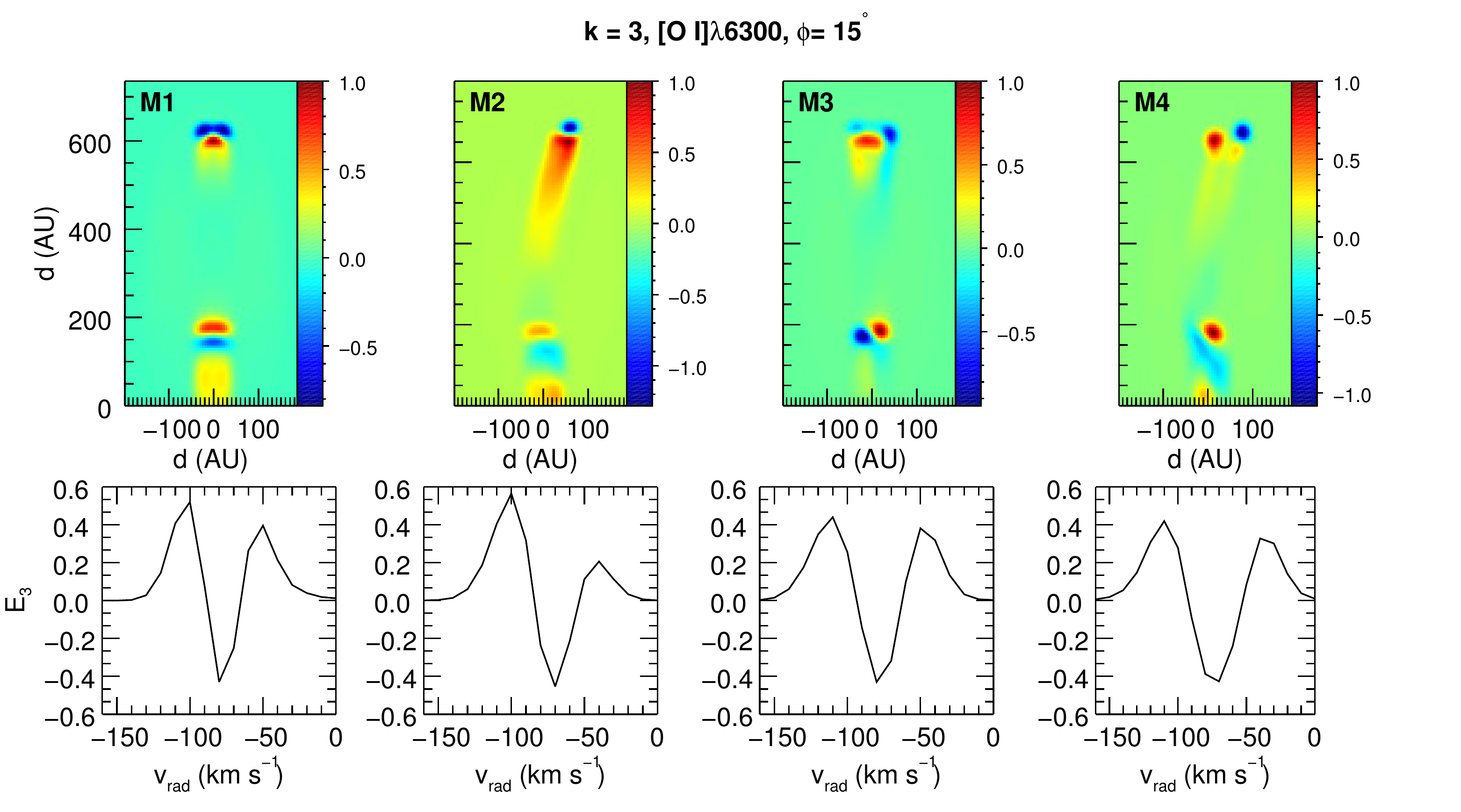}
}
\centerline{
\includegraphics[scale=0.30]{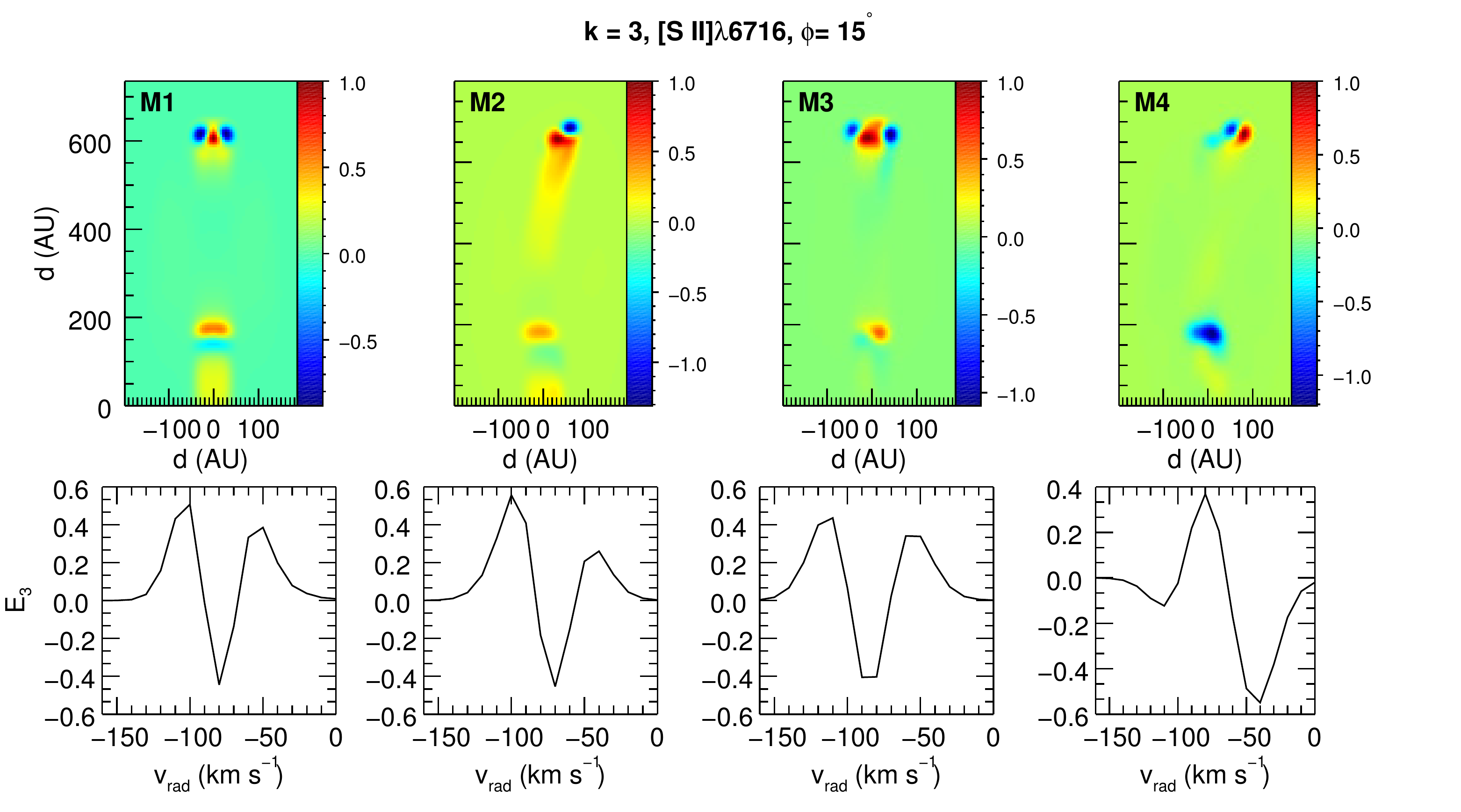}
}
\caption{The same as in Figure \ref{fig05}, but for
$k = 3$ (the third tomogram, ${\bf T}_{i,j,k}$, for $k = 3$,
and its respective eigenspectrum, $E_k$, for $k=3$). }\label{fig07}
\end{figure}

The tomograms for the M2 (precessing) model (panels in the
second column of the Figure \ref{fig04}) are considerably different
for the two emission lines. In the case of the [\ion{O}{1}]$\lambda$6300,
the tomogram combined with its eigenspectrum clearly traces the
precession of the jet axis.  The eigenspectrum has anti correlated
wings: the red wing of the line, associated with positive weights
and positive regions in the tomogram, traces the receding portion
of the jet (coincident with the internal working surface near to
the jet inlet), while the blue wing of the line, associated with
negative weights and negative regions in the tomogram, traces the
approaching jet. We note that there is also a positive gradient,
from the internal working surface (hereafter, IWS) up to the jet
head, in the tomogram's intensity along the jet. The eigenspectrum
for the [\ion{S}{2}]$\lambda$6716 emission line also depicts the
same anti correlation in the wings of the line, but here the
interpretation is not as straightforward as before. In the
eigenspectrum, there is only one peak for each positive and negative
weights. This suggests that all blue regions in the tomograms are
blueshifted with respect to the red ones. According to this
interpretation, the IWS is receding and there is a smooth gradient
in blue regions from the IWS towards the jet head suggesting the
precession. However, at the jet head seen in the tomogram we have
a strong gradient in radial velocity, and the anti correlation seen
in the eigenspectra in this case may be due to a combination of
these two features, and could not be attributed only to the precession.
We note that a signature for the precession can be found in all
tomograms until $k = 4$ (not shown here).

The second eigenvector for M3 (rotating) model accounts for
6.7645\% and 3.9785\% of the variance in the dataset for
[\ion{O}{1}]$\lambda$6300 and [\ion{S}{2}]$\lambda$6716 emission
lines, respectively (Table \ref{taba1}). Their tomogram/eigenspectrum
are in the third column of the Figure \ref{fig04}. The eigenspectrum
for the [\ion{O}{1}]$\lambda$6300 emission line indicates again an
anti correlation between both wings of the line. Negative weights
are associated with negative regions (blue regions) in the tomogram,
indicating a receding jet (it is in the red wing of the line), while
positive weights, in the blue wing of the line, indicate that the
red regions in the tomogram are approaching. This is consistent
with the sense of the rotation imposed at the beginning of our
simulation. This eigenvector for [\ion{O}{1}] traces, then, the
rotation of the jet. Conversely, the same tomogram and the eigenspectrum
for the [\ion{S}{2}]$\lambda$6716 emission line are completely
symmetric with respect to the jet axis and to the zero radial
velocity, respectively.  There is no trace at all for the presence
of the rotation in this tomogram for this emission line. It will
appear only in the third eigenvector (not show here).

\begin{figure}
\centerline{
\includegraphics[scale=0.30]{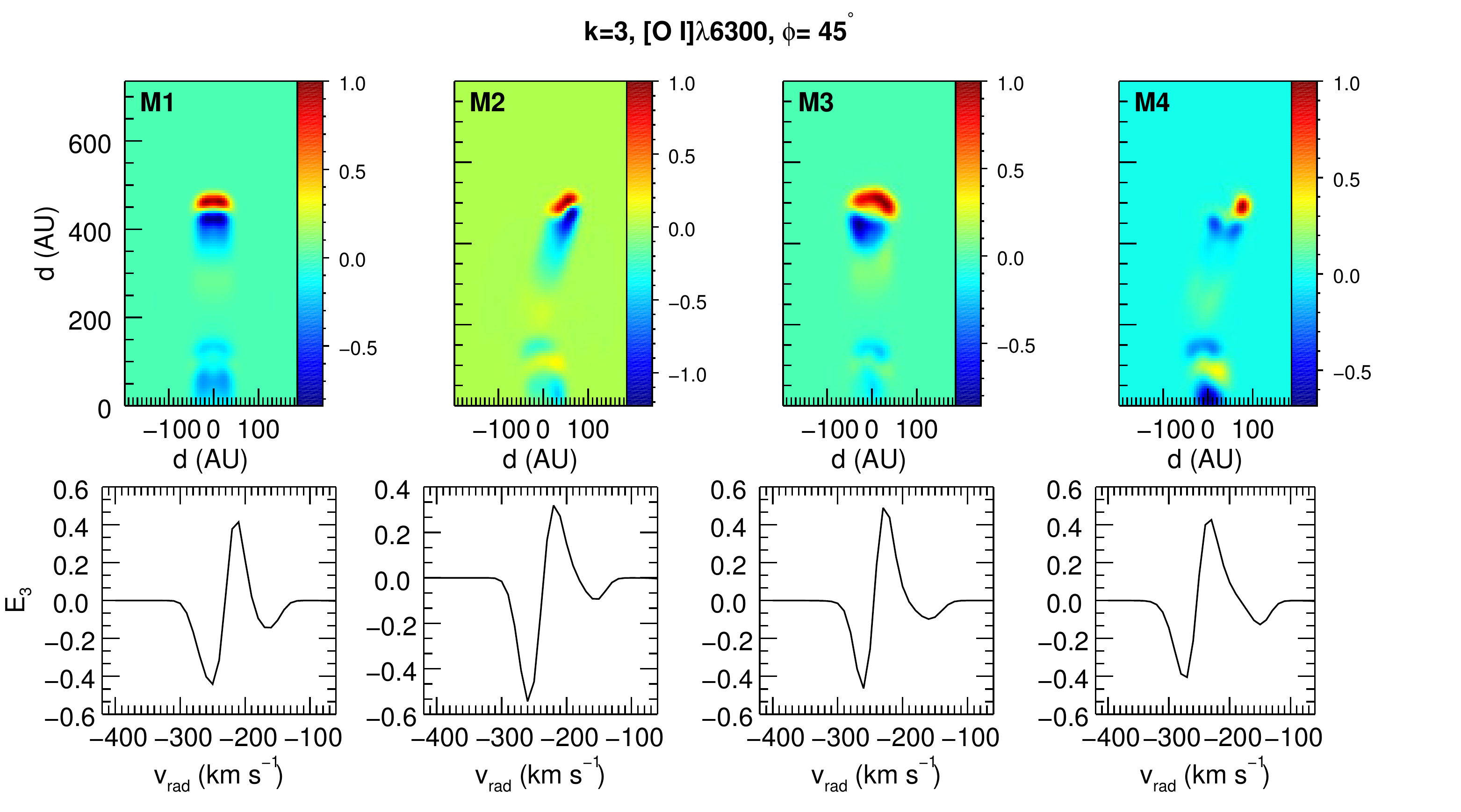}
}
\centerline{
\includegraphics[scale=0.30]{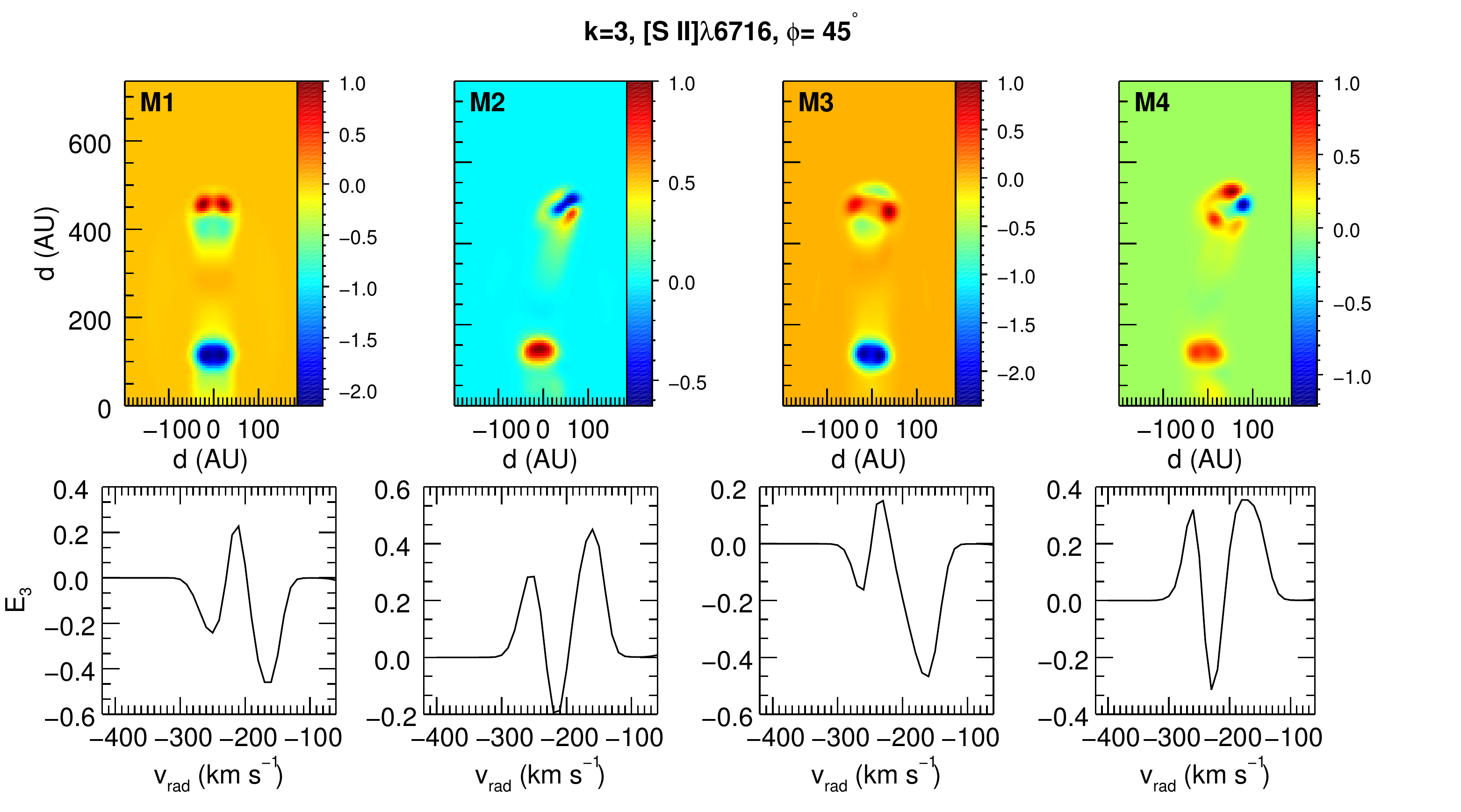}
}
\caption{Third tomogram (${\bf T}_{i,j,k}$, for $k = 3$)
and eigenspectrum ($E_k$, for $k=3$) for models M1 to M4 (from left
to right, respectively), for [\ion{O}{1}]$\lambda$6300 (two
topmost panel's sequence) and [\ion{S}{2}]$\lambda$6716 (two bottommost
panels) emission lines at $\phi = 45^{\circ}$. In each tomogram
the distance from the jet inlet, in AU, is indicated. The color bar
indicate the level of the (normalized) intensity.
The eigenspectrum of each model is plotted below  the respective
tomogram.}\label{fig08}
\end{figure}

The second eigenvector for M4 (rotating and precessing) model
accounts for 18.007\% and 12.7514\% of the variance of the data,
for the [\ion{O}{1}]$\lambda$6300 and [\ion{S}{2}]$\lambda$6716
emission lines, respectively. Their tomogram/eigenspectrum are in
the fourth column of the Figure \ref{fig04}. The tomogram and the
eigenspectrum for the [\ion{O}{1}]$\lambda$6300 emission line shows
anti correlation between both sides of the jet axis near to the jet
inlet only. Positive regions, red in the tomogram, are correlated
with the positive, red wing of the line, while negative regions,
blue in the tomogram, are correlated with negative weights in the
blue wing of the line. This is consistent with the interpretation
that the jet is rotating, and the correct jet rotation sense is
recovered. From the IWS until the jet head the jet is dominated by
blue, negative regions. The gradient seen in the tomogram can suggest
a precession.  For the [\ion{S}{2}]$\lambda$6716 emission line the
situation is less clear.  Some asymmetries with respect to the jet
axis can be detected near to the jet inlet, but it is not obvious
that rotation is involved.  The eigenspectrum seems to reflects the
gradient in radial velocity at the jet head, which is in turn very
similar to the one detected in the M2 model for this same tomogram.

In the following sections, we will not show the first tomograms
anymore (i.e., tomograms for $k = 1$): they are always similar to
an image in the real space of the integrated emission line and they
do not contribute to our present purposes, which is the search for
evidence of rotation and precession in jets.

\begin{figure}
\centerline{
\includegraphics[scale=0.30]{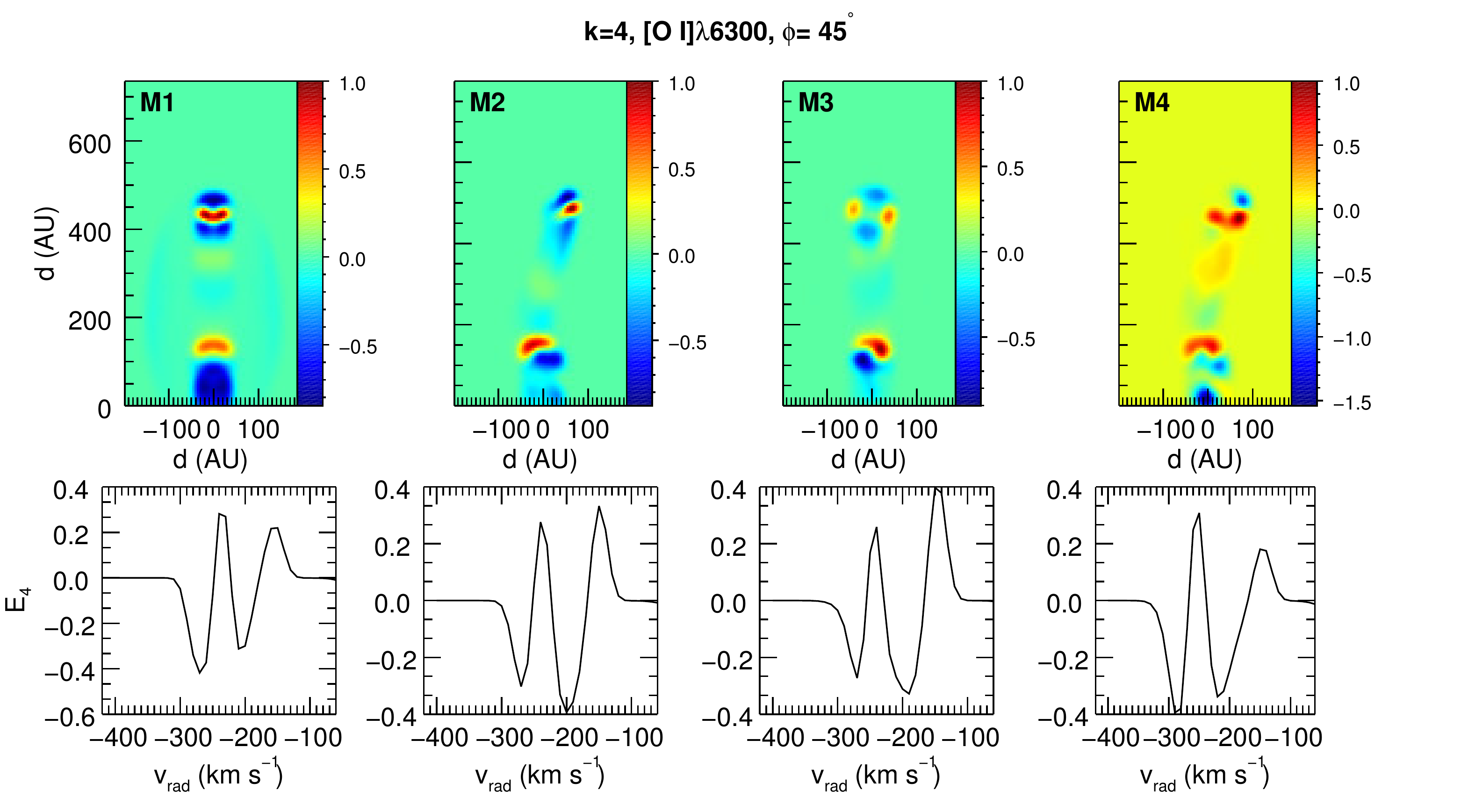}
}
\centerline{
\includegraphics[scale=0.30]{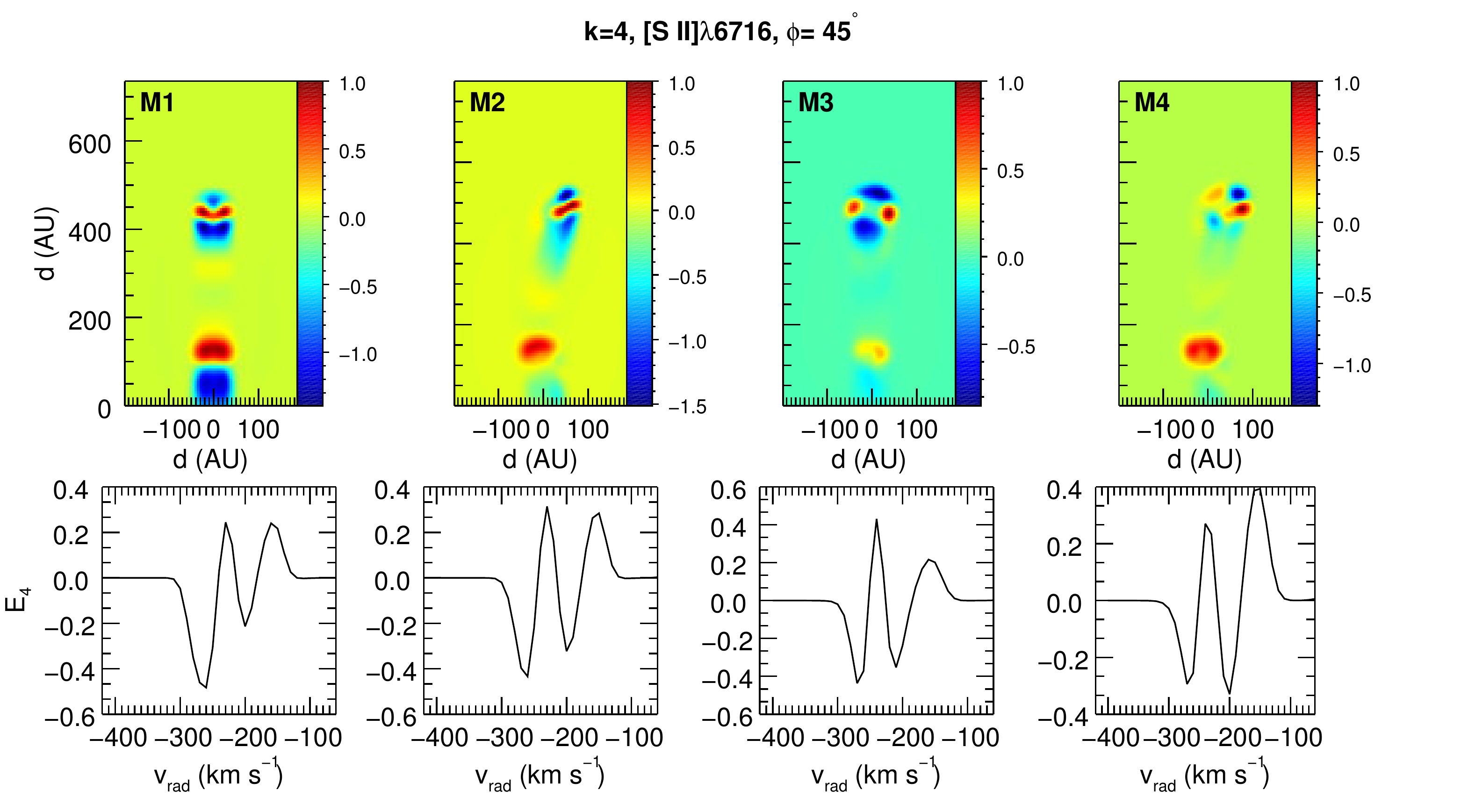}
}
\caption{The same as in Figure \ref{fig08}, but for
$k = 4$ (the fourth tomogram, ${\bf T}_{i,j,k}$, for $k = 4$,
and its respective eigenspectrum, $E_k$, for $k=4$).}\label{fig09}
\end{figure}

\subsubsection{Tomograms for moderately inclined systems ($\phi =
15^{\circ}$)} \label{phi15}

In Figure \ref{fig05} we show the second tomogram ($k = 2$)
and its associated eigenspectrum for models M1 to M4 (from left to
right), for [\ion{O}{1}]$\lambda$6300 (top) and [\ion{S}{2}]$\lambda$6716
(bottom) emission lines, at $\phi = 15^{\circ}$. All eigenspectra
are blueshifted in comparison with the models in previous section,
as expected for jets that are pointing towards the observer. They
also present anti correlation between the red and blue wings of the
line, no matter if we are looking for a simple, non-precessing and
non-rotating (M1) model or a complex (M4) one, and this fact deserves
a careful analysis.

\begin{figure}
\centerline{
\includegraphics[scale=0.3]{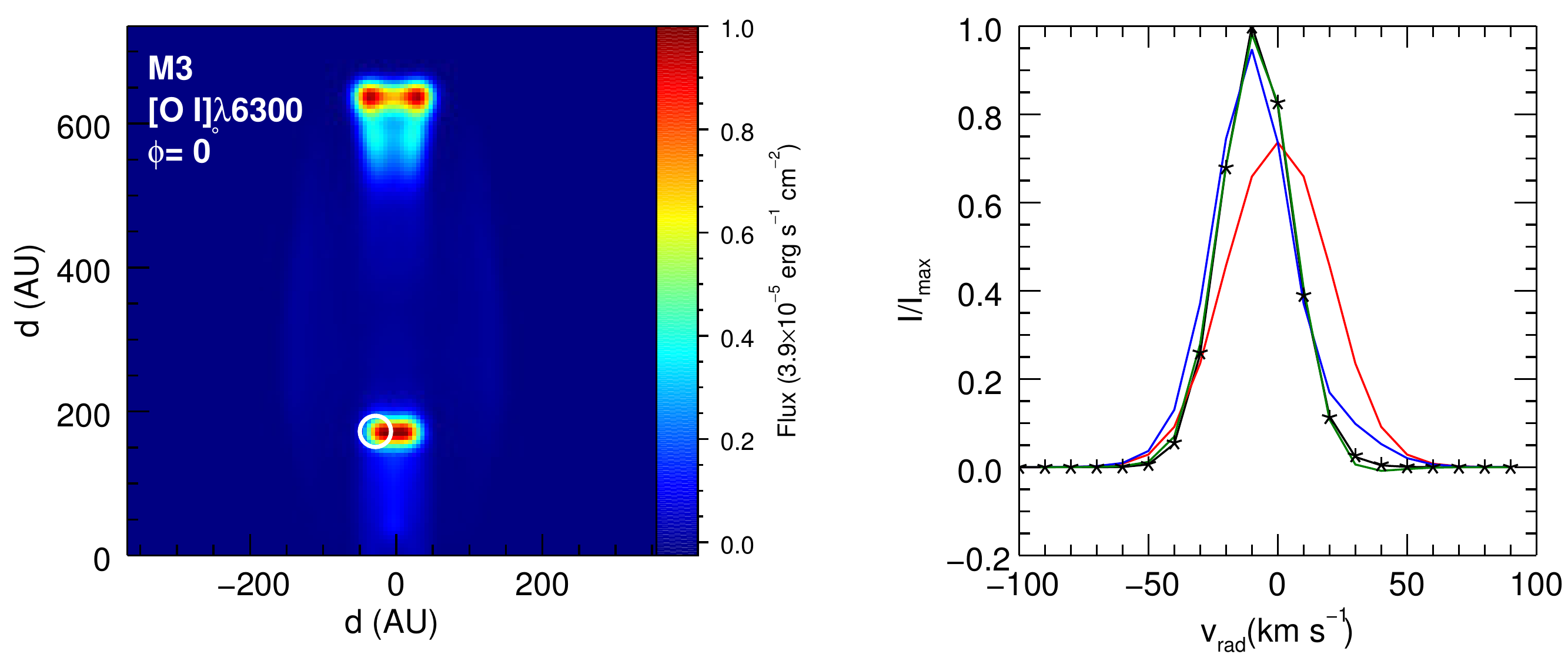}
}
\centerline{
\includegraphics[scale=0.3]{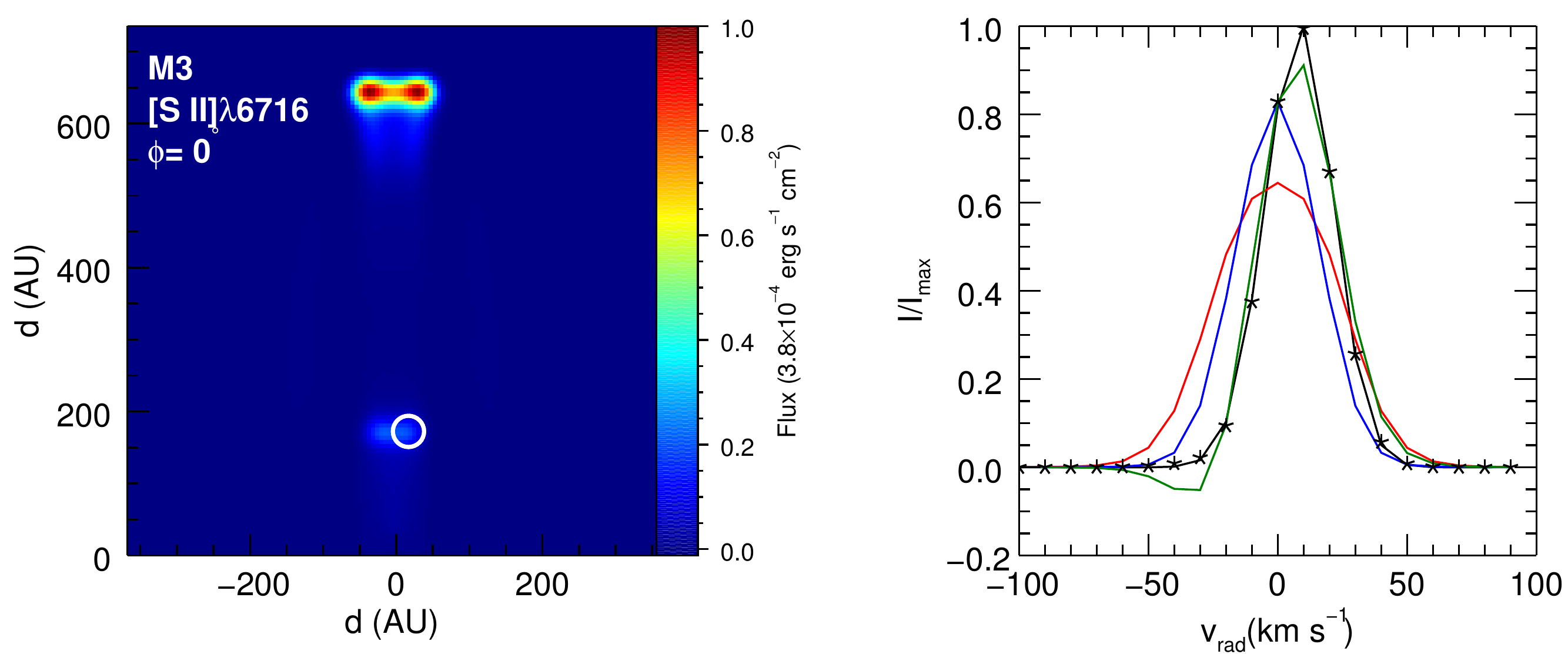}
}
\caption{Images (left) of the collapsed original datacube
for the M3 (rotating) model around the [\ion{O}{1}]$\lambda$6300
(top) and [\ion{S}{2}]$\lambda$6716 emission lines.  Spectra (right)
have been extracted from the area indicated by the white circles
in the images using the original datacube (black solid line), and
the reconstructed datacubes considering eigenvectors until
$k_{\rm max} = 1$ (red), $k_{\rm max} = 2$ (blue), $k_{\rm max} = 3$ (green) and
$k_{\rm max} = 4$ (black stars). Fluxes are normalized to unity,
and each maximum is indicated
in the right side of the color bars. In the extracted spectra,
intensities are also normalized with respect to the maximum of the
integrated spectra (black solid line). The circles in the images
have a diameter of 4 pixels, or 22.9 AU. This corresponds to
$0\farcs16$ considering the distance to Taurus (140 pc), for example.
The origin of the coordinate system corresponds to the position of
the jet inlet, and distances from the origin are indicated in AU.}
\label{fig10}
\end{figure}

In the case of the reference model M1, the structures
(positive/negative regions) in the tomogram are symmetric with
respect to the jet axis. The portions of the jet immediately behind
the working surfaces (at $d(y) \le 150$ AU and at 550 AU $\le d(y)
\le$ 600 AU) are correlated with the negative weights of the
eigenspectra for both emission lines (see the first column of Figure
\ref{fig05}), and then, they are correlated with the blue wing of
the emission line. Beyond these limits (that is, for $d(y) \sim
180$ AU and $d(y) \sim 630$ AU), the tomograms display positives
(see the color bar in each figure) regions, that are correlated
with the red wing of the line. We interpret these structures at the
jet working surfaces as a signature of the backflowing post-shocked
material, that escapes laterally at the Mack disk (or jet shock),
favoring lower (absolute) radial velocities on the top half of the
jet cross section, and, analogously, higher (absolute) radial
velocities on the bottom half of the jet cross section.  In Figure
\ref{fig06} (a) we present a cartoon summarizing the interpretation.
We note that an inclined bow shock could also account for the
blue/red asymmetry in the line and cannot be ruled out (see the
sketch presented also in Figure \ref{fig06}-b), particularly for
the case of an IWS, for which the bow shock propagates into a
non-stationary ambient medium. In this case this effect can be more
noticed since we expect high bow shock velocities in comparison
with the leading bow shock.

The precessing model M2 also show an asymmetric eigenspectrum.
In the [O I]$\lambda$6300 emission line, the tomogram essentially
traces the IWS near the jet inlet. It is a positive region (red in
the color bar), correlated with the positive weights that are present
in the red wing of the line, as we can see in the eigenspectrum.
There is a clear contrast between positive (internal emission knot)
and negative regions (the jet head), suggesting a precession.
However, positive weights dominate the eigenspectrum (it has a
maximum that is more than three times the negative peak at $v_{rad}
= -90$ km s$^{-1}$), which is evident also in the tomogram. For the
[S II]$\lambda$6716 emission line, the contrast between positive
and negative weights is less intense and the structures more clear.
There is a gradient in the negative region that goes from the IWS
and peaks at the jet head. There, we have also a positive region,
suggesting that the same effect explored in Figure \ref{fig06} may
also be present. However, the precession breaks the symmetry with
respect to the jet axis.

\begin{figure}
\centerline{
\includegraphics[scale=0.32]{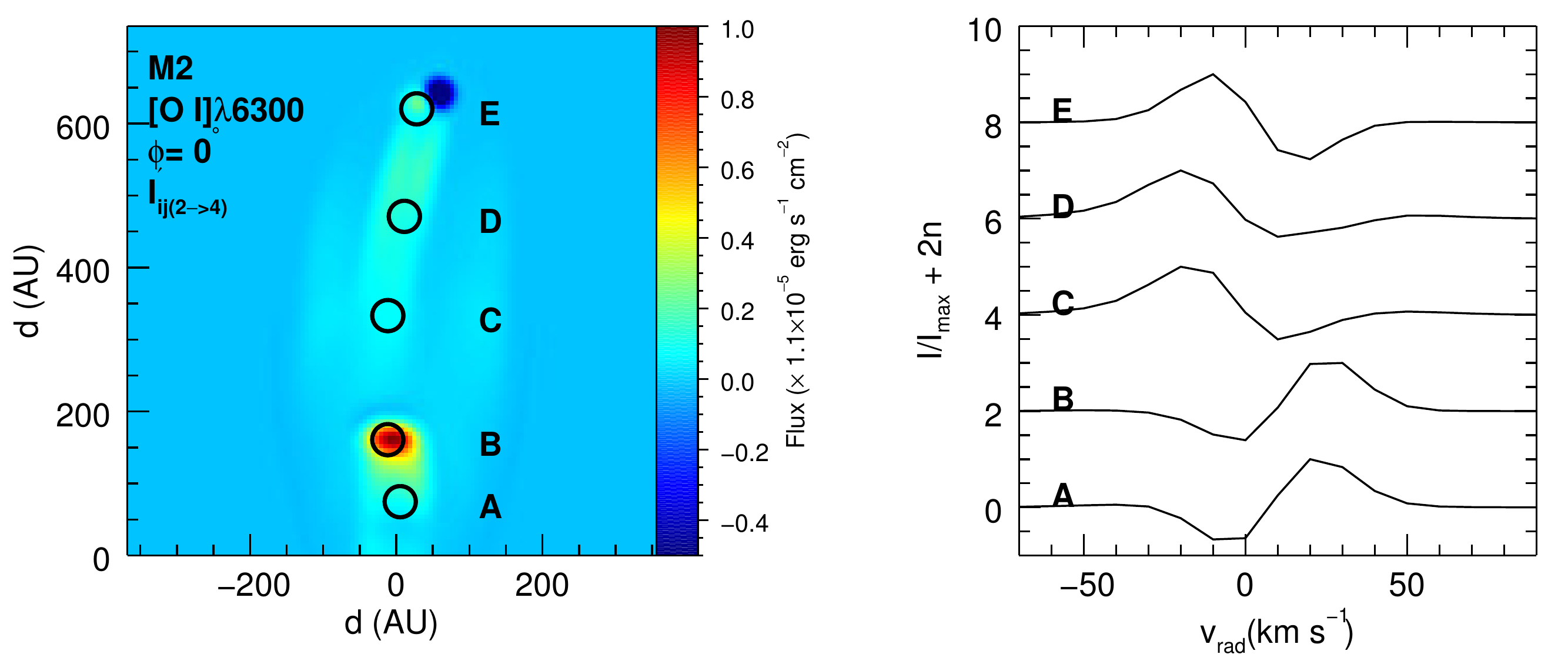}
}
\centerline{
\includegraphics[scale=0.32]{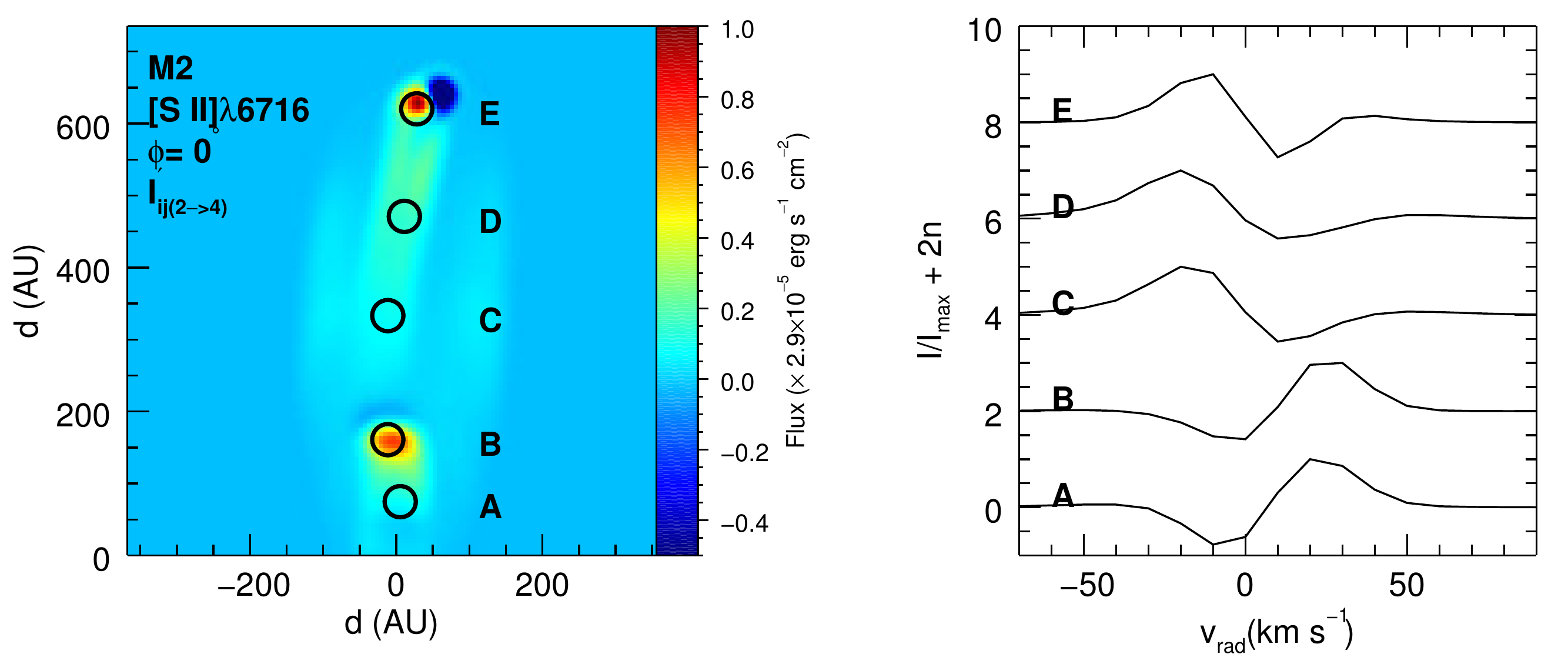}
}
\caption{Images (left) and spectra (right) from the reconstructed
datacube for the precessing model M2 at $\phi = 0^{\circ}$, for the
[\ion{O}{1}]$\lambda$6300 (top) and [\ion{S}{2}]$\lambda6716$
(bottom) emission lines.  Eigenvectors from 2 to 4 were used in the
reconstruction process, since they have signatures for the jet
precession. Circles labeled from A to E mark the positions where a
spectrum has been extracted.  They are in the right side panels,
normalized to unit. We added a constant 2$n$, where $n=0-4$ to
the spectra at positions A to E, respectively, in order
to separate them in different levels, to improve the visualization.
The images are integrated in wavelength.}
\label{fig11}
\end{figure}

For the M3 (rotating) model (third column in Figure \ref{fig05}),
the second tomogram shows positive (red) and negative (blue) regions
that are distorted with respect to the jet axis. The region that
extends from the jet inlet until the IWS is blue in the left side
of the jet axis for both emission lines (although more clearly seen
in the tomogram associated with the [O I]$\lambda$6300 emission
line). At the tip of the IWS, the red (positive) region is distorted
toward the right side of the jet axis, indicating that this side
of the jet is more redshifted than the other one (since positive
weights in the eigenspectra are in the red wing of the line; see
the eigenspectrum associated with this model in the third column
of Figure \ref{fig05}). At the jet head, the rotation signature is
evident only in the case of the [S II]$\lambda$6716 emission line:
we can interpret the blue/red region in the jet head in this tomogram
as in the case of model M1, and attributes to the rotation the cause
for the observed asymmetry with respect to the jet axis.


The second eigenvector for the model M4 contributes to 33.2961\%
and 14.5136\% of the variance in the dataset for [O I]$\lambda$6300
and [S II]$\lambda$6716 emission lines, respectively, at $\phi =
15^{\circ}$. The tomogram and eigenspectrum can be seen in the last
column of the Figure \ref{fig05}. Again, the eigenspectra for both
emission lines show blue/red wing asymmetry. The tomograms are,
however, different. For the [O I]$\lambda$6300 emission line,
positive regions in the tomogram are concentrated in the IWS. It
has a peak on the right side of the jet axis, showing an asymmetric
distribution in intensity with respect to the jet axis (as in the
case of the M3 model). All other parts of the jet are mostly dominated
by blue, negative regions, with a maximum at the jet head.  There
is a negative gradient from the jet inlet to the IWS, and then a
positive gradient from the IWS up to the jet head, which is suggest
the precession. Although the eigenspectra are similar for the [S
II]$\lambda$6716 emission line in comparison with the one for the
[O I]$\lambda$6300 emission line, their tomograms are not.  Both
positive and negative regions peaks at the jet head, suggesting
that the anti correlation in the wings of the line are related to
gradients in velocity at the jet region.

In Figure \ref{fig07} we show the tomograms (first and third
rows of panels) and eigenspectra (second and fourth rows of panels)
for models M1 to M4 (from left to right) at $\phi = 15^{\circ}$ and
$k = 3$.  As before, the two topmost panels refer to the [O
I]$\lambda$6300 emission line, while the two bottommost panels
concern the [S II]$\lambda$6716 emission line. There is a pattern
in the eigenspectra that is almost the same for all cases\footnote{The
only exception is the model M4, in the [S II]$\lambda$6716 emission
line (see below)}: there is an interval in radial velocity of
negative weights, bracketed by two peaks of positive weights, which
make the interpretation far less obvious. In particular, a negative
region in a given tomogram can be either red- or blue-shifted with
respect to a positive one.  The same ``color" in a tomogram can,
in this case, be indicative of different kinematics, and the
interpretation is actually difficult.

M1 model (leftmost column in Figure \ref{fig07}) has tomograms
that show negative/positive regions that peak at the jet head (for
both emission lines). The positive (red) regions at the jet head
are correlated now with positive weights in the eigenspectra, since
the maximum in both tomograms is localized in this region.  Then,
the negative region just above it in the case of [O I]$\lambda$6300
emission line, and the negative regions symmetrically displaced
with respect to the jet axis in the case of [S II]$\lambda$6716
emission line are redshifted in comparison with the red regions at
the jet head, as we can see in the eigenspectrum. This interpretation
is still compatible with the scenario proposed in Figure \ref{fig06}.
The same is true for the IWS, which correlates with the positive/negative
peaks at $v_{rad} = - 50$ and -80 km s$^{-1}$, respectively. The
yellow (positive) region in the jet base can be only correlated
with the positive weights at the bluemost wing of the line in the
eigenspectra.

M2 model (panels in the second column of Figure \ref{fig07})
shows that the highest values for the positive and negative regions
in the tomogram occur at the jet head. Negative regions are
concentrated at the jet tip, while the positive regions are immediately
behind of it, with a negative gradient of intensities from the jet
head until the IWS. The peak in the positive weights of both
eigenspectra (i.e., for both emission lines) is at $v_{rad} = -100$
km s$^{-1}$, that are in the blue wing of the lines. This means
that the negative blue region on top of these positive regions at
the jet head are redshifted with respect to them, since the negative
weights in the eigenspectra are at lower (absolute) radial velocities
($\sim -60$ km s$^{-1}$).

\begin{figure*}
\centerline{\includegraphics[scale=0.5]{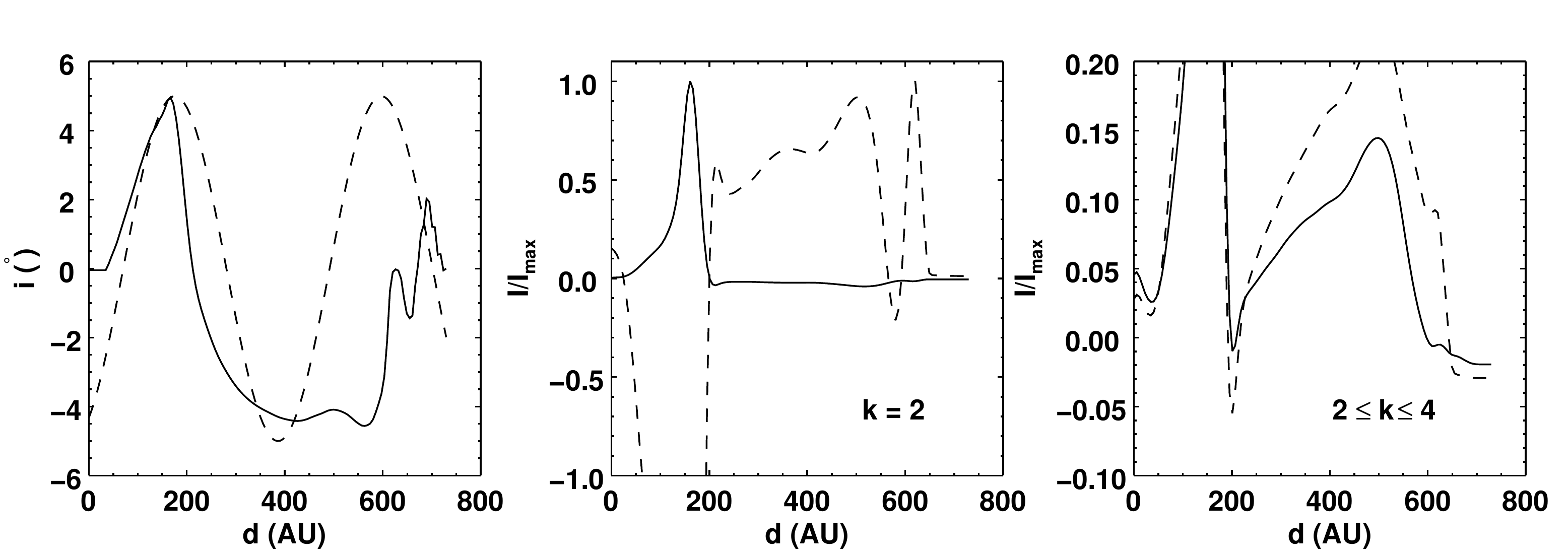}}
\caption{{\em Left:}) The inclination of the jet axis with respect
to the plane of the sky, here defined as $i$, as a function of the
vertical distance from the jet inlet for the model M2 (precessing jet),
estimated from the simulation (solid line), using velocities taken
in a straight line along the $y-$coordinate, starting at the jet
inlet at $x =  y = 0$. The dashed line is an arbitrary cosine
function.  {\em Middle:}) Normalized fluxes taken from the reconstructed
datacube considering only the second eigenvector ($k=2$)
(solid line: [\ion{O}{1}]$\lambda$6300; dashed line:
[\ion{S}{2}]$\lambda$6716).  {\em Right:}) The same as in the middle
panel but using the eigenvectors $2 \le k \le 4$ to
reconstruct the datacube.  The fluxes have been normalized to their
maximum value in each curve.  In the rightmost panel, we have limited
the ordinate to 0.2.}
\label{fig12}
\end{figure*}

The third tomogram for the M3 and M4 models (third and fourth
columns of panels in Figure \ref{fig07}, respectively) at $\phi =
15^{\circ}$ show a twisted pattern in the regions of positive and
negative values, which is more evident for the [O I]$\lambda$6300
emission line in comparison with the [S II]$\lambda$6716 emission
line\footnote{It becomes also evident in high order tomograms for
the [S II]$\lambda$6716 emission line. We should note that it it
also present in high order tomograms for the non-inclined, $\phi=
0^{\circ}$, rotating models.}.  We can interpret this behavior
saying that negative regions in the tomograms can be both redshifted
or blueshifted with respect to positive regions placed in a symmetric
position with respect to the jet axis. This is because the negative
peaks in the eigenspectra for both models are bracketed by two
positive peaks, and that it is an effect of the rotation, that
changes the tomograms with respect to the non-rotating M1 and M2
models even if the eigenspectra have the same shape/form.  However,
by analyzing only this eigenvector, the presence of such a twisted
feature in the tomograms does not allow us to say/conclude anything
about the rotation sense of the jet.

Unlike the case of $\phi = 0^{\circ}$ shown in the previous
section, evidences for the rotation here are more subtle. The mere
fact that we are ``observing" the jet at a non-null inclination
angle makes the analysis of the principal components more difficult.

\subsubsection{Tomograms for highly inclined systems ($\phi = 45^{\circ}$)}
\label{phi45}

At $\phi = 45^{\circ}$, the dataset becomes even less redundant\footnote{In
the sense that the number of relevant eigenvectors, defined by the
scree test for example, is higher in comparison with a more redundant
dataset.  In other words: if just a few eigenvectors can be used
to describe the dataset and the majority of them can be discarded
for this purpose, then it is a highly redundant dataset.} with an
appreciable amount of variance distributed in higher order eigenvectors.
In Figures \ref{fig08} and \ref{fig09} we show as an example the
third and fourth tomograms/eigenspectra\footnote{The first two
tomograms in these cases traces mainly the IWS and the leading
working surfaces and will not be discussed here.}, respectively,
for models M1 to M4 at $\phi = 45^{\circ}$. As before, the results
for the [O I]$\lambda$6300 (topmost two panels) and [S II]$\lambda$6716
(bottommost two panels) emission lines are shown.  Some of the
patterns already discussed can be seen here too. Eigenspectra becomes
more complex, in part due to gradients in velocity at the jet head.
The clumpy, alternating pattern observed there specially in models
M3 and M4 (for $k = 3$ and $k=4$) makes the kinematic analysis, and
its interpretation, more difficult. The presence of the rotation,
for instance, is only suggested near the jet inlet in the $k=4$
tomograms of model M3 (and it is not so evident for the M4 model;
see Figure \ref{fig09}). It is present also in higher order tomograms
(not shown here).

We conclude that it is not possible to express, or to describe,
the precession and/or the rotation with a single eigenvector, unless
if the precession/jet axis lies in the plane of the sky. We have
actually a collection of eigenvectors for which the signature for
the precession and/or for the rotation can be found. In general,
the higher the inclination angle $\phi$, the higher will be the
eigenvector's order in which they will manifest. In this Section
we have discussed the most relevant eigenvectors. Furthermore, the
interpretation of higher order tomograms becomes less and less
evident, with eigenspectra showing several negative and positive
successive peaks.

\section{PCA signatures for precession and rotation from reconstructed
datacubes}
\label{sec:res}

In this section we will present images from reconstructed
datacubes, obtained using the equation (\ref{steiner8}) for a
collection of eigenvectors that are representative of a given feature
(see Appendix \ref{apendb}). We will also discuss the properties
of spectra extracted at fixed spatial positions obtained from
reconstructed datacubes. We note that all images presented in this
section are actually integrated in wavelength.

Before entering into the details of our results, and as a test
for the method, we can choose a region in a given image and
reconstruct, step by step, the line profile using a certain number
of eigenvectors.  In Figure \ref{fig10} we show the integrated (non
treated by the PCA) images of the [\ion{O}{1}]$\lambda$6300 (top
panel) and [\ion{S}{2}]$\lambda$6716 emission lines (bottom panel)
for the M3 (rotating) model.  Superimposed on the images we show
circles (of 22.9 AU of diameters; which corresponds to $0\farcs16$
at 140 pc of distance) to indicate the position where spectra have
been extracted.  The spectra are on the right panels (black, solid
lines).  We also show line profiles obtained from the reconstructed
datacube, keeping in the reconstruction procedure an increasing
number of eigenvectors from 1 to 4 (see equation \ref{steiner8}):
$k_{\rm max}=1$ (red curve), $k_{\rm max} = 2$ (blue curve), $k_{\rm max} = 3$ (green curve)
and $k_{\rm max} = 4$ (black stars).  In this example, the reconstructed
profiles converge very well to the original one for both  emission
lines (the properties and features of the profiles will be discussed
in the next Section), using only 4 eigenvectors.

\subsection{Jets in the plane of the sky}
\label{sec:plane1}

\begin{figure}
\centerline{
\includegraphics[scale=0.32]{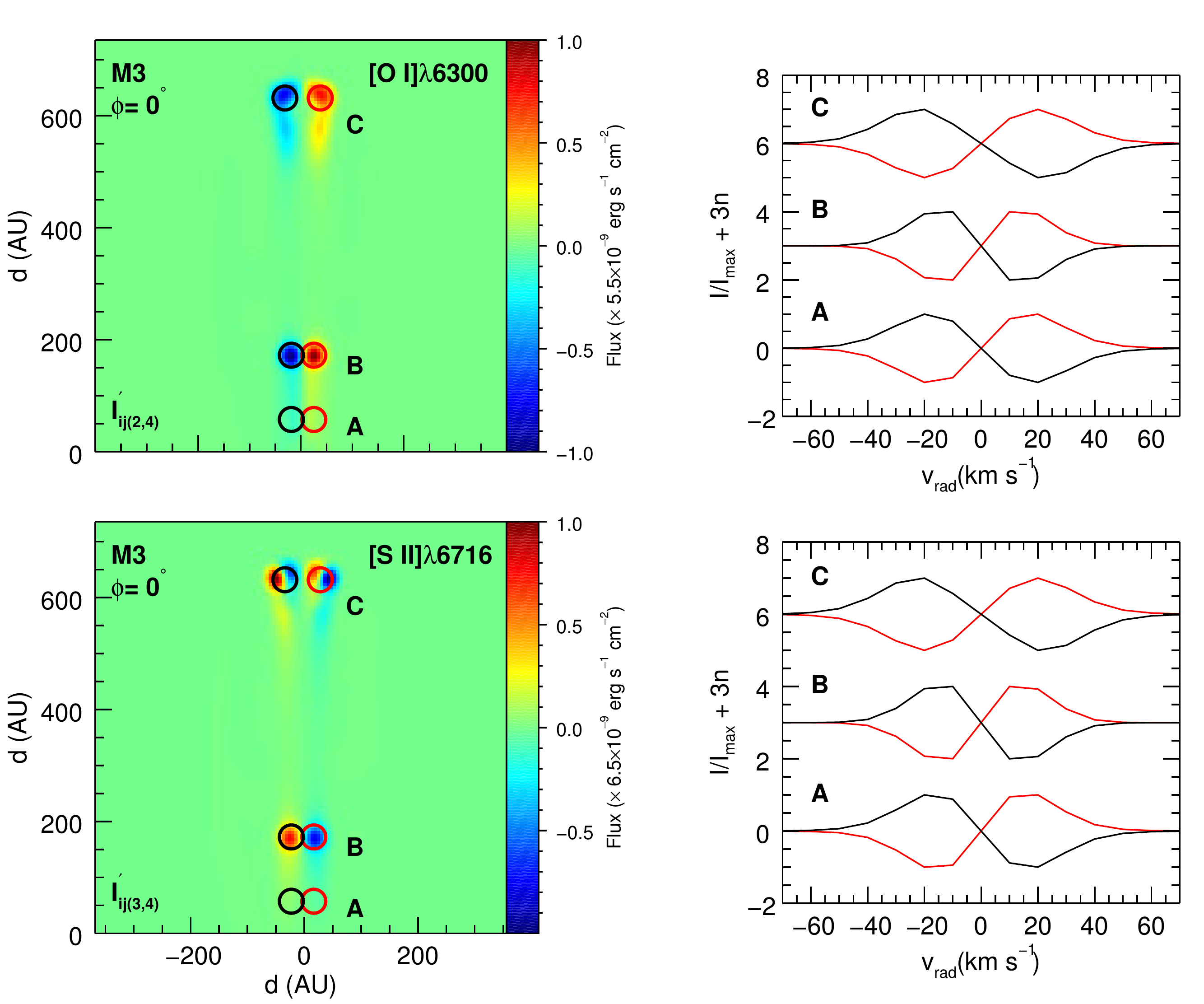}
}
\caption{Reconstructed images (left) and extracted spectra
(right) for Model M3 at $\phi = 0^{\circ}$ and for
[\ion{O}{1}]$\lambda$6300 (top) and [\ion{S}{2}]$\lambda$6716
(bottom) emission lines. Different eigenvectors have been considered
to reconstruct the datacube: $k = 2$ and 4 for [\ion{O}{1}]$\lambda$6300 and $k =
3$ and 4 for [\ion{S}{2}]$\lambda$6716 lines, as indicated in the bottom left
part of each image.  Spectra were extracted at symmetrical positions,
labeled A, B and C, with respect to the jet axis.  The colors of
the solid lines (right panels) corresponds to the color of the
slits. Distances from the jet inlet are indicated in AU in images.
Spectra are normalized and we have added to their intensity a constant 3$n$, with $n
=$ 0, 1 and 2 at the positions A, B and C, respectively.
The images are integrated in wavelength.}
\label{fig13}
\end{figure}

In Figure \ref{fig11} we show the image (left panels) and
spectra (right panels) obtained from the reconstructed datacube for
the model M2 (precessing model) at $\phi = 0^{\circ}$ (which means
that the axis of precession is coincident with the $y-$axis).  As
discussed in Section \ref{apenda} (Section \ref{phi0}), we have
identified the precession in eigenvectors' orders $2 \le k \le 4$
among the eigenvectors suggested by the scree test (see Table
\ref{taba5}). The datacube was then reconstructed considering these
three eigenvectors\footnote{At each image obtained from a reconstruction
process we indicate inside the parenthesis of the expression
$I_{ij()}^{\prime}$, placed at the left-bottom corner of the images,
the selected eigenvectors used in the reconstruction process.}. For
the sake of clarity and to allow the comparison between them, the
results for the [\ion{O}{1}]$\lambda$6300 and [\ion{S}{2}]$\lambda$6716
emission lines are shown. The emission line is indicated in the
top-left corner of each image.  We choose different positions along
the jet to extract a spectrum. These regions are indicated by circles
superimposed in the image (as in Figure \ref{fig10}, the size of
the circle is representative of the collected area which means
$0\farcs16$ at a distance of 140 pc, for example). Viewing from the
base of the jet (region A) up to its tip (region E), each spectrum
peaks at different radial velocities, suggesting clearly a precession
pattern. Assuming as an upper limit for the jet velocity $ v_{j,lim}
< 400$ km s$^{-1}$ (see equation \ref{jetvel}), we can estimate the
radial velocity to be of the order of $\vert v_{rad} \vert \le
v_{j,lim} {\rm sin}\theta$, or $\vert v_{rad} \vert \le 34$ km
s$^{-1}$. The emission peaks fall well within this limit.

\begin{figure}
\centerline{
\includegraphics[scale=0.32]{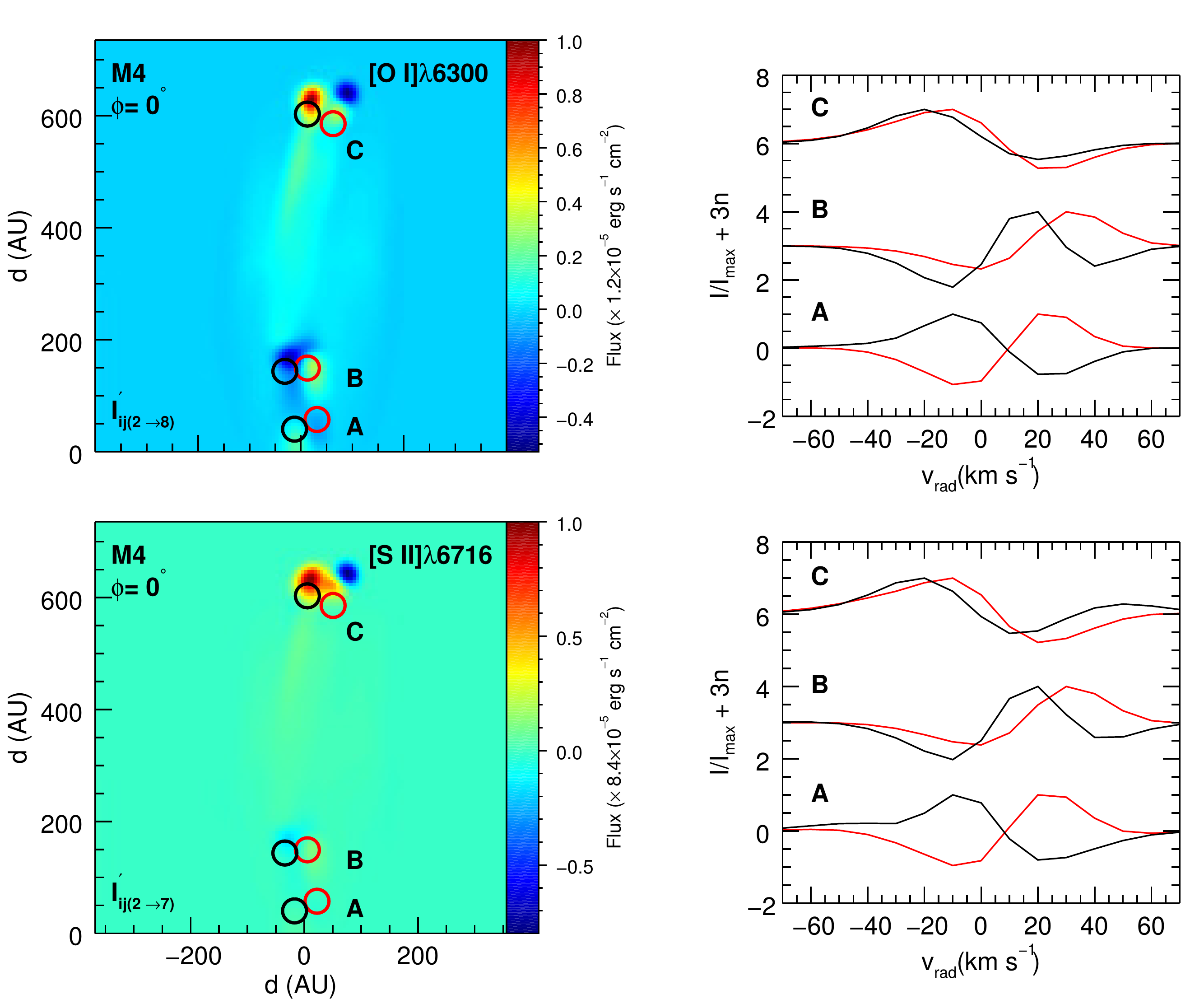}
}
\caption{Reconstructed images (left) and extracted spectra
(right) for Model M4 at $\phi = 0^{\circ}$ and for
[\ion{O}{1}]$\lambda$6300 (top) and [\ion{S}{2}]$\lambda$6716
(bottom) emission lines. Different eigenvectors have been considered
to reconstruct the datacube: $k = 2$ to 8 for the oxygen and $k =
2$ to 7 for the sulfur lines, as indicated in the bottom left part
of each image.  Spectra were extracted at symmetrical positions,
labeled A, B and C, with respect to the jet axis.  The colors of
the solid lines (right panels) corresponds to the color of the
slits. Distances from the jet inlet are indicated in AU.
Spectra are normalized and we have added to their intensity a constant 3$n$, with $n
=$ 0, 1 and 2 for spectra at the positions A, B and C, respectively.
The images are integrated in wavelength.}
\label{fig14}
\end{figure}

Figure \ref{fig11} indicates that there is a variation in the
flux along the jet length that can be due to changes of the viewing
angle. In Figure \ref{fig12} we try to correlate these quantities.
For a fixed $x-$coordinate at the jet inlet ($x=0$), we compute
${\rm tan}^{-1}(v_{rad}/v_{y})$ along the jet length, which is an
estimate for the angle $i$ that the jet axis makes with the precession
axis that lies in the plane of the sky (the $y$ axis at $x=0$).  In
this equation, $v_{rad}$ and $v_y$ are the velocities parallel and
perpendicular to the line of sight, respectively, taken from the
simulation.  This quantity is plotted in the leftmost panel in
Figure \ref{fig12} (solid line), where we show also an arbitrary
cosine function (dashed line).  In the middle and right panels we
show the flux in the [\ion{O}{1}]$\lambda$6300 (solid line) and
[\ion{S}{2}]$\lambda$6716 (dashed line) emission lines, taken along
the precession axis from the reconstructed datacube, considering
the eigenvectors with $k = 2$ (middle panel) and $2 \le
k \le 4$, respectively.  The abrupt variation of the intensity along
the jet seems to correlates with the variation in the inclination
angle when we consider the combined eigenvectors (rightmost
panel)\footnote{The mean spectra, $Q_{\lambda}$, has not been summed
up in the reconstruction process, since we took just a few eigenvectors
to reconstruct the datacube.}.

\begin{figure}
\centerline{
\includegraphics[scale=0.32]{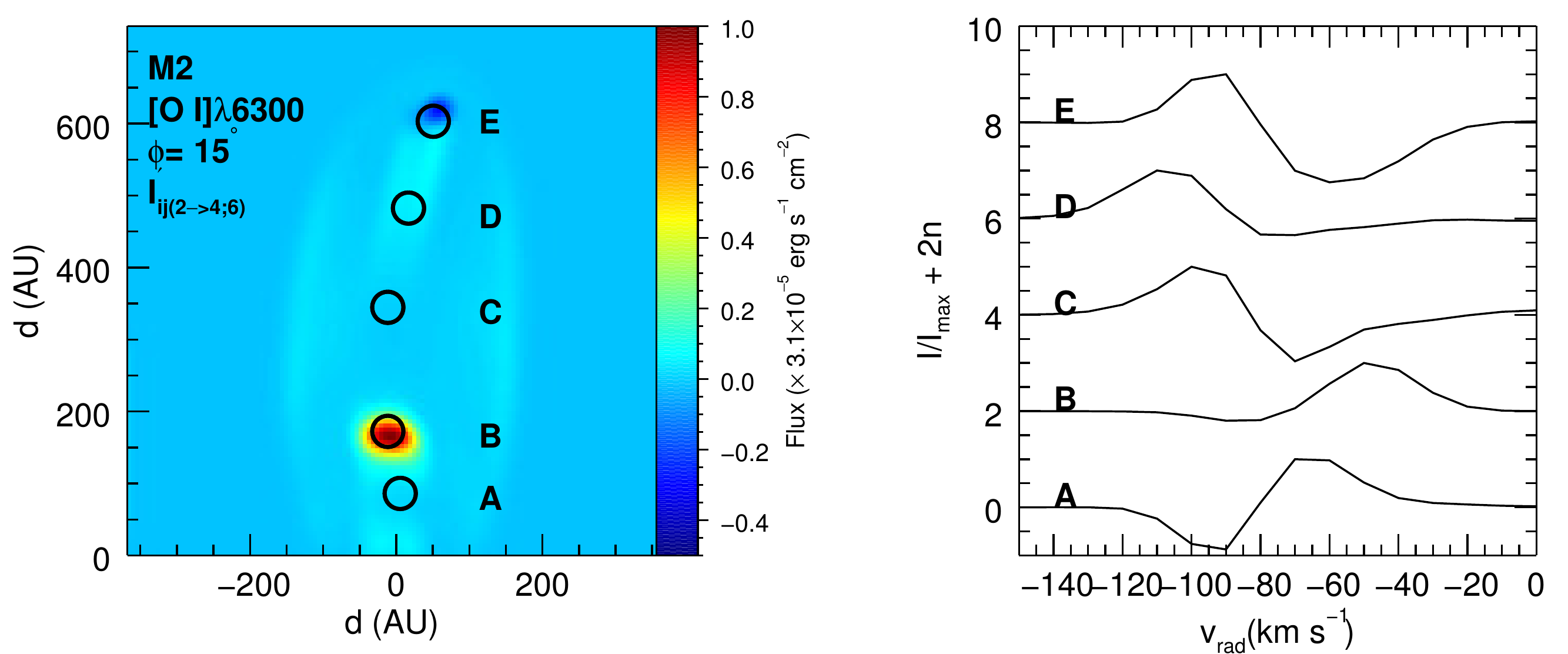}
}
\centerline{
\includegraphics[scale=0.32]{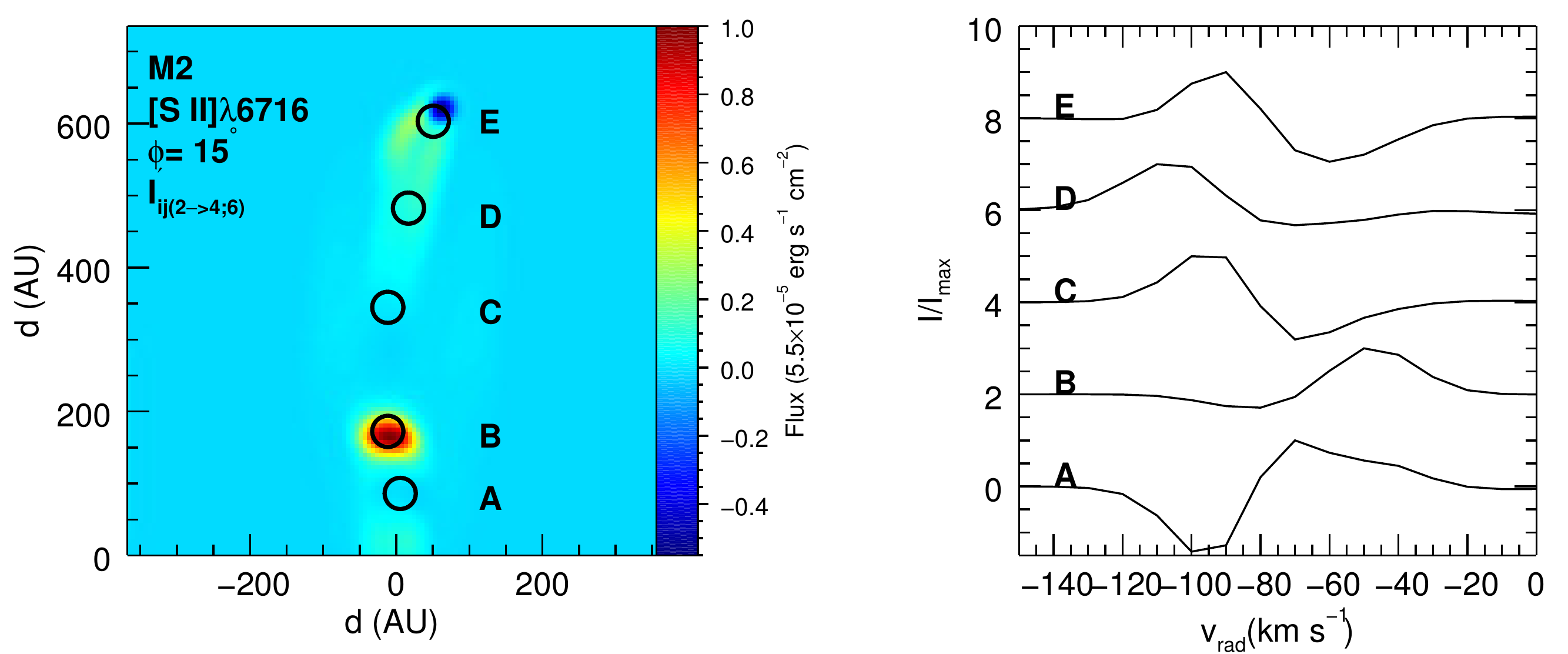}
}
\centerline{
\includegraphics[scale=0.32]{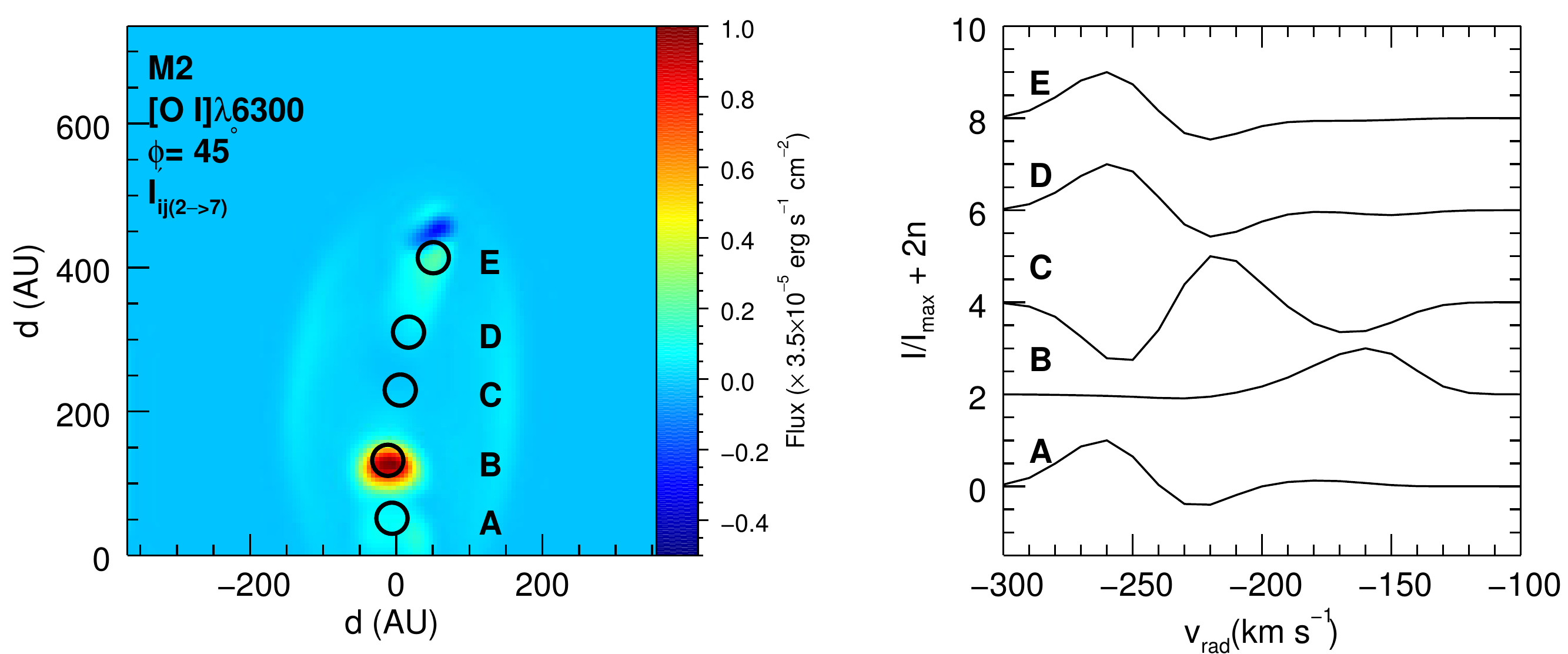}
}
\centerline{
\includegraphics[scale=0.32]{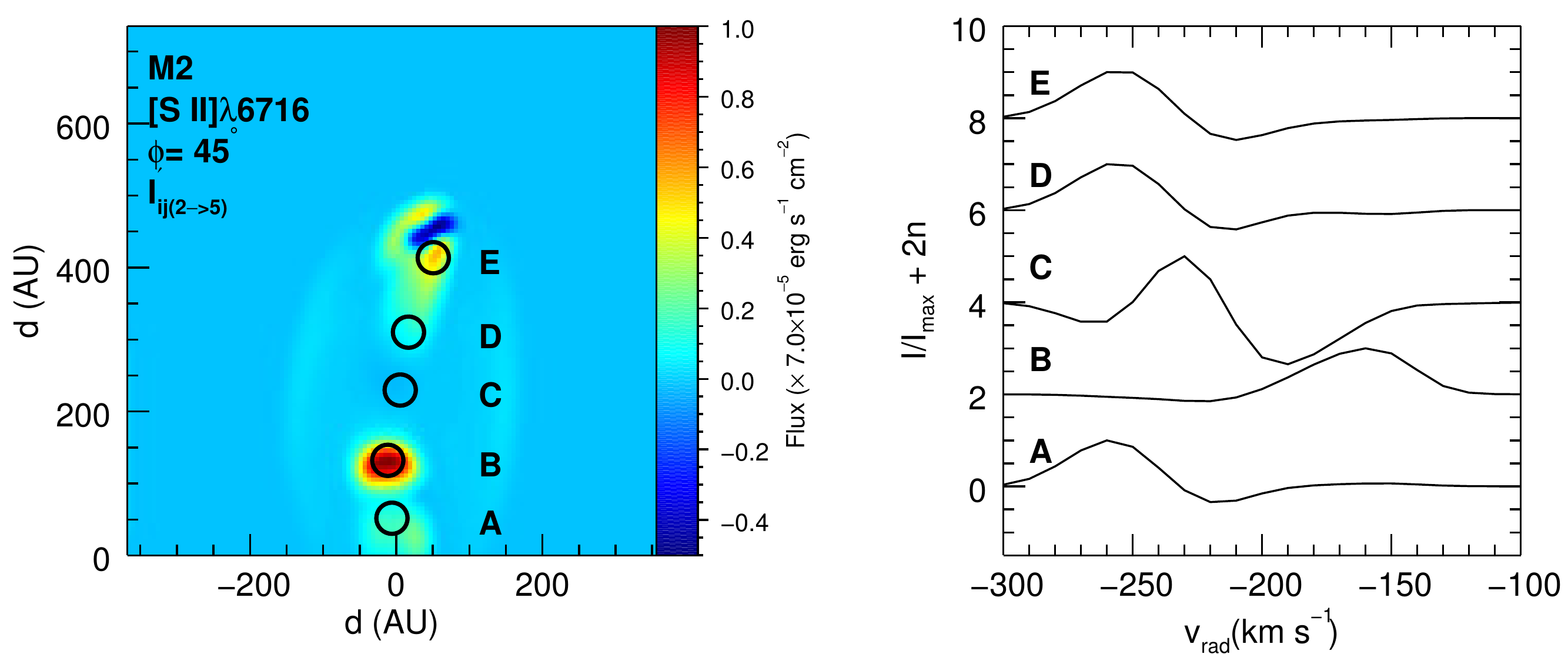}
}
\caption{Reconstructed images (left) and spectra (right) for
the precessing model M2 for $\phi = 15^{\circ}$ (first and second
panels, from top to bottom) and $\phi = 45^{\circ}$ (last two bottom
panels), for [O I]$\lambda$6300 (first and third rows) and [S
II]$\lambda$6716 (second and fourth rows) emission lines. Different
eigenvectors have been used in the reconstruction procedure.  Their
range is show at the top left of each image.  Circles labeled from
A to E mark the positions where a spectrum has been extracted.  They
are in the right side panel, normalized to unit (we added a constant
2$n$, $n=0-4$ from A to E, in each spectra in order to separate
them in different levels).
The images are integrated in wavelength.}
\label{fig15}
\end{figure}

In Figure \ref{fig13} we show images (left) and spectra (right)
from the reconstructed datacube for the [\ion{O}{1}]$\lambda$6300
(top) and [\ion{S}{2}]$\lambda$6716 (bottom) emission lines, for
the rotating M3 model observed at $\phi = 0^{\circ}$.  For the
[\ion{O}{1}]$\lambda$6300 line, we have found evidence for rotation
in the eigenvectors 2, 4, 7, 9, 12 and 14 (this is not a comprehensive
list). For the [\ion{S}{2}]$\lambda$6716 line, the eigenvectors
associated with the rotation are 3, 4, 7, 9, 12 and 14. The scree
test (see Section \ref{apenda}) suggests for this model, however,
that the relevant eigenvectors are those with $k \le 6$ (see Table
\ref{taba5} and Figure \ref{fig02}).  We have, then, restricted the
relevance of the eigenvectors in the reconstruction procedure up
to this limit.  In these two images we placed {\em artificial slits}
at increasing distances from the jet inlet, at positions labeled
A, B and C. The slits are symmetrically displaced in pairs with
respect to the jet axis. The extracted spectra are plotted in the
right panels (the colors of solid lines used to draw each spectra
are associated with the color of the ``apertures" in the FoV). The
peaks of red/black curves suggest that the jet material at the right
side of the jet axis is receding, while the left side is approaching,
recovering the sense of the jet rotation imposed initially, independent
of the position along the jet axis A, B or C.

\begin{figure}
\centerline{
\includegraphics[scale=0.32]{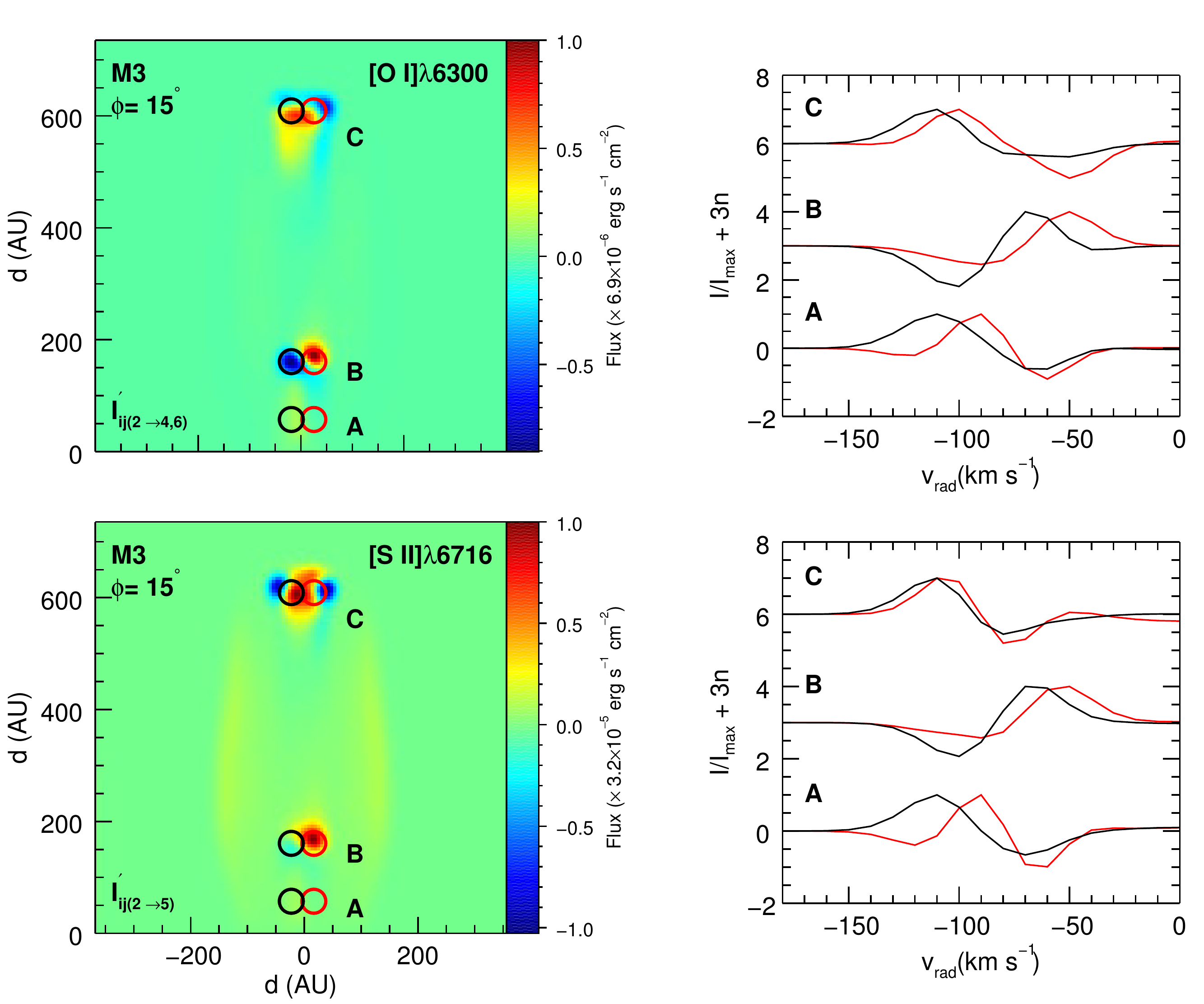}
}
\caption{Reconstructed images (left) and extracted spectra
(right) for Model M3 at $\phi = 15^{\circ}$ and for
[\ion{O}{1}]$\lambda$6300 (top) and [\ion{S}{2}]$\lambda$6716
(bottom) emission lines.  To reconstruct the datacube we have used
the eigenvectors $k = 2$, 3, 4 and 6 for oxygen and $k = 2$ to 5
for the sulfur emission lines, as indicated in the bottom left
part of each image.  Spectra were extracted at symmetrical positions
with respect to the jet axis, labeled A, B and C.  The colors of
the solid lines (right panels) corresponds to the color of the
slits. Distances from the jet inlet are indicated in AU.
Spectra are normalized and we have added to their intensity a constant 3$n$, with $n
=$ 0, 1 and 2 at the positions A, B and C, respectively.
The images are integrated in wavelength.}
\label{fig16} \end{figure}

In Figure \ref{fig14} we show images (left) and spectra (right)
from the reconstructed datacube for the (rotating/precessing) model
M4, in the [\ion{O}{1}]$\lambda$6300 (top) and [\ion{S}{2}]$\lambda$6716
emission lines. For this case, we have evidence for rotation and
precession in several eigenvectors with order higher than 2. We
then keep the eigenvectors from 2 to 8 and from 2 to 7 in the
reconstruction process of the datacube for oxygen and sulfur emission
lines, respectively, obeying the limits suggested by the scree test
(see Table \ref{taba5}).  We have extracted spectra in pairs that
do not follow a straight line along the $y-$axis but instead, we
place the slits along a ``suggested" jet symmetry axis (see Figure
\ref{fig14}). We see in the spectra that there is evidence for the
precession: radial velocities peaks for slit pairs changes from
positive to negative values from B to C, indicating that jet material
in B is receding while in C it is approaching, as expected.  There
is also an indication for rotation, since the red curves peaks are
always redshifted with respect to the black curve, as we should
expect considering the sense of the jet rotation\footnote{We have
also placed the slits parallel to the $y$-axis, and take into account
the lateral shifts in $x$ in order to ``follow" the jet precession.
The results are qualitatively similar to that presented in Figure
\ref{fig14}, the only difference being a small shift in the radial
velocity at position A.}.

\subsection{Inclined systems}

In Figure \ref{fig15} we show the images  (left) and spectra
(right) for the (precessing) model M2 at $\phi = 15^{\circ}$ (two
upper panels) and at $\phi = 45^{\circ}$ (two bottom panels), for
different emission lines, as indicated in each figure. Each
image/spectrum has been obtained from a reconstructed datacube,
using those eigenvectors that indicate the presence of the precession
in each case: $k = 2$, 3, 4 and 6 for the $\phi = 15^{\circ}$ (first
and second rows); $2 \le k \le 7$ for the oxygen line at $\phi =
45^{\circ}$ (third row of panels) and $2 \le k \le 5$ for sulfur
line at $\phi = 45^{\circ}$ (bottom panel).  We can see that the
peaks of the emission lines indicate the presence of precession,
and that it is consistent with the pattern of the precession obtained
for model M2 (for $\phi = 0^{\circ}$; left panel in Figure \ref{fig12}).
As we increase the inclination angle, the amplitude of the variation
of the radial velocity increases (bottom panels).  The scenario is
independent of the emission line, although maps from [O I] and [S
II] may differ considerably.

\begin{figure}
\centerline{
\includegraphics[scale=0.32]{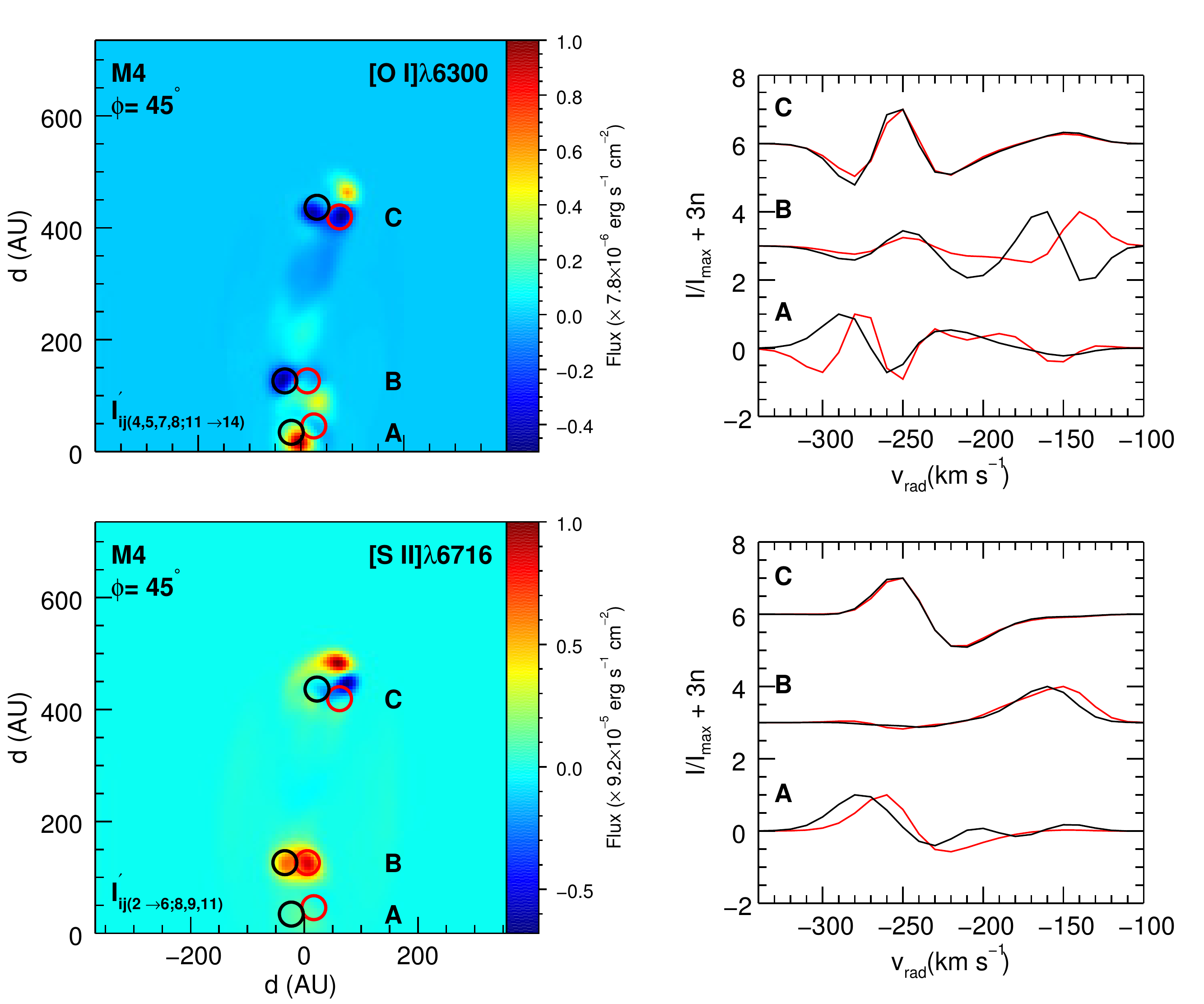}
}
\caption{Images (left) and spectra (right) for model M4 at
$\phi = 45^{\circ}$, for [\ion{O}{1}]$\lambda$6300 (top) and
[\ion{S}{2}]$\lambda$6716 (bottom) emission lines.  In the
reconstruction process of the datacube we have used eigenvectors
$k = 4$, 5, 7, 8, 11, 12, 13 and 14 for the oxygen emission line
and $k = 2$ to 6, and $k = 8$, 9 and 11 for the sulfur emission
line, as indicated in the bottom left part of each image.  Spectra
were extracted in pairs, along and across the jet length, at different
positions labeled A, B and C. The colors of the solid lines (right
panels) corresponds to the color of the slits. Distances from the
jet inlet are indicated in AU in images.  Spectra are normalized
and we have added to their intensity a constant 3$n$, with $n =$ 0, 1 and 2
at the positions A, B and C, respectively.
The images are integrated in wavelength.}\label{fig17}
\end{figure}

Eigenvectors 2, 3, 4 and 6 have been used to reconstruct the
datacube for the (rotating) M3 model, at $\phi = 15^{\circ}$ for
the [O I]$\lambda$6300 emission line, while eigenvectors from 2 to
5 were used to reconstruct the datacube around the sulfur line for
the same model and inclination angle. In Figure \ref{fig16} we show,
as in the previous figures, the reconstructed image and several
spectra, extracted along and across the jet axis.  As in the case
of $\phi=0^{\circ}$ (see Figure \ref{fig13}), spectra from red slits
(right side of the jet axis) are always redshifted in comparison
with spectra from black slits (positioned on the left side of the
jet axis), suggesting that the jet is rotating and recovering the
initial rotating sense. These peaks occur, however, at different
radial velocities. While in A and C the spectra peaks at $v_{rad}
\sim -100$ km s$^{-1}$, in B they peaks at $v_{rad} \sim -60$ km
s$^{-1}$. This difference may be due to the presence of an internal
working surface in B. Another potential source for this difference
is the fact that jet velocity is time variable (see equation
\ref{jetvel}), and this lower (absolute) value for the radial
velocity is consistent with the one expected when the jet velocity
is in its minimum, $v_{j,min} = 200$ km s$^{-1}$, when projected
at $\phi = 15^{\circ}$.

In the more complex case in which we have rotation and precession,
as in model M4, we can see the same effect (in all the emission
lines). In Figure \ref{fig17} we show images (left) and spectra
(right) for the model M4, for [\ion{O}{1}]$\lambda$6300 (top) and
[\ion{S}{2}]$\lambda$6716 emission lines.  The eigenvectors used
in the reconstruction process are indicated in the bottom-left
corner of each figure. The pattern observed in the spectra is
consistent with a rotating and velocity variable jet. However, the
spectra are substantially more fuzzy in comparison with those
obtained for this same model and emission lines at small inclination
angles (see Figure \ref{fig14}).

\begin{figure}
\centerline{
\includegraphics[scale=0.32]{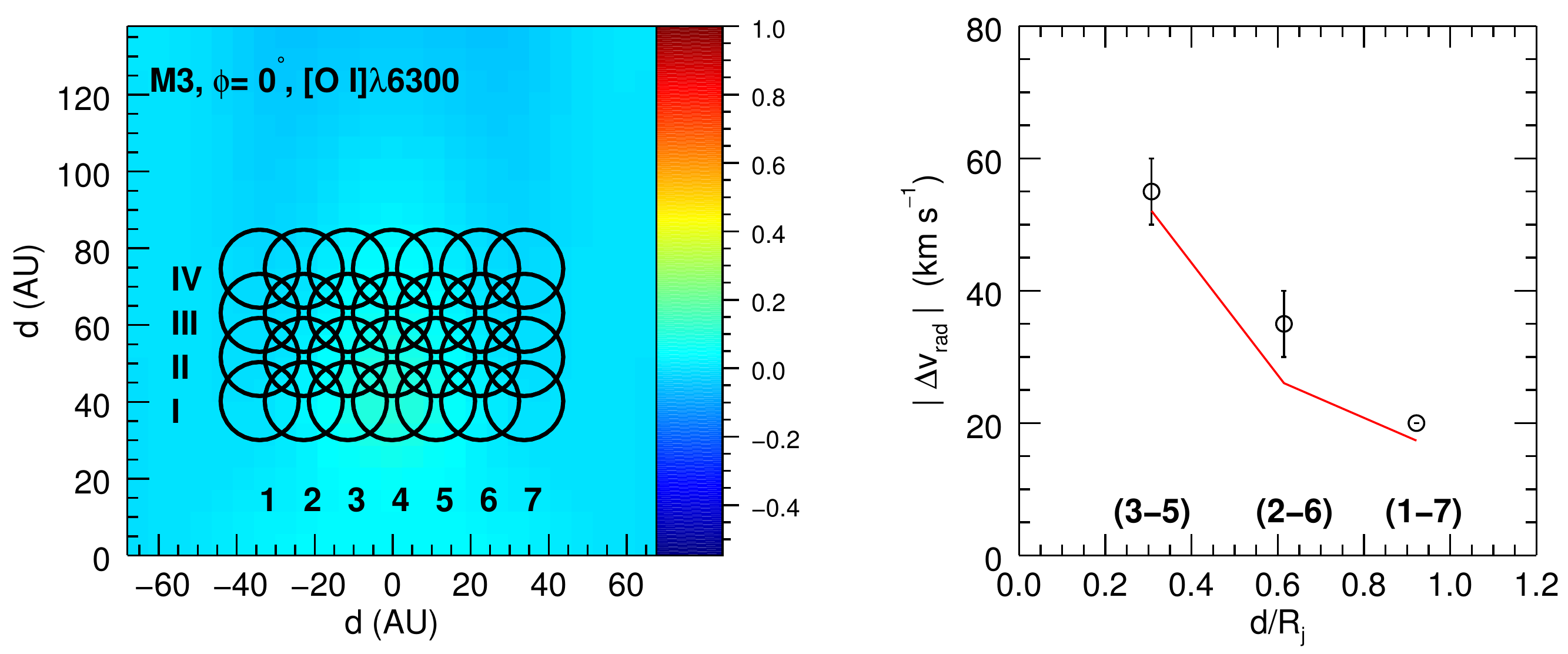}
}
\centerline{
\includegraphics[scale=0.32]{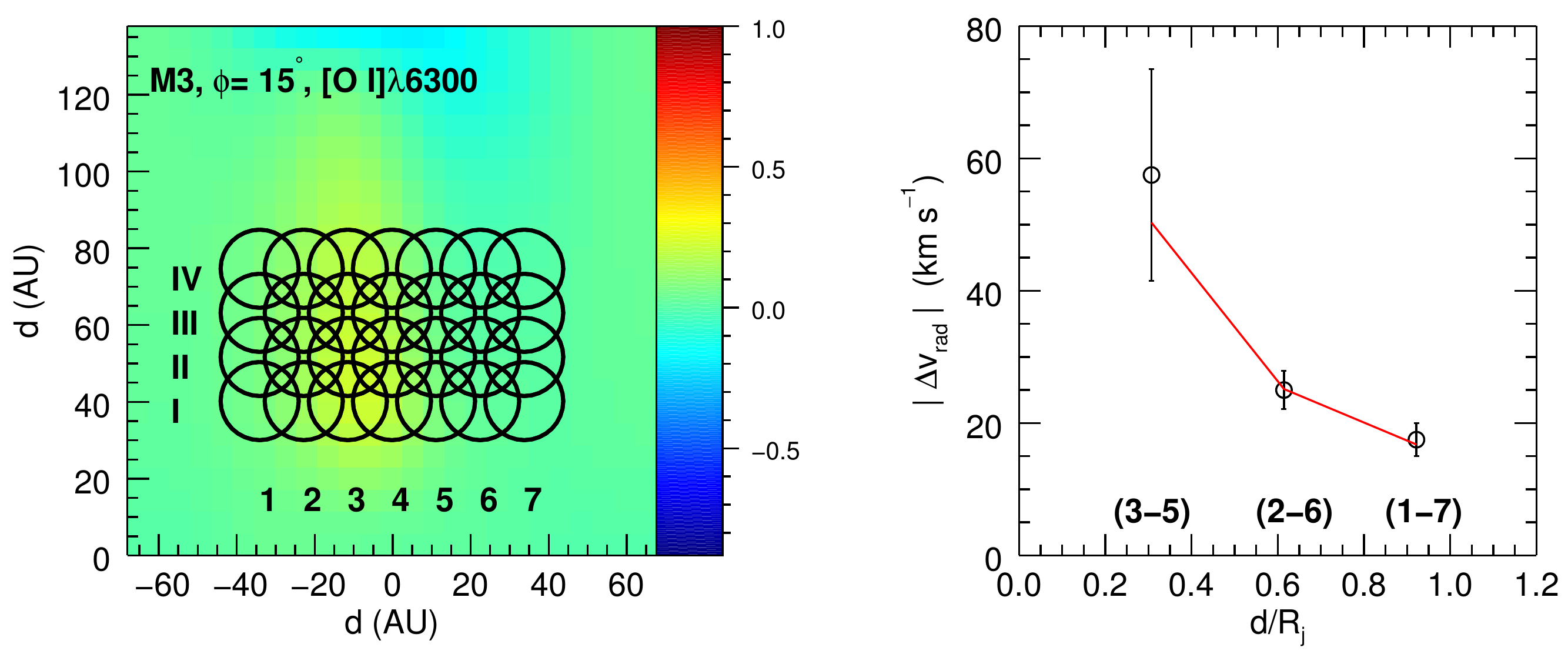}
}
\caption{Array of slits superimposed on the reconstructed image, zoomed at
the jet inlet (left panels). The array of 4$\times$7 slits disposed
across and along the jet length is equivalent to that of
\cite{cerq06}. The open circles in the right panels are the mean
values of the radial velocity shift from symmetrically disposed slits
(3-5), (2-6) and (1-7), considering the four different regions (from
I to IV).  The error bars indicate the dispersion around the mean
value. The red solid line is obtained calculating analytically
$\Delta v_{rad} = 2\times v_{rad}$, where $v_{rad} = 8 {\rm
cos}\phi \cdot R_j/R$. In the top panels we have the result for
M3 model at $\phi = 0^{\circ}$ and in the bottom panel, $\phi = 15^{\circ}$.
In all cases, the emission line is the [\ion{O}{1}]$\lambda$6300.
The images are integrated in wavelength.}\label{fig18}
\end{figure}

In Figure \ref{fig18} we attempt to recover the initially imposed
rotational profile using the reconstructed datacube. We build
an array of 4$\times$7 slits, or four regions (I to IV) at increasing
distances from the jet inlet and seven positions (1 to 7) across
the jet axis.  The size of the slits and their distribution follow
those of \cite{cerq06}. For each extracted profile we take the
radial velocity of the peak, and calculate the difference (i.e.,
the radial velocity shift) for each pair of slits displaced
symmetrically with respect to the jet axis: (3-5), (2-6) and (1-7).
Each region (from I to IV; see Figure \ref{fig18}) contributes to
one radial velocity shift for each slit pair, and the mean value
for a such radial velocity shift is plotted as a function of the
distance from the jet axis in the right panel of the Figure (open
circles; the error bar gives the dispersion around the obtained
mean value). In order to compare with the initially imposed rotational
profile, we calculate for each radial distance the expected (at
$t=0$) radial velocity shift (for $R > 0.2R_j$): $\vert \Delta
v_{rad} \vert = 2 \times 8\cdot{\rm cos}\phi (R_j/R)$, where the
${\rm cos}\phi$ accounts for the inclination (red, solid curve in
Figure \ref{fig18}).  Although the match is reasonable for both
inclinations ($\phi = 0^{\circ}$ and $\phi = 15^{\circ}$), we can
not reproduce the original profile for the case of $\phi = 45^{\circ}$.
The method applied here fails also to recover the original profile
in the case where rotating and precession are combined, as in model
M4.

\section{Conclusions}
\label{sec:con}

The PCA technique has been extensively applied in the astrophysical
context \citep[e.g.,][]{stei09, ricc11, malyshev12}.  PCA is
used to identify quickly in large datasets patterns and correlations
that otherwise would not be revealed. These patterns can then be
interpreted in terms of arising from physically uncorrelated phenomena
present in the system, that thanks to PCA can be separated, identified
and studied in detail. PCA offers also an efficient way to remove
unwanted features from the data, as noise or instrumental fingerprints,
and to compress and transmit only the relevant information present
in a large dataset. We describe briefly the method in Section
\ref{sec:2}.  The PCA method can be fruitfully applied to the
analysis of spectro-imaging data like those provided by observations
of diffuse targets with Integral Field Spectroscopy as shown in
\cite{stei09, ricc11}, who apply PCA to observations of galaxies.
In general, the interpretation of the observational data based on
PCA tomograms is highly improved with the help of three-dimensional
numerical experiments. This has been shown, for example, in
\cite{heyer97}, \cite{brunt09} and \cite{carr10}.  In this paper
we show the potential applicability of the PCA to the study of HH
jets by using the results of three-dimensional numerical simulations
of rotating, precessing jets.

We apply the PCA to the synthetic images produced from the
simulations, where we control the presence of known physical
phenomena. We use the PCA to test if signatures of these phenomena
can be isolated in the data, and produce a benchmark of PCA tomograms,
with which PCA-processed real IFU observations of jets can be
compared. In this work, we started generating from the simulations
spectro-imaging datacubes around the wavelength of relevant emission
lines for four different jet models. These are an intermittent
jet model (M1 model), produced by imposing a periodic variation
in velocity at the base of the jet, and models also including
precession (M2 model), rotation (M3 model), or both
rotation and precession (M4 model). For each of these
models we applied the PCA decomposition to find the modes of variance
with respect to the average intensity, and we discussed the tomograms
corresponding to the modes of highest variance ($k \le 4$).

The tomograms, combined with their respective eigenspectra,
give important information on the processes occurring in the jet.
The tomogram for $k = 1$ shows where the bulk of the emission varies
the most, and it is indicative of the position of internal bow
shocks generated by the imposed jet pulsation. The tomogram for $k$
= 2, in the precessing models, presents longitudinal gradients in
intensity that we have identified with the precession. 
The slow variation of the intensity along the jet beam correlates
with the jet orientation angle with respect to the observer, with
abrupt changes in sign at the point where the jet inclination varies
from directed toward the observer to away from the observer. The
second tomogram therefore, can be used to determine if precession
is present in the jet, and to quantify the precession angle. This
feature is not present if precession is not included in the model.

Tomograms 2 and 4 (for [O I]$\lambda$6300 emission line), or
3 and 4 (for [S II]$\lambda$6716 emission line) can be
identified with the rotation of the jet around its axis in a rotating
jet. We can use the reconstructed datacube to retrieve the initially
imposed rotation profile. In a more general case, in which both
precession and rotation are included, the signature for both is
mixed in a given tomogram, and usually the signature is present in
several different eigenvectors. For this reason, we are not able
to retrieve the imposed rotation profile.

Despite these promising results, it is worth to mention that there
are some caveats to be kept in mind. In order to fully explore this
method in real data, some further effects should be carefully
treated. The datacube should be as cleaned as possible and free of
systematic effects. Slit positions should be placed along the jet
beam, following the changes in the jet axis position due
to changes in the jet propagation direction.
The inclination of the jet with respect to the observer plays also
an important role.  It increases the number of relevant eigenvectors
that must be taken into account. It also turns the analysis more
complex, in the sense that several radial velocity gradients appears
in a given tomogram, making its interpretation more difficult. As
the inclination angle increases, emission coming from the jet basis
becomes more and more mixed with those from the internal working
surface, which makes prohibitive the determination of the rotation
profile. Also, the noise is expected to dominate the contribution
to the variance in the dataset at high order eigenvectors
\citep[see][]{heyer97}, and real data will only give access to the
first few eigenvectors.

Nevertheless, the PCA remains a powerful diagnostic technique
to analyze the structure seen in observation of jets in typical
emission lines. We have shown that PCA can help disentangling
precession and rotation in protostellar jets when the jet axis lies
close to the plane of the sky. In a future work we will apply the
PCA technique to the analysis of the HH 111 jet observed with the
Gemini Integral Field Unit spectrograph.


\acknowledgments

This paper was greatly improved thanks to the referee, who has
raised several and important questions during the review process.
We are deeply indebted for his/her contribution to this work. We
are thankful to T. Ricci and J.  Steiner for enlightening discussions
about the PCA technique and its applications to a datacube. AHC and
MJV thank CNPq/CAPES for financial support using the PROCAD project
552236/2011-0, and CAPES/CNPq Science without Borders program, under
grants 2168/13-8 and 2565/13-7, respectively.  We would like
also to thanks Jerome Bouvier and Jonathan Ferreira for their
kind hospitality during our stay in Grenoble (UJF/IPAG). FDC and
JRI thank the DGAPA-PAPIIT-UNAM (grants IA101413, IA103315).


{}


\appendix

\section{Eigenvalues}
\label{APP2}

In Table \ref{taba1} we show the eigenvalues for the first ten
eigenvectors, in terms of \% of the variance, for the four models
and the three inclinations. The first column displays the eigenvector
$E_k$. The second column displays the inclination angle $\phi$. The
associated eigenvalues are show for each model (M1 to M4), for the
computed emission lines [\ion{O}{1}]$\lambda$6300 (columns 3 to 6) and
[\ion{S}{2}]$\lambda$6716 (columns 7 to 8).

The number of relevant eigenvectors, following the criteria of the
scree test, is given in Table \ref{taba5}. The scree test is
a visual test used to obtain $k_{\rm max}$, which is the maximum
number of eigenvectors that we might consider in the analysis of
the dataset. In this paper, we have defined a threshold for the
eigenvalue, $\Lambda_{\rm threshold} = 10^{-3}$ \%, since we note
that most of the eigenvalues will be intercepted by a line defined
by such a value in a {\em scree plot} (that is, an eigenvalue versus
eigenvector's order plot; see Figure \ref{fig02}). This means
that the rate of change in the eigenvalue as a function of the
eigenvector's order is almost constant (and equal to zero, in
our case) for $k_{\rm max} \le k \le k_{50}$, while it can
varies a lot in the $k \le k_{\rm max}$ range. With this in
mind, we can look at Figure \ref{fig02} in order to find this
point, that can easily recognized as the first point (from left
to right) intercepted by the threshold eigenvalue function
(black dashed line).

\begin{table}[htb]
\begin{center}
\caption{Eigenvalues, in \% of the variance, of the first ten
eigenvectors.}
\label{taba1}
\bigskip
\begin{tabular}{c | c | c c c c | c c c c}
\tableline
\tableline
Eigenvector    & $\phi$ ($^{\circ}$) & \multicolumn{8}{c}{Eigenvalue (\%)}   \\
               &                     & \multicolumn{4}{c}{[O I]$\lambda$6300}  & \multicolumn{4}{c}{[S II]$\lambda$6716}  \\
$E_k$          &                     & M1      &  M2      &  M3     &  M4      & M1       &  M2     &  M3     &  M4       \\
\tableline
\tableline
               & 0                   & 99.7039 & 90.3804  & 88.3875 & 75.5315  & 99.7225  & 97.4961 & 92.6859 & 84.1669 \\
$E_{1}$        & 15                  & 89.8065 & 79.8084  & 71.6271 & 53.2939  & 97.2702  & 96.4368 & 85.6025 & 79.3238 \\
               & 45                  & 83.886  & 76.2575  & 62.3434 & 49.7390  & 94.6670  & 90.2402 & 80.3527 & 70.0142 \\
\tableline
               & 0                   & 0.2597  & 8.6249   & 6.7645  & 18.007   & 0.2655   & 1.8583  & 3.9785  & 12.7514 \\
$E_{2}$        & 15                  & 8.8301  & 18.1922  & 20.7925 & 33.2961  & 2.3652   & 2.5940  & 10.0743 & 14.5136 \\
               & 45                  & 10.7070 & 16.6092  & 22.8332 & 33.4023  & 3.6990   & 6.3589  & 14.4443 & 18.6909 \\
\tableline
               & 0                   & 0.0339  & 0.8386   & 4.5483  & 5.6351   & 0.0108   & 0.5996  & 3.0637  & 2.1859  \\
$E_{3}$        & 15                  & 0.8766  & 1.3354   & 5.1462  & 8.9036   & 0.2421   & 0.7740  & 3.5704  & 4.6916  \\
               & 45                  & 3.1367  & 4.8132   & 10.2354 & 9.4788   & 0.9499   & 2.2272  & 2.9921  & 6.8108  \\
\tableline
               & 0                   & 0.0020  & 0.1035   & 0.2100  & 0.6425   & 0.0009   & 0.0307  & 0.2266  & 0.7387  \\
$E_{4}$        & 15                  & 0.2799  & 0.3317   & 1.5756  & 3.3236   & 0.0708   & 0.0977  & 0.4867  & 1.0325  \\
               & 45                  & 0.9187  & 1.1429   & 1.9413  & 3.6313   & 0.3297   & 0.7967  & 1.513   & 2.8013  \\
\tableline
               & 0                   & 0.0002  & 0.0424   & 0.0822  & 0.1289   & O(-5)    & 0.0107  & 0.0414  & 0.1194  \\
$E_{5}$        & 15                  & 0.1300  & 0.2048   & 0.4929  & 0.5452   & 0.0298   & 0.0575  & 0.1233  & 0.2162  \\
               & 45                  & 0.7214  & 0.4886   & 1.2753  & 1.4002   & 0.1827   & 0.1442  & 0.2512  & 0.8761  \\
\tableline
               & 0                   & 0.0002  & 0.0068   & 0.0031  & 0.0326   & O(-5)    & 0.0024  & 0.0016  & 0.0232  \\
$E_{6}$        & 15                  & 0.0528  & 0.0699   & 0.2246  & 0.3954   & 0.0139   & 0.0201  & 0.0978  & 0.1262  \\
               & 45                  & 0.3256  & 0.4470   & 0.8301  & 1.0521   & 0.0921   & 0.1339  & 0.2133  & 0.2717  \\
\tableline
               & 0                   & O(-5)\tablenotemark{1}   & 0.0020   & 0.0029  & 0.0166   & O(-5)    & 0.0013  & 0.0015  & 0.0114  \\
$E_{7}$        & 15                  & 0.0133  & 0.0364   & 0.0755  & 0.1059   & 0.0041   & 0.0109  & 0.0209  & 0.0468  \\
               & 45                  & 0.1595  & 0.0909   & 0.2250  & 0.6748   & 0.0299   & 0.0430  & 0.1326  & 0.2618  \\
\tableline
               & 0                   & O(-6)   & 0.0004   & 0.0008  & 0.0027   & O(-6)    & 0.0003  & 0.0003  & 0.0015  \\
$E_{8}$        & 15                  & 0.0062  & 0.0120   & 0.0411  & 0.0811   & 0.0021   & 0.0052  & 0.0152  & 0.0297  \\
               & 45                  & 0.0732  & 0.0609   & 0.1165  & 0.2478   & 0.0213   & 0.0207  & 0.0374  & 0.1205  \\
\tableline
               & 0                   & O(-7)   & 0.0003   & 0.0001  & 0.0013   & O(-7)    & 0.0001  & (-5)    & 0.0007  \\
$E_{9}$        & 15                  & 0.0019  & 0.0045   & 0.0128  & 0.0318   & 0.0007   & 0.0017  & 0.0050  & 0.0101  \\
               & 45                  & 0.0396  & 0.0507   & 0.0765  & 0.1441   & 0.0111   & 0.0153  & 0.0253  & 0.0580  \\
\tableline
               & 0                   & O(-7)   & 0.0001   & O(-5)   & 0.0003   & O(-7)    & O(-5)   & O(-5)   & 0.0002  \\
$E_{10}$       & 15                  & 0.0014  & 0.0017   & 0.0061  & 0.0106   & 0.0003   & 0.0009  & 0.0019  & 0.0037  \\
               & 45                  & 0.0163  & 0.0220   & 0.0536  & 0.0942   & 0.0098   & 0.0111  & 0.0140  & 0.0407  \\
\tableline
\end{tabular}
\tablenotetext{1}{The order of magnitude will be indicated instead of eigenvalue 
if it is less than 10$^{-4}$. In this particular case, O(-5) implies that 
$\Lambda \sim 10^{-5}$\%.}
\end{center}
\end{table}

\begin{table}[htb]
\begin{center}
\caption{The number of eigenvector that should be retained in the data
considering the scree test.}
\label{taba5}
\bigskip
\begin{tabular}{c c c c c c }
\tableline
\tableline
Model  &                             \multicolumn{4}{c}{$k_{{\rm scree~test}}$}                         & $\phi$ ($^{\circ}$) \\
       & [\ion{O}{1}]$\lambda$6300 &  [\ion{O}{1}]$\lambda$6363  &  [\ion{S}{2}]$\lambda$6716 &  [\ion{S}{2}]$\lambda$6731 &         \\
\tableline
\tableline

       &   4   &    4   &    4    & 3    &  0       \\
M1     &   7   &    7   &    7    & 4    &  15      \\
       &  11   &   11   &   11    & 9    &  45      \\
\tableline
       &   6   &    6   &    5    & 4    &  0       \\
M2     &   8   &    8   &    7    & 6    &  15      \\
       &  11   &   11   &   11    & 9    &  45      \\
\tableline
       &   6   &    6   &    6    & 6    &  0       \\
M3     &   9   &    9   &    9    & 9    &  15      \\
       &  13   &   13   &   12    & 9    &  45      \\
\tableline
       &   8   &    8   &    7    & 7    &  0       \\
M4     &  11   &   11   &    9    & 8    &  15      \\
       &  14   &   14   &   13    &11    &  45      \\
\tableline
\end{tabular}
\end{center}
\end{table}

\section{The effect of noise and the enhancement factor}
\label{apendb}

\subsection{The effects of a noise background level in the data}

All figures in this paper have been obtained after a convolution
procedure applied in each velocity channel map (VCM) in the real
space ($i$, $j$, $k$) with a Gaussian profile, as explained in
Section \ref{sec:num}.  Compared with raw, non-convolved datacube,
the net effect of the convolution process is to alter the value of
the variance in the PCA treated datacubes. The first three columns
of Table \ref{tabb1} illustrate it, for a given model as an example
(model M3, considering the [\ion{O}{1}]$\lambda$6300 emission line
observed at an angle $\phi = 0^{\circ}$).

\begin{figure}
\centerline{
\includegraphics[scale=0.40]{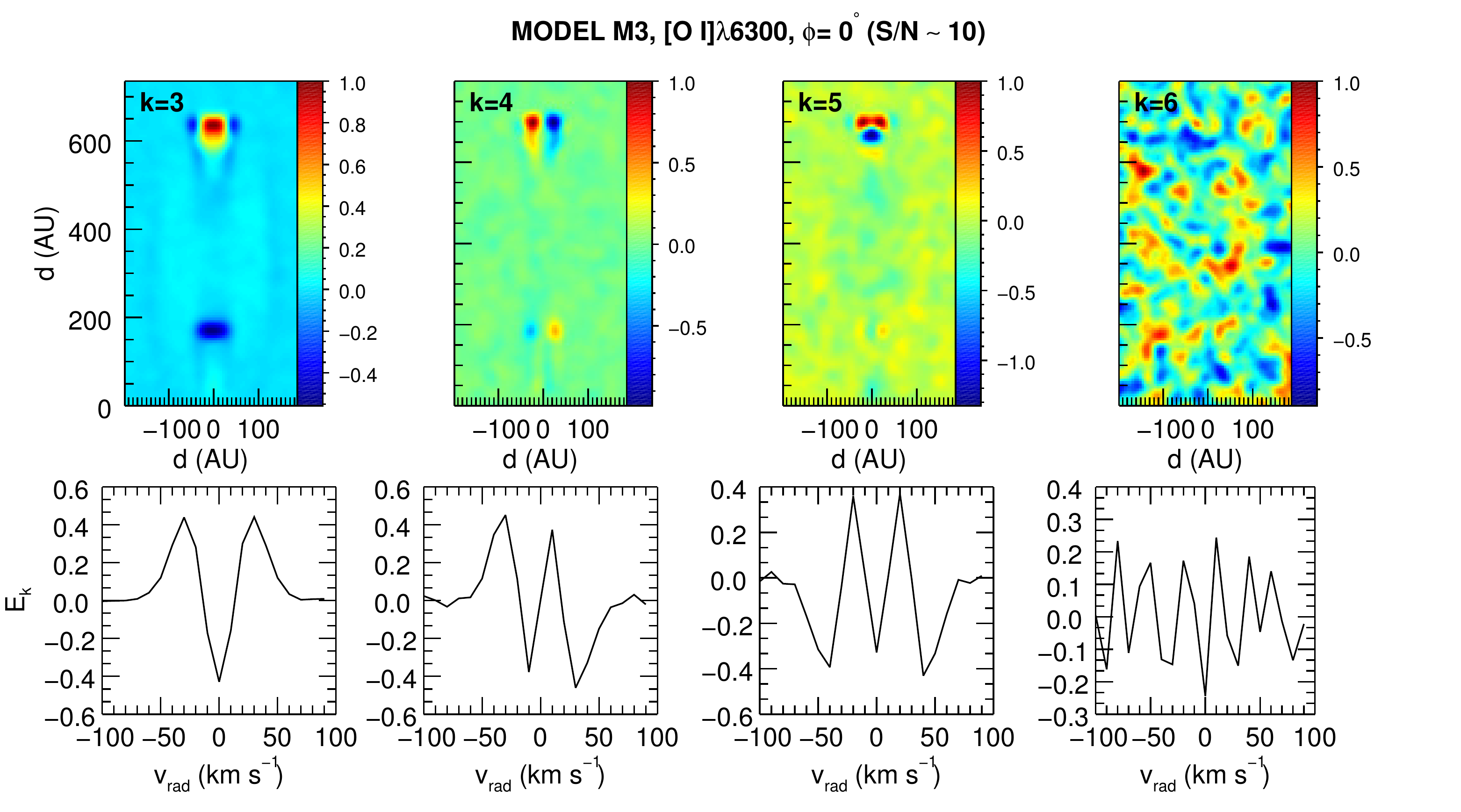}
}
\centerline{
\includegraphics[scale=0.40]{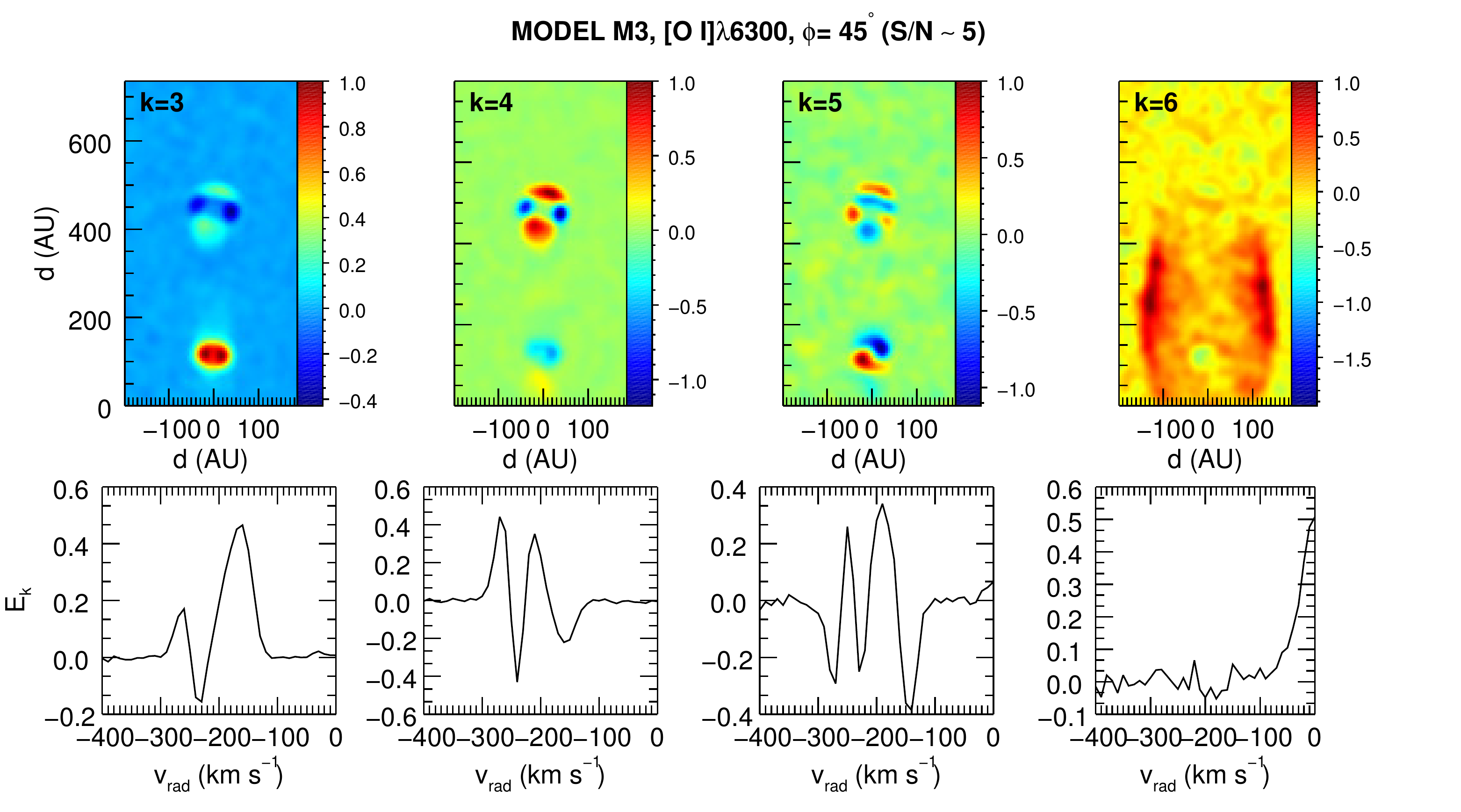}
}
\caption{Sequence from the third (left) to the sixth (right)
tomogram and eigenspectra for M3 model and [\ion{O}{1}]$\lambda$6300
emission line with a background noise.  The two topmost panels are
for S/N $\sim 10$ and $\phi = 0^{\circ}$ and the bottommost ones are
for S/N $\sim 5$ and $\phi = 45^{\circ}$. The contribution to the
variance is mainly dominated by noise fluctuation in high order
eigenvectors.}\label{fig19} 
\end{figure}

We have also degraded the data including in the simulation two
effects: the presence of a random, background level of noise and
considering also a broadened line profile\footnote{In the procedure
to build the VCM's we have artificially increased the FWHM of the
Gaussian considered to spread the line intensity in radial velocity
by a factor of $\sim$ 2, in order to worsen the spectral resolution.}.
In the fourth column of the Table \ref{tabb1} we show the eigenvalues
for the first ten eigenvectors of such a model (again, considering
the model M3, at $\phi = 0^{\circ}$ and the [\ion{O}{1}]$\lambda$6300
emission line). In Figure  \ref{fig19} we present the
tomograms/eigenspectra from $k =3$ to $k = 6$ (from left to right,
respectively). Different levels of signal to noise and inclination
angles have been used: S/N $\sim$ 10 and $\phi = 0^{\circ}$ (the
first two rows of panels) and S/N $\sim$ 5  and $\phi = 45^{\circ}$
(the last two rows of panels).

The higher order eigenvectors that are dominated by noise can
be discarded in the reconstruction process, which constitutes on the
other hand as a powerful filtering process for real data
\citep{stei09,cerq15}.

\subsection{The enhancement factor}
\label{enhan}

The tomograms represent an image in a new system of uncorrelated
coordinates. We can use the tomograms, together with their eigenspectra,
to identify different physical properties present in the data. We
can then reconstruct the datacube, omitting some undesirable property
and putting in evidence others. In particular, we want to
identify and discriminate from the tomograms calculated so far, the
presence of precession and rotation. In the Section
\ref{apenda} we have shown several tomograms and eigenspectra.

Once identified, a given property ``A" can be emphasized using the
equation (\ref{steiner8}):

\begin{equation}\label{steiner8b}
{\bf I}^{\prime}_{\beta \lambda}(A)={\bf T}_{\beta k} (k_A)
\cdot[{\bf E}_{\lambda k}(k_A)]^T \,,
\end{equation}

\noindent to reconstruct the datacube. In this process, we
keep the eigenvectors $E_{k_A}$ already identified and associated
with the property ``A", in the characteristic matrix. This can be
done following the prescription in \cite{stei09}: we can multiply
the columns of the characteristic matrix by a factor, $\Gamma_k$,
which can be 0 or 1. It acts suppressing a given eigenvector (or
$\Gamma_k = 0$) and keeping others ($\Gamma_k = 1$) in the
characteristic matrix. The end product is a matrix that has the
desired property only (feature enhancement).

In the Section \ref{apenda} we have shown the tomograms
associated with a specific eigenvector, and extracted information
about the properties they contain. In Section \ref{sec:res}, the
reconstruction process has always been guided by these findings,
where the tomograms and eigenspectra have been analyzed together
in the search for the precession and/or rotation signatures.


\begin{table}[htb]
\begin{center}
\caption{Eigenvalues, in \% of the variance, for the first 10
eigenvectors for model M3 (rotating model), for the raw, convolved
and noisy data. The [\ion{O}{1}]$\lambda$6300 line has been considered
for $\phi=0^{\circ}$. }
\label{tabb1}
\bigskip
\begin{tabular}{c c c c c c }
\tableline
\tableline
Eigenvector  &    \multicolumn{3}{c}{Eigenvalue (\%)} \\
E$_\mathrm{k}$ & raw data &  + convolution  & + noise \\
\tableline
\tableline

$E_{1}$     &   99.190     & 88.3875 &  99.182  \\
$E_{2}$     &   0.7523     & 6.7645  &  0.481  \\
$E_{3}$     &   0.0572     & 4.5483  &  0.057  \\
$E_{4}$     &   0.0033     & 0.2100  & 0.009 \\ 
$E_{5}$     &   O(-5)      & 0.0822  & 0.009 \\
$E_{6}$     &   O(-6)      &  0.0031 & 0.009  \\
$E_{7}$     &   O(-7)      &  0.0029 & 0.008 \\
$E_{8}$     &   O(-8)      & 0.0008  &  0.008 \\
$E_{9}$     &   O(-9)      & 0.0001  &  0.008 \\
$E_{10}$    &  O(-10)      & O(-5)   &  0.008 \\
\tableline
\end{tabular}
\end{center}
\end{table}


\begin{thebibliography}{}

\bibitem[Anderson et al.(2003)]{ande03} Anderson, J.M., Li, Z.-Y.,
Krasnopolsky, R. et al. 2003, \apjl, 590, L107

\bibitem[Bacciotti (2004)]{bacc04b} Bacciotti, F. 2004, \apss, 293,
37

\bibitem[Bacciotti et al.(2002)]{bacc02} Bacciotti, F., Ray, T.P.,
Mundt, R. et al. 2002, \apj, 576, 222

\bibitem[Bacciotti et al.(2003)]{bacc03} Bacciotti, F., Ray, T.P.,
Eisl\"offel, J. et al. 2003, \apss, 287, 3

\bibitem[Bacciotti et al.(2004)]{bacc04} Bacciotti, F., Ray, T.P.,
Coffey, D. et al. 2004, \apss, 292, 651

\bibitem[Blandford \& Payne(1982)]{blan82} Blandford, R.D., \&
Payne, D.G. 1982, \mnras, 199, 1982

\bibitem[Brunt, Heyer \& Mac Low(2009)]{brunt09} Brunt, C.M., Heyer,
M.H. \& Mac Low, M.-M. 2009, A\&A, 504, 883

\bibitem[Cabrit et al.(2006)]{cabr06} Cabrit, S., Pety, J., Pesenti,
N., et al. 2006, A\&A, 452, 897

\bibitem[Cai et al.(2008)]{cai08} Cai, M.~J., Shang, H., Lin, H.-H.,
\& Shu, F.~H.\ 2008, \apj, 672, 489

\bibitem[Carrol, Frank \& Blackman(2010)]{carr10} Carrol, J.A.,
Frank, A. \& Blackman, E.G.  2010, \apj, 722, 145

\bibitem[Casse \& Ferreira(2000a)]{cass00a} Casse, F., \& Ferreira,
J. 2000a, A\&A, 353, 1115

\bibitem[Casse \& Ferreira(2000b)]{cass00b} Casse, F., \& Ferreira, J. 2000b, A\&A, 361, 1178

\bibitem[Chrysostomou, Lucas \& Hough(2007)]{chry07} Chrysostomou,A.,
Lucas, P.W., \& Hough, J.H. 2007, Nature, 450, 71

\bibitem[Cerqueira et al.(2015)]{cerq15} Cerqueira, A.H., Vasconcelos, M.J.,
Raga, A.C. et al.  2015, AJ, 149, 98

\bibitem[Cerqueira et al.(2006)]{cerq06} Cerqueira, A.H.,  Vel\'azquez,
P.F., Raga, A.C. et al. 2006, A\&A, 448, 2006

\bibitem[Codella et al.(2007)]{code07} Codella, C., Cabrit, S.,
Gueth, F., et al. 2007, A\&A, 462, L53

\bibitem[Coffey et al.(2004)]{coff04} Coffey, D., Bacciotti, F.,
Woitas, J., et al. 2004, \apj, 604, 758

\bibitem[Coffey et al.(2007)]{coff07} Coffey, D., Bacciotti, F.,
Ray, T.P., et al. 2007, \apj, 663, 350

\bibitem[Coffey, Bacciotti \& Podio (2008)]{coff08} Coffey, D.,
Bacciotti, F., \& Podio, L.  2008, \apj, 689, 1112

\bibitem[Coffey et al.(2011)]{coff11} Coffey, D., Bacciotti, F.,
Chrysostomou, A., et al. 2011, A\&A, 526, 40

\bibitem[Coffey et al.(2012)]{coff12} Coffey, D., Rigliaco, E.,
Bacciotti, F. et al. 2012, \apj, 749, 139

\bibitem[Coffey et al.(2015)]{coff15} Coffey, D., Dougados, C.,
Cabrit, S. et al. 2015, ApJ, 804, 2

\bibitem[Correia et al.(2009)]{corr09} Correia, S., Zinnecker, H.,
Ridgway, S.T., et al 2009, A\&A, 505, 673

\bibitem[Davis et al.(2000)]{davi00} Davis, C.J., Berndsen, A.,
Smith, M.D., et al. 2000, \mnras, 314, 241

\bibitem[Dougados et al.(2003)]{doug03} Dougados, C., Cabrit, S.,
Lopez-Martin, L. et al. 2003, \apss, 287, 135

\bibitem[Dougados et al.(2004)]{doug04} Dougados, C., Cabrit, S.,
Ferreira, J. et al. 2004, \apss, 292, 643

\bibitem[Fendt(2011)]{fend11} Fendt, C. 2011, \apj, 737, 43

\bibitem[Ferreira(1997)]{ferr97} Ferreira, J. 1997, A\&A, 319, 340

\bibitem[Ferreira(2008)]{ferr08} Ferreira, J. 2008, New Astronomy
Review, 52, 2008

\bibitem[Ferreira, Dougados \& Cabrit(2006)]{ferr06} Ferreira, J.,
Dougados, C., \& Cabrit, S. 2006, A\&A, 453, 785

\bibitem[Hartmman(2008)]{hart08} Hartmman, L. 2008, in {\it Accretion
Process in Star Formation}, Cambridge University Press (2nd Edition)

\bibitem[Heyer \& Schloerb(1997)]{heyer97} Heyer, M.H. \& Schloerb, F.P.
1997, \apj, 475, 173

\bibitem[Kenyon, Dobrzycka \& Hartmann(1994)]{ken94} Kenyon, S.J.,
Dobrzycka, D., \& Hartmann, L. 1994, \aj, 108, 1872

\bibitem[Krasnopolsky, Li \& Blandford(2003)]{kras03} Krasnopolsky,
R., Li, Z.-Y., \& Blandford, R.D. 2003, \apj, 595, 631

\bibitem[Malyshev(2012)]{malyshev12} Malyshev, D.\ 2012, arXiv:1202.1034

\bibitem[Melnikov et al.(2009)]{meln09} Melnikov, S.Yu, Eisl\"offel,
J., Bacciotti, F., et al. 2009, A\&A, 506, 763

\bibitem[Menezes et al.(2014)]{men14} Menezes, R.~B., Steiner,
J.~E., \& Ricci, T.~V.\ 2014, \apjl, 796, LL13

\bibitem[Mundt et al.(1990)]{mund90} Mundt, R., Buehrke, T., Solf,
J., Ray, T.~P., \& Raga, A.~C.\ 1990, \aap, 232, 37

\bibitem[Pesenti et al.(2004)]{pese04} Pesenti, N., Dougados, N.,
Cabrit, S. et al. 2004, A\&A, 416, L9

\bibitem[Pety et al.(2006)]{pety06} Pety, J., Gueth, F., Guilloteau,
S., \& Dutrey, A.\ 2006, A\&A, 458, 841

\bibitem[Pudritz(2004)]{pudr04} Pudritz, R.E. 2004, \apss, 292, 471

\bibitem[Pudritz \& Norman(1983)]{pudr83} Pudritz, R.E., \& Norman,
M.L., 1983, \apj, 274, 677

\bibitem[Pudritz \& Norman(1986)]{pudr86} Pudritz, R.E., \& Norman,
M.L., 1986, \apj, 301, 571

\bibitem[Pudritz et al.(2012)]{pudr12} Pudritz, R.E., Hardcastle,
M.J., \& Gabuzda, D.C., 2012, Space Science Reviews, 169, 27

\bibitem[Raga et al.(2000)]{raga00} Raga, A.~C., Navarro-Gonz{\'a}lez,
R., \& Villagr{\'a}n-Muniz, M.\ 2000, RMxAA, 36, 67

\bibitem[Ricci et al.(2011)]{ricc11} Ricci, T., Steiner, J., \&
Menezes, R.B. 2011, \apj, 734, L10

\bibitem[Sauty et al.(2012)]{saut12} Sauty, C., Cayatte, V., Lima,
J.J.G., et al. 2012, \apjl, 759, L1

\bibitem[Smith \& Rosen(2007)]{smit07} Smith, M.D., \& Rosen, A.,
2007, \mnras, 378, 691

\bibitem[Solf \& B\"ohm(1993)]{solf} Solf, J., \& B\"ohm, K.H. 1993,
\apj, 410, L31

\bibitem[Spruit(1996)]{spru96} Spruit, H.C. 1996, in {\it Evolutionary
Processes in Binary Stars} (R.A.M.J. Wijers et al., eds.), pp.
249-286, Kluwer, Dordrecht

\bibitem[Staff et al.(2010)]{staf10} Staff, J.E., Niebergal, B.P.,
Ouyed, R., et al. 2010, \apj, 722, 1325

\bibitem[Steiner et al.(2009)]{stei09} Steiner, J., Menezes, R.B.,
Ricci, T. et al. 2009, \mnras, 396, 788

\bibitem[Testi et al.(2002)]{testi02} Testi, L., Bacciotti, F.,
Sargent, A.I. et al. 2002, A\&A, 394, L31

\bibitem[Vasconcelos et al.(2005)]{vasc05} Vasconcelos, M.J.,
Cerqueira, A.H., Plana, H., Raga, A.C., \& Morisset, C. 2005, \aj,
130, 1707

\end{thebibliography}
\end{document}